\documentclass{article}
\usepackage{amssymb}
\usepackage{bm}

\usepackage{graphicx}
\usepackage{amsmath}
\usepackage{verbatim}

\def\nsection#1{\section{#1}\setcounter{equation}{0}}

\newcommand{\qq}{\begin{eqnarray}}
\newcommand{\qqq}{\end{eqnarray}}
\newcommand{\ee}{{\rm e}}

\newcommand{\HJ}{H\hspace{-0.03cm}J}

\begin{document}
	
\title{Perturbative calculation of quasi-potential in
non-equilibrium diffusions: a mean-field example}

\author{Freddy Bouchet$^{1}$ \and Krzysztof Gaw\c{e}dzki$^{1,2}$\and 
Cesare Nardini$^{1,3,4}$\cr
$^1$\small{\,Laboratoire de Physique de l'Ecole Normale Sup\'erieure 
de Lyon, Universit\'e de Lyon and CNRS,}\\\small{46, All\'ee d'Italie, 
F-69007 Lyon, France}\\
$^2${\,\small chercheur \'em\'erite}\\
$^3$\small{\,SUPA, School of Physics and Astronomy, University of Edinburgh,}\\
{\small Peter Guthrie Tait Road, Edinburgh EH9 3FD, United Kingdom}\\ 
$^4$\small{\,DAMTP, Centre for Mathematical Sciences, University of Cambridge,}\\
\small{Wilberforce Road, Cambridge CB3 0WA, United Kingdom}\\ \\
\small{Corresponding author: cesare.nardini@gmail.com}}
\date{}
\maketitle

\begin{abstract}
In stochastic systems with weak noise, the logarithm of the stationary 
distribution becomes proportional to a large deviation rate function called the 
quasi-potential. The quasi-potential, and its characterization through a variational 
problem, lies at the core of the Freidlin-Wentzell large deviations theory
~\cite{freidlin1984}.
In many interacting particle systems, the particle density is described by fluctuating 
hydrodynamics governed by Macroscopic Fluctuation Theory
~\cite{bertini2014},
which formally fits within Freidlin-Wentzell's framework with a weak noise 
proportional to $1/\sqrt{N}$, where $N$ is the number of particles. The quasi-potential 
then appears as a natural generalization of the equilibrium free energy to 
non-equilibrium particle systems. A key physical and practical issue is to actually 
compute quasi-potentials from their variational characterization for 
non-equilibrium systems for which detailed balance does not hold. We discuss how 
to perform such a computation perturbatively in an external parameter $\lambda$, 
starting from a known quasi-potential for $\lambda=0$. In a general setup, 
explicit iterative formulae for all terms of the power-series expansion of 
the quasi-potential are given for the first time. The key point is a proof 
of solvability conditions that assure the existence of the perturbation expansion 
to all orders. We apply the perturbative approach to diffusive particles interacting 
through a mean-field potential. For such systems, the variational characterization 
of the quasi-potential was proven by Dawson and Gartner
~\cite{dawson1987,dawson1987b}. 
Our perturbative analysis provides new explicit results about the quasi-potential 
and about fluctuations of one-particle observables in a simple example of mean 
field diffusions: the Shinomoto-Kuramoto model of coupled rotators
~\cite{shinomoto1986}. 
This is one of few systems for which non-equilibrium free energies can be computed 
and analyzed in an effective way, at least perturbatively.
\end{abstract}



\vskip 1cm

\nsection{Introduction}

Large deviations theory studies the exponential decay of probabilities 
of large fluctuations in stochastic systems. Such probabilities are important 
in many fields, including physics, statistics, finance or engineering, 
as they often yield valuable information about extreme events 
far from the most probable state or trajectory of the system. 
\cite{touchette2009,dembo1998,ellis1985}.

Weak noise large deviation theory has been developed in the $'70$s 
by Freidlin and Wentzell \cite{freidlin1984} in a mathematical framework 
and by Graham and collaborators \cite{graham1989} with physicists' 
perspective. It concerns the study of large fluctuations in dynamical systems 
subject to weak random noise. In this framework, the stationary probability 
to observe some state $x$ of the system obeys a large deviation asymptotics
\qq\label{eq:Chap1-LD-invariant-measure}
P_{\infty}(x)\asymp \exp\left( -\frac{F(x)}{\epsilon}\right),
\qqq
where $\epsilon$ denotes the noise strength squared. The function $F$ is called 
the quasi-potential. It generalizes the notion of (free) energy to general 
finite-dimensional systems where detailed balance does not hold.but noise is 
weak. If known, the quasi-potential permits to calculate, at leading order 
in $\epsilon$, important statistical quantities such as the probability 
to observe an arbitrary large fluctuation of the system or the mean residence 
time that the system spends close a metastable state.

In the last $15$ years, Jona-Lasinio and collaborators developed a framework 
to study large deviations of macroscopic quantities (like particle density 
or current) in a class of many-body systems. Their approach, known as the Macroscopic 
Fluctuation Theory, was mainly applied to stochastic lattice gases, see 
\cite{bertini2014} for a recent review. Without stress on mathematical rigor, 
the MFT can be understood as a generalization of the Freidlin-Wentzell theory 
to the fluctuating diffusive hydrodynamics where a weak noise is added to 
non-linear diffusion equations. Indeed, it is easily obtained by employing 
a saddle-point approximation in a path integral formalism. Even if this 
approach is only formal because a mathematical meaning of non-linear 
fluctuating hydrodynamics is still lacking, the results obtained are in 
complete agreement with the ones obtained through rigorous probabilistic 
methods in all cases where a comparison is possible. In this paper, a further 
example of such an agreement will be discussed in chapter \ref{sec:mean-field}.

The Freidlin-Wentzell theory provides a variational characterization of 
the quasi-potential through dynamical large deviations. However, its explicit 
computation is typically very difficult and this strongly restricts the 
practical applicability of the theory. This is true both for 
finite-dimensional and infinite-dimensional systems. Only in few cases 
the quasi-potential can be evaluated analytically, the most important one 
being, of course, the case where detailed balance holds.

The main focus of the present paper is to answer a very natural but still 
open question. Let us consider a system depending on a control parameter 
$\lambda$ and suppose that we are able to calculate the quasi-potential for 
$\lambda=0$. Can we build a perturbation theory to calculate the 
quasi-potential for small but finite $\lambda$?
Surprisingly, only few works in the literature discuss this question 
\cite{graham1983,graham1984,graham1984a,graham1985a,jauslin1986,gang1989,maier1993,tel1989,dykman1990,poquet2014}. All of them focus on specific examples, 
perform only the $1^{\rm st}$-order analysis and most of them 
consider only finite-dimensional 
dynamical systems (see however \cite{smelyanskiy1997} for an example where 
an infinite dimensional system is analyzed, and the case of transition rates between two basins of attraction is studied). We obtain here a precise answer 
to the above question, giving an explicit iterative formula for computing each 
order in the power series expansion of the quasi-potential in $\lambda$.

We first analyze the perturbation theory for finite dimensional systems, 
where the discussion can be made quite precise. It is well known that, given 
the quasi-potential function, one can deduce a simple $1^{\rm st}$-order equation for 
the instanton (the corresponding variational problem minimizer). The converse 
is also obviously true: the values of the quasi-potential (the minima) can be 
easily computed from the minimizers. It is however often difficult to compute 
either the instantons or the quasi-potential without the knowledge of the 
other. We show that this loop can be broken in a perturbative setting: 
the quasi-potential at any order may be computed just from the knowledge 
of the instanton dynamics of the unperturbed problem and the quasi-potential 
at previous orders. This gives a very natural iterative procedure. We also 
explain that an equivalent simple recursive scheme appears when starting from 
the perturbative expansion for solutions of the Hamilton-Jacobi equation. 
A key point is to prove that solvability conditions hold at each order, 
assuring the existence of the perturbative expansion to all orders. We show 
that such conditions are related to the behavior close to the attractor, which 
gives a simple proof that they are satisfied at all orders.

In the second part of the paper, we consider a particular class of 
non-equilibrium many-body systems described by the Macroscopic Fluctuation 
Theory \cite{bertini2014}, namely diffusive particles interacting through 
mean-field potential and driven by an external non-conservative force. 
Our approach is based on the fact that for such diffusions, it is possible 
to derive a fluctuating hydrodynamics describing the evolution of the 
empirical density for large but finite number of particles $N$. This 
evolution equation was first obtained by Dean in \cite{dean1996} and 
it was thereafter called the Dean equation. Although its mathematical 
status is uncertain, the Dean equation allows to treat the mean-field 
diffusions, at least at a formal level, as a dynamical system in an 
infinite dimensional space perturbed by a weak noise whose strength 
is proportional to $1/\sqrt{N}$. In the $N\to \infty$ limit, similarly 
to the law of large numbers, the dynamical system becomes a deterministic 
equation known under the name of the McKean-Vlasov or the Vlasov-Fokker-Planck 
one, as was proven together with the propagation of chaos in 
\cite{mckean1966,sznitman1991,meleard1996}.

We subsequently move our attention to the large deviations around the 
$N\to\infty$ behavior. The Dean equation is a formal random dynamical
system to which we apply the Martin-Siggia-Rose formalism \cite{martin1973}. 
Using the saddle point approximation we then end up with an 
infinite-dimensional generalization of the Freidlin-Wentzell theory. 
This result has been previously obtained rigorously in the mathematical 
literature by Dawson and Gartner \cite{dawson1987,dawson1987b}. The formal 
approach based on the Dean equation makes an explicit connection with 
the Macroscopic Fluctuation Theory and adds another class of systems 
to the ones covered by the latter theory.

Explicit results, original to the best of our knowledge, 
are discussed in the case of the Shinomoto-Kuramoto model, a specific stochastic
system describing coupled planar rotators \cite{shinomoto1986}. This model may 
be also viewed as a non-equilibrium version of the dynamical mean-field XY 
model. We discuss in detail both its $N\to\infty$ behavior and the large 
deviations around it. For what concern the $N\to\infty$ limit, we 
are able to fully describe analytically the phase diagram of the 
Shinomoto-Kuramoto model, deriving self-consistent equations for the stationary 
states that are easily solved numerically.

The perturbation theory developed in the first part of the paper is subsequently applied 
to the Dean equation corresponding to the Shinomoto-Kuramoto model, resulting
in an explicit calculation of the quasi-potential close to the known cases: around
the free particle dynamics and in the vicinity of stationary states of the $N=\infty$
theory. This turns out to be a rather simple numerical task. We show that 
explicit results could be obtained to any order but, for the sake of brevity, 
we only present, the $1^{\rm st}$ order calculations of the quasi-potential.

The paper is organized as follows. After Introduction, we begin with 
a review of Freidlin-Wentzell theory results in Sec.\,\ref{sec:1}. Our aim 
is to present the known results that can be found in \cite{freidlin1984} with 
a physicist perspective, but nevertheless being precise on many important points 
that are typically overlooked in the physics literature. In particular, we 
discuss under which hypothesis the quasi-potential can be seen as a 
solution of the Hamilton-Jacobi equation that forms the basis for the 
perturbative treatment that is developed in Sec.\,\ref{sec:Chap2}.
In the context of finite-dimensional systems, the discussion of
the perturbation theory for quasi-potential is made quite precise. 
We also show that the Taylor expansion of the quasi-potential 
close to an attractor of the deterministic dynamics is a particular 
case of our perturbative analysis. We conclude the section by observing 
that close to a codimension-one bifurcation of the deterministic dynamics, 
fluctuations diverge with the mean-field critical exponent. Final 
Sec.\,\ref{sec:mean-field} is about $N$-body systems and is central 
to this paper. It concentrates on the case of overdamped diffusions 
with mean-field interaction that are introduced in 
Sec.\,\ref{sub:Chap3-mean-field}. In Sec.\,\ref{sec:Chap3-Empirical-density}, 
we discuss the fluctuating hydrodynamics for such diffusions, deriving 
the Dean equation making the original argument \cite{dean1996} 
more precise. The McKean-Vlasov equation is described as the limit of
the Dean equation when $N\to\infty$. In 
Sec.\,\ref{sec:Chap3-Kuramoto-typical}, we introduce the Shinomoto-Kuramoto 
model \cite{shinomoto1986} as a particular example of the mean-field 
diffusions and we analyze its McKean-Vlasov limit and the corresponding
phase diagram. Sec.\,\ref{sub:Chap3-LD} derives formally, starting from
the Dean equation, an extension of the Freidlin-Wentzell theory to 
the case of mean-field diffusions, and discusses the infinite-dimensional 
version of the Hamilton-Jacobi equation for the quasi-potential and,
briefly, the large deviations for empirical currents.
In Sec.\,\ref{sub:Chap3-perturbative}, we adapt the perturbative approach 
developed for finite dimensional systems to the diffusions with mean-field
interaction, concentrating on the power series in the mean-field coupling 
constant and the Taylor expansion close to stationary solutions 
of the McKean-Vlasov dynamics. A discussion of explicit results for the 
Shinomoto-Kuramoto model obtained by combining analytical and numerical 
treatments is presented. Sec.\,\ref{sec:conclusions} summarizes the results 
and discusses the perspectives for future work. Two Appendices contain 
some additional material.

\nsection{Freidlin-Wentzell theory: a brief summary}\label{sec:1}

We consider in this section a finite dimensional random dynamical system 
defined by the Ito stochastic differential equation\footnote{Here and below, 
we use physicists' notation for stochastic equations rather than 
mathematicians' one with differentials.}
\qq\label{eq:dynamics-Chap1}
\dot{x} = K(x) + \sqrt{2\epsilon}\,g(x)\,\eta_t\,,
\qqq
where $x\in \mathbb{R}^d$, $g$ is a $d\times m$ matrix and  $\eta_t\in 
\mathbb{R}^m$ is a vector of white in time Gaussian noises $\eta^i_t$ with 
zero mean and covariance
\qq
\mathbb{E}[\eta^i_t\,\eta^j_{t'}] = \delta^{ij} \delta(t-t')\,.
\qqq
Throughout the paper it is supposed that $K$ and $g$ are smooth and
the Ito stochastic convention is employed, if not stated otherwise. We use 
the notation $Q(x)=g(x)\,g^T(x)$ with the superscript $T$ indicating the matrix 
transposition and we demand that $Q(x)$ be positive definite for all $x$. 
Generalizations to the case in which $Q$ is semi-positive definite are 
possible but we do not consider this situation here.

We assume that the stochastic process solving Eq.\,(\ref{eq:dynamics-Chap1}) 
has a unique invariant measure. Then its density $P_{\infty}(x)$ is smooth 
and solves the stationary Fokker-Planck equation
\qq\label{eq:Chap1-FP}
\sum_{i=1}^d\frac{\partial}{\partial x^i} \bigg[\Big(-K^i(x) +\epsilon 
\sum_{j=1}^d\frac{\partial}{\partial x^j} Q^{ij}\Big)P_{\infty}(x)\bigg]=0\,.
\qqq
We are interested in the behavior of the above stochastic dynamical system 
in the small noise 
limit $\epsilon\ll 1$. In this limit, the stationary 
measure obeys the large deviation principle 
\eqref{eq:Chap1-LD-invariant-measure} \cite{freidlin1984}, 
where the symbol $\asymp$ stands for the asymptotic logarithmic equivalence:
\qq\label{eq:Chap1-QP}
F(x) = -\lim_{\epsilon\to 0} \epsilon \ln P_{\infty}(x)\,.
\qqq
The rate function $F(x)$ is called quasi-potential associated to the random 
dynamical system (\ref{eq:dynamics-Chap1}). It is of central importance to 
this paper that $F$, under suitable hypothesis that will be specified in 
Sec.\,\ref{subsec:Chap1-HJ}, is the unique solution of the Hamilton-Jacobi 
equation
\qq\label{eq:HJ-Chap0}
\nabla F\cdot \left[ Q\cdot \nabla F+ K\right](x) = 0\,.
\qqq
At an informal level, this can be guessed by looking for solutions of the 
stationary Fokker-Planck equation (\ref{eq:Chap1-FP}) obeying Ansatz
\qq
P_{\infty}(x)\simeq \exp\left[ -\frac{F(x)}{\epsilon} + Z(x) +\,\dots\right]
\qqq
and retaining only terms of order $1/\epsilon$. The reader may consult 
\cite{graham1989} where also the lower-order equation for $Z(x)$ is derived.

The present introductory section is devoted to an informal discussion of 
the Freidlin-Wentzell theory, with the particular emphasis on how 
the quasi-potential can be described from different points of view. 
This knowledge will be used in the subsequent section. We 
refer the reader to \cite{freidlin1984} for a mathematical presentation, 
to Chapter 6 of \cite{graham1989} for a treatment more oriented towards 
the physics community and to  \cite{Heymann2008} for a general review oriented 
to computational aspects. In Sec.\,\ref{subsec:FW-Fbarx}, we discuss dynamical 
large deviations, that is, the probability that a solution of 
Eq.\,(\ref{eq:dynamics-Chap1}) is close to a given path in 
the limit $\epsilon\to 0$. This permits to define quasi-potential $F_A$ 
relative to an attractor $A$ of the deterministic dynamics $\dot{x}=K(x)$. 
In Sec.\,\ref{subsec:FW-Transversal}, we describe the properties of solutions 
of the Hamilton-Jacobi equation (\ref{eq:HJ-Chap0}) that we denote $F_{\HJ}$. 
The connection between $F_{\HJ}$, $F_A$, and $F$ and the conditions under 
which they coincide are discussed in Sec.\,\ref{subsec:Chap1-HJ},
concentrating on the case of attractive points. A proof of the local 
existence and uniqueness of $F_{\HJ}$ around a stable fixed point of 
the deterministic dynamics $\dot x=K(r)$, and of its local regularity, is sketched in 
Sec.\,\ref{subsec:Chap1-Hamiltonian-picture}. 


\subsection{Freidlin-Wentzell action and quasi-potential relative to an 
attractor}\label{subsec:FW-Fbarx}

Freidlin and Wentzell considered the probability for a trajectory 
of the stochastic process $x(\cdot)$ defined by Eq.\,(\ref{eq:dynamics-Chap1}) 
to be arbitrarily close to a given continuous path $\hat{x}(\cdot)$ on the 
time interval $[t_i,t_f]$. They showed rigorously that  
\qq\label{eq:Chap1-FW-action}
\lim_{\delta\downarrow0}\liminf_{\epsilon\downarrow0}\,\epsilon\ln\mathbb{P}
\Big[\sup_{t_i\leq t\leq t_f}|x(t)-\hat{x}(t)|<\delta\Big]=
\lim_{\delta\downarrow0}\limsup_{\epsilon\downarrow0}\,\epsilon\ln\mathbb{P}
\Big[\sup_{t_i\leq t\leq t_f}|x(t)-\hat{x}(t)|<\delta\Big]=-\mathcal{A}[\hat{x}(\cdot)],
\qqq
where $|\cdot|$ denotes the norm of a vector in $R^d$ and
\qq\label{eq:FW-action}
\mathcal{A}[x(\cdot)] = \frac{1}{4} \int_{t_i}^{t_f}
\left[\dot{x}(t)-K(x(t))\right]\,
\cdot\, Q(x(t))^{-1}\,[\dot{x}(t)-K(x(t))]\,dt
\qqq
is the so-called Freidlin-Wentzell action functional, see 
Theorem 2.3 in Chapter 3 of \cite{freidlin1984}. Observe that 
such a functional vanishes on solutions of the deterministic equation
\qq 
\dot{x} = K(x)\,.
\label{eq:det-dyn}
\qqq
It measures the difficulty for a trajectory to deviate, 
due to a weak noise, from the deterministic behavior.

The above result may be elucidated in the framework of the formal path-integral
approach going back to Onsager and Machlup 
\cite{machlup1953,stratonovich1962,martin1973,dykman1979,zinnjustin2002}. 
In this approach, the transition 
probability from a state $x_i$ at time $t=t_i$ to a state $x_f$ at time $t=t_f$ 
is written as the path integral
\qq\label{eq:Chap1-transition-probability}
\mathbb{P}[x_i,t_i;x_f,t_f] \asymp \frac{1}{Z_A}\int
\exp\left[-\frac{\mathcal{A}[x(\cdot)]}{\epsilon}\right] \,
\mathcal{D}[x(\cdot)],
\qqq
where the functional integration is restricted to the paths $[t_i,t_f]\ni t
\mapsto x(t)$ such that $x(t_i)=x_i$ and $x(t_f)=x_f$. 
$Z_A$ is the normalization factor. The exponential factor 
$\exp\left[-\frac{\mathcal{A}[x(\cdot)]}{\epsilon}\right]$ plays then for small
$\epsilon$ the role of the probability density in the space of paths.\\

Let us now consider an attractor $A$ of the deterministic dynamical system 
(\ref{eq:det-dyn}) and a point $x_0\in A$. The quasi-potential relative 
to $x_0$ is defined as
\qq\label{eq:Fbarx-variational}
F_{x_0} (x) = \min_{\{\hat{x}(\cdot)\,|\, \hat{x}(-\infty)=x_0\,,\, \hat{x}(0)=x\}}\,
\mathcal{A}[\hat{x}(\cdot)]\,,
\qqq
where the minimum is over all absolutely continuous paths starting from $x_0$ 
at time $t=-\infty$ and ending in $x$ at time $t=0$. The choice of the 
time interval $[-\infty,0]$ is arbitrary and any semi-infinite interval 
$[-\infty,t_f]$ would give the same result. It is easy to see that $F_{x_0}$
does not depend on the choice of $x_0\in A$ and is constant on attractors. 
Indeed, given two points on the attractor, one can 
always find trajectories starting and ending arbitrarily close to them on 
which the action is arbitrarily small. As an example, the reader may
consider a limit cycle where any two points of the attractor can be connected 
by a solution of the deterministic evolution $\dot{x}=K(x)$. For an
attractor $A$, we shall denote by $F_A$ the quasi-potential relative to 
any $x_0\in A$.
For the sake of simplicity, we shall mainly consider in this paper single 
point attractors referring the reader to \cite{graham1989} and references 
therein for explicit calculations of quasi-potentials with respect to 
non-trivial attractors, for example, limit cycles or the Lorentz attractor.

The quasi-potential $F$ can be built from the quasi-potentials relative to 
the attractors of $\dot{x} = K(x)$. 
In the case where a fixed point $\bar{x}$ is the only attractor of the 
deterministic dynamics $\dot{x}=K(x)$ for any initial condition, $F_{\bar{x}}$ 
coincides with the quasi-potential $F$ defined through the invariant measure 
in (\ref{eq:Chap1-QP}). Formally, this may be understood by observing that 
$\mathbb{P}[x_i,t_i;x_f,t_f]$ is actually independent of $x_i$ in the 
$t_f\to \infty$ limit and thus $P_{\infty}(x)=\mathbb{P}[x_i,t_i;x,\infty]$. 
Then, one obtains the stated result by applying the saddle point approximation 
to the right hand side of (\ref{eq:Chap1-transition-probability}). The 
reader may consult Theorem 4.3 in Chapter 4 of \cite{freidlin1984} for the 
rigorous result.

If more attractors  $\{A_i\}_{1\leq i \leq I} $ are present then the quasi-potential 
$F$ may still be constructed once the quasi-potentials relative to each 
attractor $F_{A_i}$ is known. One has 
\qq\label{eq:F-variational-patching}
F(x)=\min_{i} \left(F_{A_i} (x) + C_{i}\right)-\min_i C_i
\qqq
where constants $C_{i}$ describe the ``height'' of each attractor $A_i$. 
More precisely, one has to consider the $i$-graphs $G(i)$ on the set
$\{A_1,\dots,A_I\}$ of attractors composed of arrows $j\to k$ such that 
$j\neq i$, from every $A_j\neq A_i$ starts exactly one arrow 
and there are no closed cycles. Constants $C_i$ are then defined as 
\qq\label{eq:constants_C_i}
C_i= \min_{G(i)} \sum_{\{j\to k\}} F_{A_j}(A_k)\,,
\qqq
where the minimum is over all $i$-graphs and the sum over all the arrows 
of an $i$-graph and $F_{A_j}(A_k)=F_{A_j}(x)$ for any $x\in A_k$. This rigorous 
result is discussed in detail in Chapter 6 of \cite{freidlin1984}. 
  Eq.\,\eqref{eq:F-variational-patching} balances the
contributions from different attractors with the use
of the invariant measure of the Markov chain with transition probabilities
describing the passages between different attractors. This point 
was discussed in \cite{graham1986} from physicist's perspective. In that 
reference, a computation of constants $C_i$ for few explicit  
examples was also carried out.
In the present paper, we concentrate on the calculation of quasi-potentials
$F_{\bar{x}}$ relative to attractive points $\bar{x}$ and thus we do not enter 
into further details on how the heights $C_i$ may be practically found.


\subsection{Transverse decomposition, fluctuation and relaxation dynamics}
\label{subsec:FW-Transversal}

In the previous section we saw that the quasi-potential $F$ can be obtained
by solving the variational problem given by Eqs.\,(\ref{eq:Fbarx-variational}), 
(\ref{eq:F-variational-patching}) and (\ref{eq:constants_C_i}). 
In this section, we discuss 
a different approach that will permit to obtain $F$ as a solution of the 
Hamilton-Jacobi equation (\ref{eq:HJ-Chap0}) of Sec.\,\ref{subsec:Chap1-HJ}. 

Consider an open set $D\subseteq \mathbb{R}^d$  
and its closure $\bar{D}=D\cup \partial D$, where $\partial D$ denotes 
the boundary of $D$ assumed to be smooth. We suppose in this section that 
the vector field $K$ admits a transverse decomposition in $\bar{D}$ in the 
following sense: there exists a smooth function $F_{\HJ}(x)$ 
such that 
\qq\label{eq:Chap1-K-split}
K(x)=-(Q\,\nabla F_{\HJ})(x) + G(x)
\qqq
and
\qq\label{eq:Chap1-K-split2}
\nabla F_{\HJ}(x) \cdot G(x)=0
\qqq
for all $x\in \bar{D}$. The existence of a transverse decomposition is 
equivalent to demanding that $F_{\HJ}$ solves the Hamilton-Jacobi equation 
(\ref{eq:HJ-Chap0}). Indeed, if $K$ admits a transverse decomposition, 
then $\nabla F_{\HJ}\cdot K = -\nabla F_{\HJ}\cdot Q\,\nabla F_{\HJ}$. 
Conversely, we can define $G=K+Q\,\nabla F_{\HJ}$ and from 
Eq.\,(\ref{eq:HJ-Chap0}) we obtain the transversality condition.
The term $Q\nabla F_{\HJ}$ may be viewed as the gradient of $F_{\HJ}$ in 
the Riemannian metric defined by the matrices $Q(x)^{-1}$ and then
Eq.\,\eqref{eq:Chap1-K-split2} states its orthogonality to $G$
with respect to the corresponding scalar product of the vector fields.
Such an interpretation is often employed in the mathematical literature
but we shall not pursue it here. 

The deterministic dynamics (\ref{eq:det-dyn}) with $K(x)
= -(Q\,\nabla F_{\HJ})(x) + G(x)$ is called relaxation dynamics. The reason 
for this name is that as the Freidlin-Wentzell action vanishes on its 
trajectories, they are in the small noise limit the most probable 
trajectories relaxing to 
an attractor. The presence of a transverse decomposition permits to define 
the so-called fluctuation or instanton dynamics 
\qq\label{eq:fluctuation-dynamics}
\dot{x}= 2(Q\nabla F_{\HJ})(x)+K(x)=(Q\nabla F_{\HJ})(x)+G(x)\,\equiv\,K_r(x)
\qqq
which plays a fundamental role in what follows. We shall also be interested in 
the Freidlin-Wentzell action functional corresponding to the stochastic 
dynamics  
\qq\label{eq:fluctuation-dynamics-stochastic}
\dot{x} = K_r + \sqrt{2\epsilon}\,g(x)\,\eta_t\,,
\qqq
which is 
\qq\label{eq:fluctuation-dynamics-stochastic-action}
\mathcal{A}_r[x(\cdot)]= \frac{1}{4} \int_{t_i}^{t_f} 
[\dot{x}(t)-K_r(x(t))]\,\cdot\, Q(x(t))^{-1}\,[\dot{x}(t)-K_r(x(t))]\,dt\,.
\qqq

The fluctuation dynamics is  connected to time-reversal of the stochastic
equation (\ref{eq:dynamics-Chap1}). To understand this point, 
let us consider the diffusion process defined by (\ref{eq:dynamics-Chap1}) 
in the time interval $[0,T]$.
It was shown in \cite{haussmann1986} that its time-reversal corresponds 
to the stochastic dynamics
\qq
\dot{x} = -\widetilde{K}(x)  + \sqrt{2\epsilon}\,g(x)\,\eta_t\,,
\qqq
where $\widetilde{K}(x)=K(x) -2\epsilon\,P(x,T-t)^{-1}\nabla \cdot 
\left[Q(x) P(x,T-t)\right]$ and  $P(x,t)$ is the solution to the 
time-dependent Fokker-Planck equation associated to 
(\ref{eq:dynamics-Chap1}). If we consider 
Eq.\,(\ref{eq:dynamics-Chap1}) with initial condition distributed accordingly 
to $P_{\infty}(x)$ then it follows from 
Eq.\,(\ref{eq:Chap1-LD-invariant-measure}) that 
$\lim_{\epsilon\to0}\widetilde{K}(x)= K(x) + 2(Q\,\nabla F)(x)$. Under 
suitable hypothesis given in Sec.\,\ref{subsec:Chap1-HJ}, $F$ and $F_{\HJ}$ 
coincide and then $\lim_{\epsilon\to0}\widetilde{K}(x)=K_r(x)$.

We now discuss some properties of $F_{\HJ}$ in connection to relaxation 
and fluctuation dynamics. The stationary points of $F_{\HJ}$ correspond 
to zeros of the vector field $K$. Indeed
\qq\label{eq:Chap1-critical-points-1}
(K\cdot Q^{-1} K)(x)=(\nabla F_{\HJ} \cdot Q\, \nabla F_{\HJ})(x)
+(G \cdot Q^{-1} G)(x) \geq 0\,,
\qqq 
as it can be directly proven by inserting Eq.\,(\ref{eq:Chap1-K-split}) 
into the left hand side and using the transverse 
decomposition. Then, if $\bar{x}$ is such that $K(\bar{x})=0$, the above 
expressions imply that $\nabla F_{\HJ}(\bar{x})=0=G(\bar{x})$.
The converse is also true if we suppose that the Hessian matrix of $F_{\HJ}$, 
denoted by $(\nabla \nabla F_{\HJ})$,  is invertible at $\bar{x}$. Indeed, 
by taking the gradient of the transversality condition and evaluating it 
at $\bar{x}$, we infer that
\qq
(\nabla \nabla F_{\HJ}G)(\bar{x})=0\,.
\qqq
By invertibility of the Hessian matrix, 
this implies that $G(\bar{x})=0$ and hence $K(\bar{x})=0$.
Moreover, $F_{\HJ}$ is a Lyapunov function for both the relaxation dynamics 
and the time-reverse of the fluctuation dynamics
\qq\label{eq:fluctuation-dynamics-time-reversal}
\dot{x}= - K_r(x) \,.
\qqq
Indeed, if $\dot{x}=K(x)$ then 
\qq
\frac{dF_{\HJ}(x)}{dt} = -(\nabla F_{\HJ} \cdot Q \,\nabla F_{\HJ})(x)\leq 0\,,
\qqq
and analogously for Eq.\,(\ref{eq:fluctuation-dynamics-time-reversal}).
From the above properties, we conclude that the relaxation and fluctuation 
dynamics have the same stationary points but attractors are transformed 
into repellers and vice-versa. Thus the relaxation and the time-reversal 
of the fluctuation dynamics have the same attractors and  $F_{\HJ}$ as 
a Lyapunov function. 

\begin{figure}
\centerline{\includegraphics[scale=0.8]{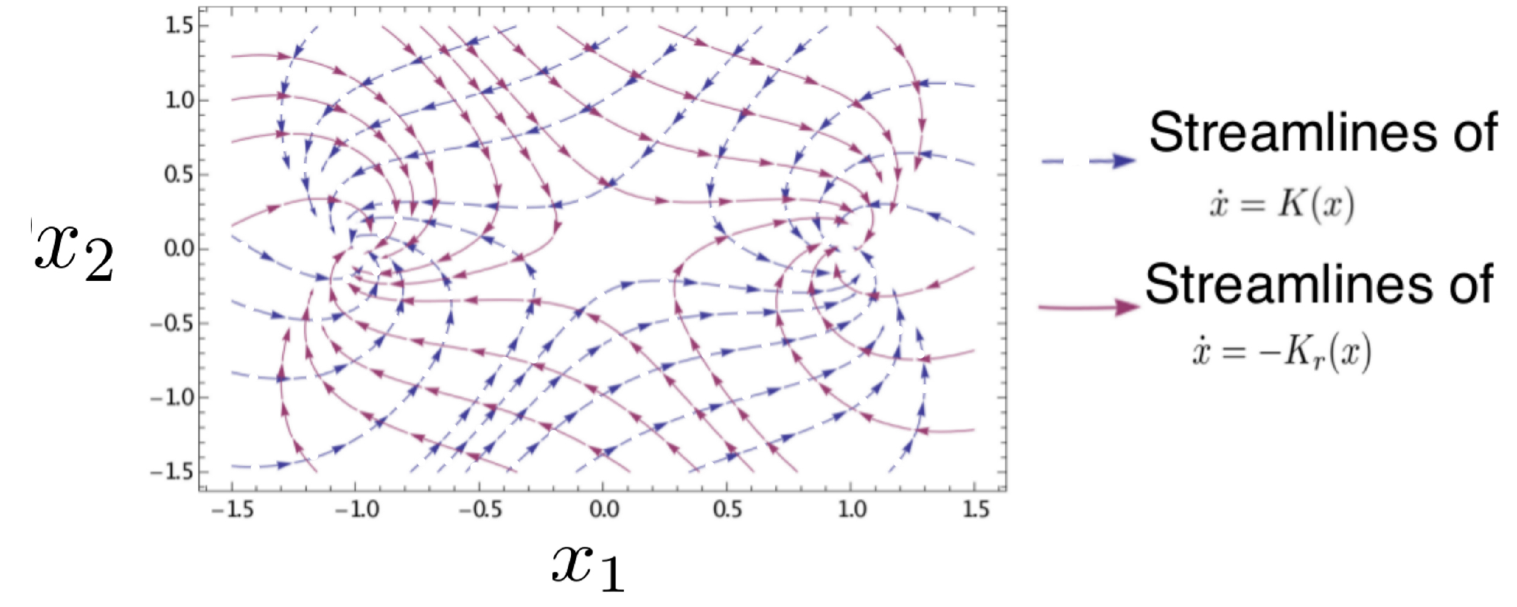}}
\caption{Example showing a simple case where the basins of attraction of 
the relaxation and of the time-reversal of the fluctuation dynamics do not 
coincide. Here, $F_{\HJ}(x_1,x_2)=-\frac{x_1^2}{2}+\frac{x_1^4}{4}
+\frac{x_2^2}{2}$, $Q=\mathbb{I}$, $G=(\partial F_{\HJ}/\partial x_2, 
-\partial F_{\HJ}/\partial x_1)$, where $\mathbb{I}$ is the $2\times 2$ 
unit matrix. The dashed blue arrows indicate the direction of the relaxation 
dynamics $\dot{x} = K(x) = -Q\,\nabla F_{\HJ}(x) + G(x)$ and the continuous red lines the 
direction of the time-reversal of the fluctuation dynamics 
$\dot{x} = -K_r(x) = -Q\,\nabla F_{\HJ}(x) - G(x)$. 
\label{fig:different-attractors}}
\end{figure}

One could be led to an incorrect conclusion that basins of attraction 
of those two dynamics are the same which, however, is not true because 
the transverse parts of dynamics can push the systems to different 
attractors. This can be checked, for example, in a very simple two-dimensional 
bistable system, see Figure\nobreak\ \ref{fig:different-attractors}.


\subsection{Quasi-Potential as a solution of the Hamilton-Jacobi equation}
\label{subsec:Chap1-HJ}

Let us suppose that $F_{\HJ}$ is a smooth
solution of the Hamilton-Jacobi  equation 
\qq\label{eq:HJ-chap-1}
\nabla F_{\HJ}\cdot \left[ Q\,\nabla F_{\HJ}+ K\right](x) = 0\,,
\qqq
in $\bar{D}=D\cup \partial D\subseteq \mathbb{R}^d$ containing in the interior
a fixed point $\bar{x}$ such that $F_{\HJ}(\bar{x})=0$ and $F_{\HJ}(x)>0$,  
$\nabla F_{\HJ}(x)\not=0$ for $x\in\bar{D}$, $x\not=\bar{x}$. We shall assume
that $D$ is bounded and connected.  
We want to understand the relation between $F_{\bar{x}}$ and $F_{\HJ}$, see 
Theorem 3.1 in Chapter 4 of \cite{freidlin1984} for more details.

For any path $[-\infty,0]\ni t\mapsto\hat x(t)\in\bar{D}$ subjected to the 
boundary conditions $\hat x(-\infty)=\bar{x}$ and $\hat x(0)=x$, a simple manipulation shows that
\qq\label{eq:reverse-action}
\mathcal{A}[\hat{x}(\cdot)] = \mathcal{A}_r[\hat{x}(\cdot)]+
\int_{-\infty}^0 \frac{d\hat{x}(t)}{dt}\cdot\nabla F_{\HJ}(\hat x(t))\,dt\,
=\mathcal{A}_r[\hat{x}(\cdot)]+ F_{\HJ}(x)\geq F_{\HJ}(x)
\label{eq:comparison}
\qqq
where $\mathcal{A}_r[\hat{x}(\cdot)]$ was defined in 
(\ref{eq:fluctuation-dynamics-stochastic-action}) and we used the 
transversality condition. The inequality 
\qq
\mathcal{A}[\hat{x}(\cdot)]\geq F_{\HJ}(x)
\qqq
still holds if we allow the trajectory $\hat x(\cdot)$ to go out of 
the closed set $\bar{D}$ provided that 
$x\in\bar D_{\HJ}$, the closure of the open set
\qq
D_{\HJ}=\big\{x\in D \,\big|\, F_{\HJ}(x)<\min_{y\in \partial D} 
F_{\HJ}(y)\big\}.
\label{eq:DHJ}
\qqq
Indeed, writing the inequality (\ref{eq:comparison})
for the trajectory $\hat x(\cdot)$ restricted to $[-\infty,\tau]$,
where $\tau$ is the first exit time from $D$, we obtain the lower bounds
\qq
\mathcal{A}[\hat{x}(\cdot)]\geq F_{\HJ}(\hat x(\tau))\geq F_{\HJ}(x)
\qqq
with the second one resulting from the condition $x\in\bar{D}_{\HJ}$.
Action $\mathcal{A}_r[\hat{x}(\cdot)]$ attains its minimum equal to zero
on the trajectory of the fluctuation dynamics $\dot{x}=K_r(x)$. It
is easy to see that for $x\in\bar D_{\HJ}$ there exists a unique such
trajectory $[-\infty,0]\ni t\mapsto\tilde{x}(t,x)$ for which 
$\tilde x(-\infty,x)=\bar{x}$ and $\tilde x(0,x)=x$. Besides, such
trajectory lies entirely in $\bar D_{\HJ}$. This follows from the fact, 
that on each solution of the fluctuation dynamics that ends at $x$, 
function $F_{\HJ}$ decreases backward in time and such a solution may 
be infinitely extended in negative time direction until it reaches 
$\bar{x}$ at $t=-\infty$. The resulting trajectory saturates the 
inequality (\ref{eq:comparison}). 
One infers that for $x\in\bar D_{\HJ}$,
\qq
F_{\HJ}(x)\,=\,\min_{\{\hat{x}(\cdot)\,|\, \hat{x}(-\infty)=\bar{x}\,,\, \hat{x}(0)=x\}}
\mathcal{A}[\hat{x}(\cdot)]\,=\,F_{\bar{x}}(x)\,,
\qqq
see (\ref{eq:Fbarx-variational}). 
$D_{\HJ}$ is again a bounded open connected neighborhood of $\bar{x}$. 
Its boundary $\partial D_{\HJ}$ is composed of points of $x\in\bar D$ for which 
$F_{\HJ}(x)=\min_{y\in \partial D} F_{\HJ}(y)$. It is smooth since 
$\nabla F_{\HJ}\not=0$ at such points. 

We have just obtained two results. First, a smooth solution
of the Hamilton-Jacobi equation (\ref{eq:HJ-Chap0}) on $\bar D$, with the properties stated at the beginning of the section, coincides
with $F_{\bar{x}}$ on $\bar D_{\HJ}$ where $D_{\HJ}$ is the
sub-domain of $D$ containing $\bar{x}$ defined by (\ref{eq:DHJ}).
Second, on $\bar D_{\HJ}$,
\qq
F_{\HJ}(x)=F_{\bar{x}}(x)=\mathcal{A}[\tilde x(\cdot\,,x)]\,,
\qqq
where $[-\infty,0]\ni t\mapsto\tilde x(t,x)\in\bar D_{\HJ}$ is the solution to the instanton dynamics, i.e. the unique 
trajectory joining $\bar{x}$ to $x$ and satisfying  
Eq.\,(\ref{eq:fluctuation-dynamics}).


\subsection{Hamiltonian picture}
\label{subsec:Chap1-Hamiltonian-picture}

Above we have assumed the local existence of a smooth solution 
of the Hamilton-Jacobi equation around an attractive point $\bar{x}$
of the relaxation dynamics (\ref{eq:det-dyn}).
The linearization of such dynamics around $\bar{x}$ has the form
\qq
\dot{x}=Ax\,,
\qqq
where $A=(\nabla K(\bar{x}))^T$ is the matrix with entries 
$A^i_{\,j}=\nabla_jK^i(\bar{x})$.
We shall assume that all eigenvalues of $A$ have negative real parts,
which ensures the exponential convergence to $\bar{x}$ in the vicinity
of the attractor. Such attractive points will be called non-degenerate.
We shall sketch below a proof of the fact that around non-degenerate 
attractive point $\bar{x}$ there exists a unique local smooth solution 
of the Hamilton-Jacobi equation, The argument we present here is based 
on the analysis of the dynamics of extremal trajectories of the Freidlin-Wentzell 
action functional, see \cite{day1985} for a more global discussion. 

\begin{figure}
\centerline{\includegraphics[scale=0.45]{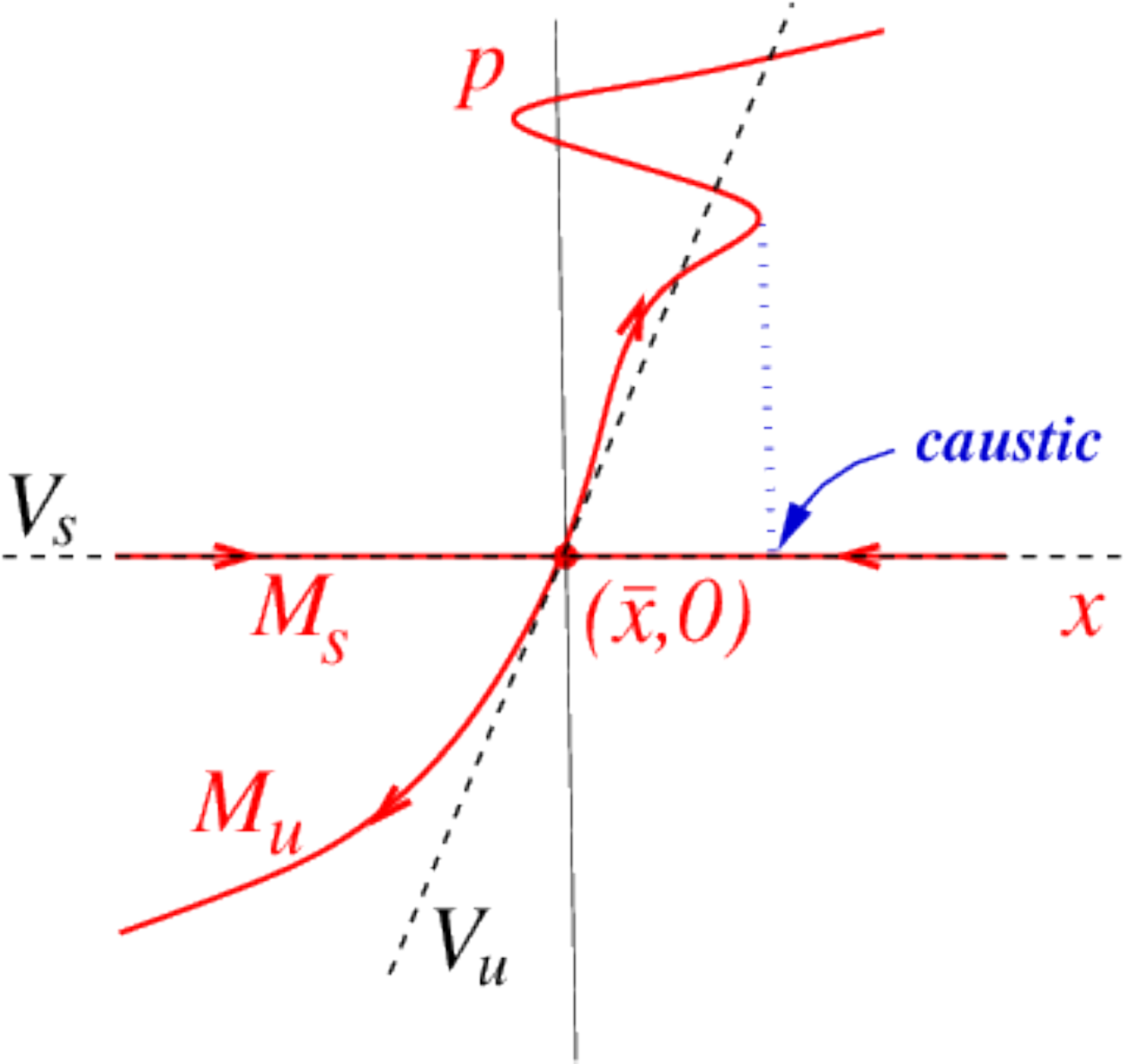}}
\caption{Hamiltonian flow around $(\bar{x},0)$ with the stable and
unstable manifolds
\label{fig:H-flow}}
\end{figure}

Let us start by considering the Hamiltonian $H(x,p)$ related by the Legendre 
transform to the Lagrangian 
\qq
L(x,\dot x)=\frac{_1}{^4}(\dot x-K(x))\cdot Q(x)^{-1}(\dot x-K(x))
\qqq
appearing in the Freidlin-Wentzell action (\ref{eq:FW-action}). One has
\qq
H(x,p)=p\cdot\dot x-L(x,\dot x)=p\cdot Q(x)p+p\cdot K(x)
\qqq
for $p=\frac{1}{2}Q(x)^{-1}(\dot x-K(x))$. The Euler-Lagrange equations 
for the extremal trajectories of the Freidlin-Wentzell action correspond
to the Hamilton equations
\qq
\dot x=\nabla_pH(x,p)=2Q(x)p+K(x)\,,\qquad
\dot p=-\nabla_xH(x,p)=-\nabla_x(p\cdot Q(x)p+p\cdot K(x))\,.
\label{eq:H-equations}
\qqq
The dynamical system (\ref{eq:H-equations}) in the phase space $\mathbb R^{2d}$
possesses a hyperbolic fixed point with $(x,p)=(\bar{x},0)$. Indeed, the right 
hand sides of Eqs.\,(\ref{eq:H-equations}) vanish at this point and 
the linearization of (\ref{eq:H-equations}) around $(\bar{x},0)$ has the form
\qq
\dot x=2Q(\bar{x})p+Ax\,,\qquad \dot p=-A^Tp\,.
\label{eq:lin-flow}
\qqq
Phase space $\mathbb R^{2d}$ may be split in a unique way into as a direct sum
$V_s\oplus V_u$ of the stable and unstable invariant subspaces of the 
linearized flow (\ref{eq:lin-flow}), 
\qq
V_s=\{(x,0)\,|\,x\in\mathbb R^d\}\,,\qquad
V_u=\{(x,Bx)\,|\,x\in\mathbb R^d\}\,,
\qqq
where 
\qq
B^{-1}=2\int\limits_0^\infty\ee^{tA}\,Q(\bar{x})\,\ee^{tA^T}\,dt\,.
\label{eq:B-1}
\qqq
is the unique positive definite matrix satisfying the relation
\qq
BA+A^TB=-2B\,Q(\bar{x})B\,.
\label{eq:ABQ}
\qqq
On $V_s$ and $V_u$ the linearized flow reduces to
\qq
(\dot x,0)=(Ax,0)\,,\qquad(\dot x,B\dot x)=(-B^{-1}A^TBx,-A^TBx)
\qqq
and is, respectively, exponentially contracting and exponentially
expanding (the eigenvalues of $-B^{-1}AB$ are the negatives of those
of $A$ and have positive real parts). 
The subspaces $V_s$ and $V_u$ are Lagrangian, i.e. the symplectic
form $\omega=dp\cdot dx$ vanishes when restricted to each of them.
It follows from a general theory of hyperbolic fixed points, see 
e.g.~\cite{teschl2012,irvin2001},
that in the vicinity of $(\bar{x},0)$ there exist
unique stable and unstable smooth $d$-dimensional submanifolds 
$M_s$ and $M_u$ composed of close points tending to $(\bar{x},0)$ under,
respectively, flow (\ref{eq:H-equations}) and its time-reversal.
$M_s$ is a local piece of $V_s$ around $\bar{x}$ and $M_u$ has 
$V_u$ as the tangent space at $(\bar{x},0)$, see Figure \ref{fig:H-flow}.
Besides, if $K$ and $Q$ are (real) analytic then the submanifolds $M_s$ and 
$M_u$ are also analytic, see Theorem 7.1 in \cite{ilyashenko2008}.
 
Both $M_s$ and $M_u$ are Lagrangian submanifolds of the phase space 
(i.e. 2-form $\omega$ vanishes when restricted to their tangent subspaces). 
This follows from the Hamiltonian nature of the flow (\ref{eq:H-equations}) 
which preserves $\omega$. Besides the Hamiltonian $H$ (conserved by the 
flow (\ref{eq:H-equations})) has to vanish both on $M_s$ and on $M_u$
since it vanishes at $(\bar{x},0)$. We may now define for $x$ 
in a small ball $D$ around $\bar{x}$
\qq
F(x)=\int_{\bar{x}}^xp\cdot dx
\qqq
with the result independent of the integration path $(x(t),p(t))$
in $M_u$ such that $x(t)$ lies in $D$. Then $M_u$ is given locally
by the equation $p(x)=\nabla F(x)$, where $(x,p(x))\in M_u$.
Clearly, $F$ is a smooth function on $D$ (which is analytic if $K$
and $Q$ are). The vanishing of the Hamiltonian $H$ on $M_u$ implies 
now that $F$ satisfies the Hamilton-Jacobi equation (\ref{eq:HJ-Chap0}) 
so that we may set 
$F_{\HJ}=F$. Note that for such $F_{\HJ}$, the Hamiltonian dynamics 
on the stable manifold $M_s$ projects to the position space to the 
relaxation dynamics and the one on the unstable manifold $M_u$ to 
the fluctuation dynamics. $F_{\HJ}$ and $\nabla F_{\HJ}$ vanish at $\bar{x}$ 
and the Hessian $\nabla\nabla F_{\HJ}(\bar{x})=B$ is positive definite.

Conversely, if $F_{\HJ}$ is a local solution of the Hamilton-Jacobi equation
around $\bar{x}$ with the latter properties then, for a small ball $D$ around 
$\bar{x}$, the sets $\{(x,0)\,|\,x\in D\}$ and $\{(x,\nabla F_{\HJ}(x))\,|\,x
\in D\}$  form, respectively, the local stable and unstable manifolds 
of the fixed point $(\bar{x},0)$ of the Hamiltonian flow (\ref{eq:H-equations}) 
so that $F_{\HJ}=F$.

The above argument shows also the local existence of a transverse 
decomposition (\ref{eq:Chap1-K-split}) and (\ref{eq:Chap1-K-split2})
of the vector field $K$ around its non-degenerate stable zeros $\bar{x}$.
The decomposition has positive definite Hessian 
$\nabla\nabla F_{\HJ}(\bar{x})$ and is uniquely determined by this property.      

The results discussed here, together with the local equality 
$F_{\HJ}=F_{\bar{x}}$, show that the quasi-potential relative to $\bar{x}$ 
is smooth (or analytic for $K$ and $Q$ analytic) in the vicinity of 
a non-degenerate attractive fixed point $\bar{x}$ of the deterministic 
dynamics (\ref{eq:det-dyn}).
It is however well known that $F_{\bar{x}}$ does not have 
to be smooth everywhere, see 
e.g.~\cite{graham1985,graham1995,kamenev2011,maier1996} and 
references therein. Such non-smoothness occurs if the unstable 
manifold of the fixed point $(\bar{x},0)$ of the Hamiltonian dynamics
(\ref{eq:H-equations}) has tangent vectors perpendicular to $V_s$ 
leading to the caustics in its projection on $V_s$, see 
Figure \ref{fig:H-flow}. In recent literature, such situations were 
connected to the so called Lagrangian phase transitions \cite{bertini2014}. 
Non-smooth quasi-potentials may be treated using viscous solutions 
of Hamilton-Jacobi equations \cite{cameron2012,fleming2006}. We leave 
such situations to a future investigation.

\nsection{Perturbative expansions of quasi-potentials}\label{sec:Chap2}
Let us consider a finite-dimensional stochastic dynamics depending smoothly
on a real external parameter $\lambda$,
\qq\label{eq:Chap2-dynamics}
\dot{x} = K^{\lambda}(x) + \sqrt{2\epsilon}\,g^{\lambda}(x)\,\eta_t\,.
\qqq
We shall use the notation $Q^{\lambda}=g^{\lambda}(g^{\lambda})^T$ for the noise 
covariance and $F^{\lambda}$ for the quasi-potential relative to the attractor 
$\bar{x}^{\lambda}$ of the dynamics $\dot{x} = K^{\lambda}(x)$ (dropping
the subscript indicating the attractor). For simplicity, we only consider 
fixed points as attractors, even if all the results in this 
section can be generalized to other kinds of attractors.

An explicit calculation of the quasi-potentials $F^{\lambda}$ for the dynamics 
of the form (\ref{eq:Chap2-dynamics}) is usually an impossible task. 
However, it is feasible in particular situations, as for example in the cases 
when (\ref{eq:Chap2-dynamics}) respects detailed balance. It is thus very 
natural to ask the following question. Supposing that we are able to calculate 
the quasi-potential for a given value of $\lambda$, say $\lambda=0$, can we 
perturbatively calculate $F^{\lambda}$ for small $\lambda$?
Some works are available in the literature containing the first-order analysis 
\cite{graham1983,graham1984,graham1984a,graham1985a,jauslin1986,gang1989,maier1993} 
and concentrating mainly on specific examples. In \cite{tel1989}, the 
first-order theory is presented in a general fashion and, recently, a rigorous 
$1^{\rm st}$ order analysis has been obtained in \cite{poquet2014}. We extend here 
the approach of  \cite{tel1989} to any order.

Before discussing in details how the perturbative approach is built in 
the rest of the section, we first summarize the main ideas. The strategy is 
to consider the quasi-potential as a solution to the Hamilton-Jacobi 
equation
\qq\label{eq:Chap2-HJ}
\nabla F^{\lambda}\cdot \left[ Q^{\lambda}\nabla F^{\lambda} 
+ K^{\lambda}\right](x) = 0\,,
\qqq
which holds under the hypotheses discussed in Sec.\,\ref{subsec:Chap1-HJ}. 
In particular, we shall assume that the vector
field $K^0$ has a non-degenerate stable zero $\bar{x}^0$. 
From the Implicit Function Theorem, it follows that for sufficiently small
$|\lambda|$ there exists a smooth family $\bar{x}^\lambda$ of 
non-degenerate zeros of $K^\lambda$. By a slight modification 
of the arguments in Secs.\,\ref{subsec:Chap1-HJ} and 
\ref{subsec:Chap1-Hamiltonian-picture}, invoking the dependence
on a parameter of the unstable manifold of a hyperbolic fixed point,
we infer that there exists a smooth family of solutions
$F_{\HJ}^\lambda(x)$ of the Hamilton-Jacobi equation (\ref{eq:Chap2-HJ})
defined in a neighborhood of $\lambda=0$ and $\bar{x}^0$ and that 
those solutions coincide with the quasi-potentials $F^\lambda$ 
for the stochastic dynamics (\ref{eq:Chap2-dynamics}) relative to 
the attractors $\bar{x}^\lambda$ on the set $D_{\HJ}^0$ 
for a sufficiently small neighborhood $D$ of $\bar{x}^0$.  
$D_{\HJ}^0$ is given by Eq.\,(\ref{eq:DHJ}) for $\lambda=0$. Besides
$F^\lambda(\bar{x}^\lambda)=0$, $\nabla F^\lambda(\bar{x}^\lambda)=0$
and the Hessians $\nabla\nabla F^\lambda(\bar{x}^\lambda)$ are positive
definite. For $K^\lambda(x)$ and $Q^\lambda(x)$ analytic in
$\lambda$ and $x$, both $\bar{x}^\lambda$ and $F_{\HJ}^\lambda(x)$ will be 
analytic for sufficiently small $|\lambda|$.

For $\bar{x}^0+y\in D_{\HJ}^0$, we shall expand the 
function $\lambda\mapsto F^\lambda(\bar{x}^\lambda+y)$, well defined
and smooth for sufficiently small $|\lambda|$, 
into the infinite Taylor series around $\lambda=0$, writing
\qq\label{eq:Chap2-centred-expansion-F}
F^{\lambda}(\bar{x}^\lambda+y)=\sum_{n=0}^{\infty} \lambda^n F^{(n)}(y)\,.
\qqq
Such a Taylor expansion is asymptotic in the smooth case but
has a finite radius of convergence in the analytic case.
An analogous notation will be used for the expansions of 
$K^{\lambda}$ and $Q^{\lambda}$ centered at $\bar x^\lambda$. We suppose that 
$F^{(0)}$ is explicitly known and attempt to calculate
the perturbative coefficients $F^{(n)}(y)$. Inserting the above expansion 
into the Hamilton-Jacobi equation (\ref{eq:Chap2-HJ}), one obtains a hierarchy 
of equations expressing $F^{(n)}$ in terms of functions $F^{(k)}$ 
with $k<n$ so that they may be solved iteratively.
We stress that in Eq.\,(\ref{eq:Chap2-centred-expansion-F}) we have moved 
attractors $\bar{x}^\lambda$ to the origin. This may seem just a detail but 
it is important in order to get a simple proof that the equations for 
$F^{(n)}$ admit a unique solution for $\bar{x}^0+y\in D_{\HJ}^0$. Recall
from Sec.\,\ref{subsec:Chap1-HJ} that $\bar{x}^0$ may be connected
to each $x\in D_{\HJ}^0$ by a unique trajectory of the
$\lambda=0$ fluctuation dynamics which, after the shift by $-\bar{x}^0$,
takes the form
\qq\label{eq:Chap2-0-order-fluct-dynamics-0}
\dot{y}= K_r^{(0)}(y)=2\left( Q^{(0)}(y)\nabla F^{(0)}(y)\right) 
+K^{(0)}(y)\,.
\qqq
The solution for each $F^{(n)}$ can be obtained with the method of  
characteristics by integrating along the trajectories of 
(\ref{eq:Chap2-0-order-fluct-dynamics-0}), see 
Eq.\,(\ref{eq:solution-ngr1}) below for the explicit expression for $F^{(n)}$. 
Such observation is of practical importance: to 
compute $F^{(n)}$, the only difficulty is to compute solutions of 
Eq.\,(\ref{eq:Chap2-0-order-fluct-dynamics-0}). This might not be doable 
analytically, but it is a simple problem for a numerical treatment.

From the  practical point of view it may be more convenient to implement 
a modified version of perturbative expansion where we replace  
Eq.\,(\ref{eq:Chap2-centred-expansion-F}) with  a Taylor expansion of
the function $\lambda\mapsto F^\lambda(x)$ at $\lambda=0$ for 
$x\in D_{\HJ}^0$\,:
\qq\label{eq:Chap2-uncentred-expansion-F}
F^{\lambda}(x)=\sum_{n=0}^{\infty} \lambda^n \hat{F}^{(n)}(x)\,.
\qqq
We shall see an example where expansion (\ref{eq:Chap2-uncentred-expansion-F}),
that we shall call direct, 
is simpler to implement than Eq.\,(\ref{eq:Chap2-centred-expansion-F}). 
The difference between the two expansions comes from the fact that here 
we do not move the $\lambda$-attractors to the origin. In this case, however,
the simplest route to prove that functions $\hat{F}^{(n)}$ may be again found 
iteratively with the method of characteristics, is to reduce 
expansion (\ref{eq:Chap2-uncentred-expansion-F}) to 
(\ref{eq:Chap2-0-order-fluct-dynamics-0}), see  
Sec.\,\ref{sec:Chap2-uncentred}.

Finally we shall show in Sec.\,\ref{subsec:Chap2-around-attractor} that, 
for a suitable choice of $K^{\lambda}$ and $Q^{\lambda}$, the expansion of 
Eq.\,(\ref{eq:Chap2-centred-expansion-F}) reduces to the Taylor expansion 
of the quasi-potential $F$ associated to Eq.\,(\ref{eq:dynamics-Chap1}) 
around the attractor $\bar{x}$.


\subsection{Expansion centered on attractors of the perturbed dynamics}
\label{sec:Chap2-centred}

This section discusses  the perturbative solution of the Hamilton-Jacobi 
equation (\ref{eq:Chap2-HJ}) in the form (\ref{eq:Chap2-centred-expansion-F}). 
We iteratively construct functions $F^{(n)}$ obtaining, in such a way, 
two results: first, we prove that the equations obeyed by $F^{(n)}$ admit a 
unique solution and, second, we give an explicit recursive formula that may 
be used to practically calculate $F^{(n)}$.

As already mentioned, we assume that $\bar{x}^0$ is a non-degenerate 
fixed point of the relaxation dynamics at $\lambda=0$. This means
that $K^0(\bar{x}^0)=0$ 
and the eigenvalues of $\nabla K^0(\bar{x}^0)$ have negative real parts. 
The assumption ensures that quasi-potential $F^0$ relative to $\bar{x}^0$
(recall that we have dropped the subscript indicating the attractor) 
is given by the solution to the Hamilton-Jacobi equation in the neighborhood 
$D_{\HJ}^0$ of $\bar{x}^0$, that $F^0(\bar{x}^0)$ and 
$\nabla F^0(\bar{x}^0)$ vanish, and that the Hessian 
$\nabla\nabla F^0(\bar{x}^0)$ is positive definite 
and $F^0(x)>0\,$ for $x\in D_{\HJ}^0$, $x\not=\bar{x}^0$, see 
Sec.\,\ref{subsec:Chap1-HJ}. 

Let us expand the drift and the noise covariance in powers of $\lambda$ after
shifting the attractor $\bar{x}^\lambda$ of the perturbed
deterministic dynamics to the origin:
\qq\label{eq:Chap2-Kn-Qn-1}
K^{\lambda}(\bar{x}^\lambda+y)=\sum_{n=0}^{\infty} \lambda^n\, K^{(n)}(y)\,,\qquad
Q^{\lambda}(\bar{x}^\lambda+y)=\sum_{n=0}^{\infty} \lambda^n\, Q^{(n)}(y)\,.
\qqq
Several properties that will be used below follow simply. 
Because $\bar{x}^\lambda$ is a fixed point for the deterministic dynamics, 
$K^{(n)}(0)=0$. Similarly, since $F^\lambda(\bar{x}^\lambda)=0$ and 
$\nabla F^\lambda(\bar{x}^\lambda)=0$, the relations 
$F^{(n)}(0)=0$ and $\nabla  F^{(n)}(0)=0$ must hold for every $n$. Moreover, 
the eigenvalues of $\nabla K^{(0)}(0)$ have negative real parts and 
$\nabla \nabla F^{(0)}(0)$ and  $Q^{(0)}$ are positive definite.

Inserting expansions (\ref{eq:Chap2-centred-expansion-F}) and
(\ref{eq:Chap2-Kn-Qn-1}) into the Hamilton-Jacobi equation 
(\ref{eq:Chap2-HJ}), we obtain the power-series identity
\qq\label{eq:provvisoria}
\sum_{n=0}^{\infty}\lambda^n\,\,\sum_{k=0}^n\,\left[\,\sum_{l=0}^{n-k} \, 
\nabla F^{(n-k-l)}\cdot Q^{(k)}\nabla F^{(l)} + \nabla F^{(n-k)}
\cdot K^{(k)}\right](y)=0\,.
\qqq
Upon equating to zero order by order, this gives a hierarchy of relations
\qq\label{eq:HJ-expanded-F0}
\hspace{-0.05cm}\nabla F^{(0)}\cdot\left[Q^{(0)}\nabla F^{(0)} + K^{(0)}\right]=0\qquad \qquad\qquad\qquad\hspace{0.47cm}{\rm for}\qquad n=0\,,
\qqq
\vskip -0.8cm
\qq\label{eq:HJ-expanded}
\nabla F^{(n)}\cdot K^{(0)}_r = S^{(n)}[F^{(0)},\dots,F^{(n-1)}]\quad \qquad\qquad
\qquad{\rm for}\qquad n>0\,,
\qqq
where 
\qq
K^{(0)}_r = 2\,Q^{(0)}\nabla F^{(0)} + K^{(0)}\,
\qqq
and $S^{(n)}$ is a functional of $F^{(0)},\dots,F^{(n-1)}$ given by 
\qq\label{eq:Chap2-Sn-1}
S^{(n)}[F^{(0)},\dots,F^{(n-1)}] &=&\hspace{-0.1cm}
-\sum_{k=1}^{n-1}\,\bigg[ \nabla F^{(n-k)}
\cdot \left(Q^{(0)}\nabla F^{(k)} + K^{(k)}\right)+ \sum_{l=0}^{n-k} 
\nabla F^{(n-k-l)}\cdot Q^{(k)}\nabla F^{(l)} \bigg]\qquad\cr
&&\hspace{-0.1cm}
-\,\nabla F^{(0)}\cdot \left[Q^{(n)}\nabla F^{(0)}  +  K^{(n)}\right].
\qqq
We have arranged Eqs.\,(\ref{eq:HJ-expanded}) in such a way that $F^{(n)}$ 
appears only on the left hand side. Eq.\,(\ref{eq:HJ-expanded-F0}) is nothing 
else but the Hamilton-Jacobi equation corresponding to the dynamics 
(\ref{eq:Chap2-dynamics}) with $\lambda=0$, once we have moved the attractor 
to the origin. We assumed that its solution $F^{(0)}(y)$ is known.
In the following, we prove that solutions to Eqs.\,(\ref{eq:HJ-expanded}) 
for $n>0$ exist and are unique. An explicit formula will permit
to obtain $F^{(n)}$ given $F^{(k)}$ for $k<n$. We start by 
looking at the properties of the $0^{\rm th}$ order fluctuation dynamics 
$\dot{y}=K_r^{(0)}(y)$ when the norm $|y|$ is small. These results will 
be useful to prove the existence of solutions to Eqs.\,(\ref{eq:HJ-expanded}).


\subsubsection{Fluctuation dynamics: exponential escape from attractor}
\label{subsubsec:Chap2-exponential-escape}

Let us consider Eq.\,(\ref{eq:Chap2-0-order-fluct-dynamics-0}) 
describing the $0^{\rm th}$ order fluctuation dynamics after the shift of 
attractor to the origin. From the results of Sec.\,\ref{subsec:Chap1-HJ}
about the trajectories of the fluctuation dynamics
it follows that for $y\in D_{\HJ}^0-\bar{x}^0\equiv D_0$ there exists
a unique solution $[-\infty,0]\ni t\mapsto\tilde y(t,y)$ of 
Eq.\,(\ref{eq:Chap2-0-order-fluct-dynamics-0}) such that
$\tilde y(-\infty,y)=0$ and $\tilde y(0,y)=y$. Besides, 
$\tilde y(t,y)$ belongs to $D_0$ for all $t$.

We now show that $\tilde y(t,y)$ escapes from the attractor $y=0$ 
exponentially fast. Indeed, we can write for $|y|$ small
\qq\label{eq:escape-reversed}
\tilde y(t,y)=\ee^{t(2Q^{(0)}B +A)}\,y\,+o\left(\Big|
e^{t(2Q^{(0)}B +A)}\,y\Big|\right).
\qqq
In the above expressions, $A$ and $B$ are defined through the small $|y|$ expansion of $K^{(0)}$ and $F^{(0)}$: 
\qq
K^{(0)}(y)=Ay+o(|y|)\,,\qquad
F^{(0)}(y)=\frac{_1}{^2}y\cdot By + o\left(|y|^2\right),
\qqq
i.e. $A=(\nabla K^{(0)}(0))^T$ and $B=\nabla\nabla F^{(0)}(0)$.
One should recall from the properties listed above for $K^{(0)}$ and 
$F^{(0)}$ that the eigenvalues of $A$ have negative real parts and $B=B^T$
is positive definite. We have encountered matrices $A$ and $B$ already
before when studying the Hamiltonian dynamics in 
Sec.\,\ref{subsec:Chap1-Hamiltonian-picture}. They are related by 
the identity 
\qq\label{eq:Lyapunov-1}
BA + A^TB + 2 B\,Q^{(0)}B=0 
\qqq
imposed by the $2^{\rm nd}$ order in $y$ contribution 
to the $0^{\rm th}$ order Hamilton-Jacobi equation (\ref{eq:HJ-expanded-F0}).
Note that identity (\ref{eq:Lyapunov-1}), which implies that
\qq
2Q^{(0)}B+A=-B^{-1}A^TB\,, 
\qqq
coincides with Eq.\,(\ref{eq:ABQ}) solved by (\ref{eq:B-1}) if we replace 
in the latter $Q(\bar{x})$ by $Q^{(0)}(0)$.  
Behavior (\ref{eq:escape-reversed}) is dictated by 
the spectrum of $(2Q^{(0)}B+A)$. Indeed, as already noticed before,
the eigenvalues of $-BA^TB$ have positive real parts.
Thus the $0^{\rm th}$-order fluctuation dynamics escapes exponentially fast from 
the attractor. This agrees with the analysis of 
Sec.\,\ref{subsec:Chap1-Hamiltonian-picture} where we showed that the 
fluctuation dynamics on the position space corresponds to the Hamiltonian 
dynamics on the unstable manifold of the hyperbolic fixed point.


\subsubsection{Iterative solution}
\label{subsec:Chap2-centred-expansion-iterative-solution}

We shall prove that the unique solution of 
Eqs.\,(\ref{eq:HJ-expanded}) on $D_0=D_{\HJ}^0-\bar{x}^0$ that satisfies $F^{(n)}(0)=0$ 
is  
\qq\label{eq:solution-ngr1}
F^{(n)}(y) = \int_{-\infty}^0\,S^{(n)}[F^{(0)},\dots,F^{(n-1)}](\tilde {y}(t,y))
\,dt\,,
\qqq
where $\tilde y(t,y)$ is the trajectory of the $0^{\rm th}$-order 
fluctuation dynamics (\ref{eq:Chap2-0-order-fluct-dynamics-0}) joining
the origin to $y$ that was discussed 
above. Moreover, the expression (\ref{eq:solution-ngr1}) is well-defined, 
as we show that the integral in this expression is convergent whenever 
$y$ belongs to $D_0$.
Eq.\,(\ref{eq:solution-ngr1}) gives an iterative solution of 
(\ref{eq:HJ-expanded}). 

Let us start by proving that if a solution of Eq.\,(\ref{eq:HJ-expanded}) 
such that $F^{(n)}(0)=0$ exists, it has to have the form 
(\ref{eq:solution-ngr1}). This is easily 
seen by taking the total time derivative of $F^{(n)}(\tilde y(t,y))$,
\qq
\frac{d}{dt} F^{(n)}(\tilde y(t,y))&=&
 \nabla F^{(n)}(\tilde y(t,y)) \cdot \frac{d\tilde y(t,y)}{dt}
=\left( \nabla F^{(n)} \cdot K^{(0)}_r\right)(\tilde y(t,y))\cr
&=&S^{(n)}[F^{(0)},\dots,F^{(n-1)}](\tilde y(t,y))\,,
\qqq
and by integrating over time. 

In the next step, we shall show that, assuming that 
the integral on the right hand side of (\ref{eq:solution-ngr1}) converges, 
the latter equation gives a function $F^{(n)}(y)$ 
that solves Eq.\,(\ref{eq:HJ-expanded}).
Indeed, on the fluctuating dynamics trajectory,  
\qq
F^{(n)}(\tilde y(s,y)) &=&  \int_{-\infty}^0S^{(n)}[\cdots]
(\tilde y(t,\tilde y(s,y)))\,dt = \int_{-\infty}^0S^{(n)}[\cdots]
(\tilde y(t+s,y))\,dt\cr
&=& \int_{-\infty}^s\,S^{(n)}[\cdots](\tilde y(t,y))\,dt\,,
\qqq
where we used the relation $\bar{y}(t,\bar{y}(s,y)) = \bar{y}(t+s,y)$ 
that holds because of the uniqueness of the solutions of 
Eq.\,(\ref{eq:Chap2-0-order-fluct-dynamics-0}). Now, deriving the 
previous expression with respect to $s$ and evaluating at $s=0$, we obtain 
Eq.\,(\ref{eq:HJ-expanded}). This means that expression 
(\ref{eq:solution-ngr1}) solves Eq.\,(\ref{eq:HJ-expanded}) provided that 
it is well defined. 

To complete the proof, we need to show that the integral appearing 
in Eq.\,(\ref{eq:solution-ngr1}) is convergent and vanishes at $y=0$. 
To show this, we have to analyze the behavior of the integrand when 
$t\to -\infty$. This corresponds to studying the behavior of the 
integrand for small $|y|$. One completes the prove combining 
Eq.\,(\ref{eq:escape-reversed}) with the observation that the expression
$S^{(n)}$ defined in Eq.\,(\ref{eq:Chap2-Sn-1}) 
is at least quadratic in $y$ for $|y|$ small if $F^{(k)}$ for $k<n$
have the same property. Under this assumption, that is true for $F^{(0)}$,
the integrand on the right hand side of Eq.\,(\ref{eq:solution-ngr1})
converges exponentially to zero when $t\to-\infty$ so that the
time integral converges. Besides, it determines function $F^{(n)}(y)$ 
that is at least quadratic in $y$ for small $|y|$ so that it may be shown
inductively  that $F^{(n)}$ starts at worst quadratically.

Let us summarize our results: we described an iterative scheme to calculate 
perturbatively the quasi-potential $F^{\lambda}$ of 
Eq.\,(\ref{eq:Chap2-dynamics}) as a power series 
(\ref{eq:Chap2-centred-expansion-F}). Once we know 
the terms $F^{(k)}$ of that expansion for $k<n$ then
$F^{(n)}$ may be obtained using Eq.\,(\ref{eq:solution-ngr1}). In that 
formula, $S^{(n)}$ is defined by Eq.\,(\ref{eq:Chap2-Sn-1}) and 
$\tilde y(t,y)$ is the solution to the $0^{\rm th}$-order fluctuation 
dynamics (\ref{eq:Chap2-0-order-fluct-dynamics-0}) that starts at the origin
at $t=-\infty$ and arrives at $y$ at time zero. Finally, we proved that this 
procedure is well defined ($F^{(n)}$ 
are finite quantities and depend smoothly, or analyticaly in the analytic
case, on $y\in D_0$). More generally, 
one could define this way $F^{(n)}(y)$ for $y$ belonging to the basin 
of attraction of the origin for the time-reversal of the fluctuation 
dynamics (\ref{eq:Chap2-0-order-fluct-dynamics-0}).

Let us conclude this section observing that the procedure described here 
may be very easily implemented numerically. Indeed, to calculate 
$F^{(n)}$ at all the orders, one only needs to compute the solution to 
the $0^{\rm th}$-order fluctuation dynamics 
(\ref{eq:Chap2-0-order-fluct-dynamics-0}). The 
scheme just described gives a powerful practical tool to compute
quasi-potentials perturbatively. 


\subsection{Direct Expansion}\label{sec:Chap2-uncentred}

In Sec.\,\ref{sec:Chap2-centred} we have presented an expansion in power 
of $\lambda$ centered on the attractors of the perturbed dynamics. We discuss 
in this section a direct expansion, which does not depend on the knowledge 
of the attractors of the perturbed dynamics. The main point we want to stress 
in that case is the appearance of non-trivial solvability conditions.

There are several reason why, in some cases, this new expansion may be simpler 
or more relevant than the one centered on the attractors. The main one is 
that the attractors of the perturbed dynamics may not be known explicitly 
and should then be computed themselves by a perturbative expansion. In that 
case, as the definition of  $K^{(n)}$ and $Q^{(n)}$ given in 
Eq.\,(\ref{eq:Chap2-Kn-Qn-1}) involves the attractor $\bar{x}^{\lambda}$ of 
the perturbed dynamics $\dot{x}=K^{\lambda}(x)$, then matrices $Q^{(n)}$ may be 
non-zero at all orders even if the covariance $Q^{\lambda}$ is independent of 
$\lambda$. An example of this kind will be encountered in 
Sec.\,\ref{sub:Chap3-perturbative}. 

We assume power series expansions for $K^{\lambda}$ and $Q^{\lambda}$
\qq\label{eq:Chap2-Kn-Qn-uncentred}
K^{\lambda}(x)=\sum_{n=0}^{\infty} \lambda^n \hat{K}^{(n)}(x)\,,\qquad
Q^{\lambda}(x)=\sum_{n=0}^{\infty} \lambda^n \hat{Q}^{(n)}(x)\,.
\qqq
Observe that $\hat{K}^{(0)}(\bar{x}^0)=0$ and  
$\nabla  \hat{F}^{(0)}(\bar{x}^0)=0$. By assumption all the eigenvalues of 
$\nabla \hat{K}^{(0)}(\bar{x}^0)$ have negative real part,
and $Q^{(0)}(\bar{x}^0)$ is a
positive definite matrix. This implies that $\nabla \nabla F^{(0)}(\bar{x}^0)$ 
is a positive definite matrix. As $\nabla \hat{K}^{(0)}(\bar{x}^0)$ is invertible, the identity $K^\lambda(\bar{x}^\lambda)=0$
permits to solve iteratively for the coefficients of the Taylor
expansion 
\qq
\bar{x}^\lambda=\sum_{n=0}^\infty\lambda^n\hat x^{(n)}\,.
\label{eq:attr-exp}
\qqq
for the perturbed attractor, where $\hat x^{(0)}=\bar{x}^0$.

Inserting Eq.\,(\ref{eq:Chap2-uncentred-expansion-F}) and 
(\ref{eq:Chap2-Kn-Qn-uncentred}) into the Hamilton-Jacobi equation 
(\ref{eq:Chap2-HJ}), we obtain a hierarchy identical to 
(\ref{eq:HJ-expanded-F0}) and (\ref{eq:HJ-expanded}). The only differences 
are, of course, that $y$ has to be replaced with $x$ and $F^{(n)}$, $K^{(n)}$ 
and $Q^{(n)}$ by $\hat{F}^{(n)}$, $\hat{K}^{(n)}$ and $\hat{Q}^{(n)}$. 
Explicitly, we have
\qq\label{eq:HJ-expanded-F0-naive}
\nabla \hat{F}^{(0)}(x)\cdot \left[\hat{Q}^{(0)}(x)\nabla \hat{F}^{(0)}(x) 
+ \hat{K}^{(0)}(x)\right]=0\quad\qquad\qquad\qquad{\rm for}\qquad n=0\,,
\qqq
\vskip -0.6cm
\qq\label{eq:HJ-expanded-naive}
\nabla \hat{F}^{(n)}(x)\cdot \hat{K}^{(0)}_r(x) = \hat{S}^{(n)}[\hat{F}^{(0)},
\dots,\hat{F}^{(n-1)}](x)\quad\qquad\qquad\quad\  
\hspace{-0.09cm}{\rm for}\qquad n\neq 0\,,
\qqq
where 
\qq
\hat{K}^{(0)}_r = 2\,\hat{Q}^{(0)}\nabla \hat{F}^{(0)} + \hat{K}^{(0)}
\qqq
and $\hat{S}^{(n)}$ is the functional of $\hat{F}^{(0)},\dots,\hat{F}^{(n-1)}$ 
given by Eq.\,(\ref{eq:Chap2-Sn-1}) with $F^{(n)}$, $K^{(n)}$ and 
$Q^{(n)}$ replaced by the hatted quantities.

We first remark that, as $\hat{K}^{(0)}_r(x^0)=0$, 
Eq.\,(\ref{eq:HJ-expanded-naive}) implies 
\qq\label{eq:Chap-2-naive-solvability}
\hat{S}^{(n)}[\hat{F}^{(0)},\dots,\hat{F}^{(n-1)}]\Big|_{x=\bar x^0}= 0\,,
\qqq
which appears as a solvability condition for Eq.\,(\ref{eq:HJ-expanded-naive}).
It is possible to prove directly this solvability condition by induction, 
however this involves subtle cancellations that are tedious to prove  
to all orders. In order to bypass this proof, we rather argue that we know 
{\it a priori\,} that the series expansion exists. Then 
Eq.\,(\ref{eq:HJ-expanded-naive}) is a consequence of the existence of 
the series expansion and this implies that the solvability condition 
(\ref{eq:Chap-2-naive-solvability}) is satisfied. The existence of the series 
expansion follows from the existence of the series expansion around 
the attractors of the perturbed dynamics, discussed in the previous section. 
Indeed, $\hat F^{(n)}(x)$ may be directly found using the expansion from
the previous section by comparing order by order both sides of the identity 
\qq
\sum_{n=0}^{\infty} \lambda^n \hat{F}^{(n)}(x)=F^{\lambda}(x) = 
\sum_{n=0}^{\infty} \lambda^n F^{(n)}(x-\bar{x}^\lambda)\,,
\qqq
where on the right hand side one inserts the Taylor expansion
(\ref{eq:attr-exp}).

We now explain how to solve Eq.\,(\ref{eq:HJ-expanded-naive}) using 
(\ref{eq:Chap-2-naive-solvability}). Let us consider the $0^{\rm th}$-order 
fluctuation dynamics
\qq\label{eq:fluct-equation-naive}
\dot{x}=\hat{K}^{(0)}_r(x)\,,
\qqq
and its trajectory $\tilde x(t,x)$ lying in $D_{\HJ}^0$  such that 
$\tilde x(-\infty,x)=\bar{x}^0$ and $\tilde x(0,x)=x$.
With the same argument as in 
Sec.\,\ref{subsec:Chap2-centred-expansion-iterative-solution}, we can show 
that 
\qq\label{eq:solution-ngr2}
\hat{F}^{(n)}(x) = C^{(n)} + 
\int_{-\infty}^0\,\hat{S}^{(n)}[\hat{F}^{(0)},\dots,\hat{F}^{(n-1)}](\tilde x(t,x))
\,dt\,.
\qqq
has to hold for a solution of Eq.\,(\ref{eq:HJ-expanded-naive}), where 
$C^{(n)}$ are (for the moment arbitrary) constants. 
This is analogous to Eq.\,(\ref{eq:solution-ngr1}). To obtain
$\hat F^{(n)}(x)$ from this equation, we have to  prove the convergence 
of the integral on the right hand side of (\ref{eq:solution-ngr2}) and to 
fix the constants $C^{(n)}$. The first task requires the control of 
the behavior of $\hat S^{(n)}[\hat{F}^{(0)},\dots,\hat{F}^{(n-1)}](x)$ around 
$x=\bar{x}^0$. It may be achieved by induction using 
(\ref{eq:Chap-2-naive-solvability}) and the exponential relaxations of  
$\tilde x(t,x)$ to $x^0$ when $t$ goes to $-\infty$.
The second task is easier and may be accomplished iteratively since 
the normalization $F^\lambda(\bar{x}^\lambda)=0$ leads upon Taylor expending 
to the relations that allow to express $C^{(n)}=\hat F^{(n)}(\bar{x}^0)$ by
the values at ${\bar x}^0$ of functions $\hat F^{(k)}$ and their derivatives 
and by the coefficients $\hat x^{(k)}$ of the Taylor expansion
(\ref{eq:attr-exp}), all for $k<n$. Note, however, that the choice
of $C^{(k)}$ for $k<n$ in not relevant for the calculation of $\hat F^{(n)}$,
except when it comes to the choice of $C^{(n)}$. Indeed, these are 
the gradients of $\hat F^{(k)}$ for $k<n$ that enter $\hat{S}^{(n)}$.


\subsection{Taylor expansion of the quasi-potential around the attractor 
of the unperturbed dynamics}
\label{subsec:Chap2-around-attractor}

Consider now the stochastic evolution (\ref{eq:dynamics-Chap1})
without an external parameter. Let $\bar x$ be a non-degenerate attractive 
point of the deterministic dynamics (\ref{eq:det-dyn}). We shall be interested 
here in the Taylor expansion of the quasi-potential in a neighborhood of 
$\bar{x}\,$:
\qq\label{eq:moments-F-expansion}
F(\bar{x}+y) = \sum_{n=2}^{\infty} \,\left(\nabla^{(n)} F\right)y^{(n)}\,,
\qqq
where $\nabla^{(n)} F$ and $y^{(n)}$ are rank-$n$ tensors with the components 
\qq
\left(\nabla^{(n)} F\right)_{i_1\cdots i_n} = \frac{1}{n!}\,
\frac{\partial^{(n)}F(\bar{x})}{\partial x^{i_1}\hspace{-0.1cm}\cdots
\partial x^{i_n}}\,,
\qquad(y^{(n)})^{i_1\cdots i_n}=y^{i_1}\cdots y^{i_n}
\qqq
and on the right hand side of (\ref{eq:moments-F-expansion}) the 
contraction of all indices is implied. We shall show here that expansion 
(\ref{eq:moments-F-expansion}) can be viewed as a particular case of the 
perturbative expansion considered in Sec.\,\ref{sec:Chap2-centred}. 
Then, all the results obtained there may be applied to 
(\ref{eq:moments-F-expansion}) providing a method to calculate tensors 
$\nabla^{(n)} F$.

With the same notation as in (\ref{eq:moments-F-expansion}), we introduce 
the expansions of $K$ and $Q$ around $\bar{x}\,$: 
\qq\label{eq:moments-KQ-expansion}
K(\bar{x}+y) = \sum_{n=1}^{\infty} \left(\nabla^{(n)} K\right)y^{(n)}\,,\qquad
Q(\bar{x}+y) = \sum_{n=0}^{\infty} \left(\nabla^{(n)} Q\right)y^{(n)}\,.
\qqq
Since $\bar{x}$ is a non-degenerate attractive fixed point of the relaxation 
dynamics,  $\nabla^{(1)}K=\nabla K(\bar{x})$ has eigenvalues with negative 
real parts and $ \nabla^{(2)} F=\frac{1}{2}\nabla\nabla F(\bar{x})$ is 
a positive definite matrix.

Let us introduce a new $\lambda$-dependent system defined by
\qq\label{eq:moments-equivalent-perturbed-system}
K^{\lambda}(\bar{x}+y)\equiv \frac{_1}{^\lambda} K\left(\bar{x} 
+\lambda y\right)\,,\qquad
Q^{\lambda}(\bar{x}+y)\equiv  Q\left(\bar{x} + \lambda y\right)
\qqq
that reduces for $\lambda=1$ to the previous one and depends smoothly
on real $\lambda$. Note that point $\bar{x}$ is a non-degenerate stable 
attractive zero of $K^\lambda$ for all $\lambda$ and that 
\qq
\label{eq:Flambda}
F^{\lambda}(\bar{x}+y) \equiv \frac{_1}{^{\lambda^2}}F\left(\bar{x} + 
\lambda y\right)
\qqq
satisfies the Hamilton-Jacobi equation (\ref{eq:Chap2-HJ}) for all 
$\lambda$ including $\lambda=0$. The scaling with $\lambda $ was 
introduced in such a way that the $\lambda=0$ case gives a non-trivial 
contribution. It allows to align the notations to those of 
Sec.\,\ref{sec:Chap2-centred}. With our choice, we indeed have 
\qq
&&K^0(\bar{x}+y)=(y\cdot\nabla)K(\bar{x})\equiv A y\,,
\qquad Q^0(\bar{x}+y)=Q(\bar{x})\,,\label{eq:nablaK0}\\
&&F^0(\bar{x}+y)=\frac{_1}{^2}(y\cdot\nabla)^2F(\bar{x})\equiv
\frac{_1}{^2} y\cdot By\,.
\label{eq:Hessian0}
\qqq
Moreover
\qq
K^{(n)}(y) =  \Big(\nabla^{(n+1)} K\Big)y^{(n+1)}\,,\qquad
Q^{(n)}(y) = \Big(\nabla^{(n)} Q\Big)y^{(n)}\,,\qquad
F^{(n)}(y) =\Big(\nabla^{(n+2)} F\Big)y^{(n+2)}
\label{eq:KQF}
\qqq
in the notation of Eq.\,(\ref{eq:Chap2-Kn-Qn-1}) and 
(\ref{eq:Chap2-centred-expansion-F}). It is worth stressing that, in this 
context, the role of the unperturbed $\lambda=0$ stochastic dynamics is played 
by the linear dynamic
\qq\label{eq:Chap1-moments-linearized}
\dot{y}=Ay + 
\sqrt{2\epsilon}\,g(\bar{x})\,\eta_t\,
\qqq
with matrix $A$ given by (\ref{eq:nablaK0}), for which the unperturbed 
quasi-potential is the quadratic approximation of $F$ around the attractor.
This is not surprising since the stochastic
equation (\ref{eq:Chap1-moments-linearized}) defines a Gaussian
process of the Orstein-Uhlenbeck type whose invariant
measure is Gaussian with the covariance equal to $\epsilon B^{-1}$,
where $B$ is the Hessian matrix of $F$ at $\bar{x}$, \,see 
(\ref{eq:Hessian0}). The Hamilton-Jacobi equation
reduces for $\lambda=0$ to the identity (\ref{eq:ABQ}) 
with the solution given by (\ref{eq:B-1}). The relaxation and the
fluctuation dynamics are (after the shift of the attractor to the origin), 
respectively,
\qq
\dot{y}=Ay\qquad{\rm and}\qquad\dot{y}=-B^{-1}A^TBy\,.
\qqq
In particular, for $\lambda=0$ the trajectories of the 
fluctuation dynamics that start at $t=-\infty$ from the origin have a particularly 
simple form:
\qq
\tilde y(t,y)=\ee^{-tB^{-1}A^TB}y
\qqq
and Eq.\,(\ref{eq:solution-ngr1}) reduces to the iterative solution 
\qq\label{eq:solution-ngr1-moments}
F^{(n)}(y) = \int_{-\infty}^0\,S^{(n)}[F^{(0)},\dots,F^{(n-1)}](\ee^{-tB^{-1}A^TB}y)\,dt
\qqq
for the Taylor coefficients of $F$ at $\bar{x}$ with $S^{(n)}$ given
Eq.\,(\ref{eq:Chap2-Sn-1}).


\subsection{Codimension-one bifurcations: the critical exponent}
\label{subsec:Chap2-crot-cal-exponents}

Let us return to a family of dynamical systems parameterized by $\lambda$ as 
in Eq.\,(\ref{eq:Chap2-dynamics}) and let us suppose that the deterministic dynamics
$\dot{x}=K^{\lambda}(x)$ has an attractive fixed point 
$\bar{x}^\lambda$ for $\lambda\leq\lambda_c$ that is non-degenerate for
$\lambda<\lambda_c$ and undergoes a codimension-one 
bifurcation at $\lambda=\lambda_c$. Large deviations for normal forms 
corresponding to codimension-one and codimension-two bifurcations were 
discussed in a series of papers in the '70s and '80s. We refer to 
\cite{mangel1979,dykman1979,dykman1980,knobloch1983,graham1987} and 
to the review \cite{tel1989} for a detailed analysis. Here we want only 
to remark that the critical exponent in this framework is equal to that 
of the mean-field theory.

The assumed scenario implies that $\nabla K^{\lambda}(x^\lambda)$ has 
a simple real eigenvalue $\alpha^\lambda$ that is negative for $\lambda<\lambda_c$
such that
\qq 
\alpha^{\lambda_c}=0\qquad{\rm and}\qquad
\frac{d\alpha^\lambda}{d\lambda}\Big|_{\lambda=\lambda_c}\not=0
\qqq
(the saddle-node bifurcation) or a complex eigenvalue
$\alpha^\lambda$ with negative real part for $\lambda<\lambda_c$,
and its conjugate, such that 
\qq
{\rm Re}\,\alpha^{\lambda_c}=0,\qquad
{\rm Im}\,\alpha^{\lambda_c}\not=0\qquad
{\rm and}\qquad 
\frac{d\alpha^\lambda}{d\lambda}\Big|_{\lambda=\lambda_c}\not=0
\qqq
(the Hopf bifurcation). Moreover all the other eigenvalues 
of $\nabla K^{\lambda}(\bar{x}^\lambda)$ have strictly negative
real parts for $\lambda\leq\lambda_c$. 

Let us consider the covariance 
\qq\label{eq:2-moment}
(C^\lambda)^{ij}=\int x^ix^j\,P^\lambda_\infty(x)\,dx\,-\,\Big(\int x^i\,P^\lambda_\infty(x)
\,dx\Big)\Big(\int x^j\,P^\lambda_\infty(x)\,dx\Big)
\qqq
of the invariant measure of the stochastic dynamics (\ref{eq:Chap2-dynamics}).
If the minimum of the quasi-potential $F^\lambda$ describing the behavior
(\ref{eq:Chap1-LD-invariant-measure}) of $P^\lambda_\infty$ in the limit 
of small noise is attained at $\bar{x}^\lambda$ then the saddle-point analysis
of (\ref{eq:2-moment}) implies that
\qq\label{eq:chap2-variance-taylor}
\lim_{\epsilon\to0}\ \frac{_1}{^\epsilon} C^\lambda\,=\,
(B^\lambda)^{-1}=\,2\int_0^\infty e^{tA^\lambda}
Q^\lambda(\bar{x}^\lambda)\,\ee^{t(A^\lambda)^T}dt
\qqq
where $B^\lambda$ is the Hessian matrix of $F^\lambda$ at $\bar{x}^\lambda$
and the last equality with $A^\lambda=(\nabla K^\lambda(\bar{x}^\lambda))^T$
follows from Eq.\,(\ref{eq:B-1}). Hence $(B^\lambda)^{-1}$ may be viewed
as the small-noise limit of the stationary equal-time connected 2-point 
function of the stochastic process solving Eq.\,(\ref{eq:Chap2-dynamics}). 
Let us examine the  behavior of the latter limiting quantity when
$\lambda\nearrow\lambda_c$. Any vector $y\in \mathbb R^d$  may be 
decomposed as
\qq
y=\beta^\lambda v^\lambda+\overline{\beta^\lambda v^\lambda}+y^\lambda
\qqq
where $\beta^\lambda\in\mathbb C$, $v^\lambda$ is the eigenvector 
of $(A^\lambda)^T$ with the eigenvalue $\alpha^\lambda$ and 
$y^\lambda\in\mathbb R^d$ belongs to the invariant subspace of 
$(A^\lambda)^T$ corresponding to the other eigenvalues so that   
\qq
|\ee^{t(A^\lambda)^T}y^\lambda|\leq\ee^{-\delta t}
\qqq
in the vicinity of $\lambda_c$ for some $\delta>0$. Then
the 2-point function
\qq
y\cdot(B^\lambda)^{-1}y\,=\,-2|\beta^\lambda|^2({\rm Re}\,\alpha^\lambda)^{-1}
\,\overline{v^\lambda}\cdot Q^\lambda(\bar{x}^\lambda)\,v^\lambda
-2\,{\rm Re}\left[(\beta^\lambda)^2(\alpha^\lambda)^{-1}\,v^\lambda\cdot 
Q^\lambda(\bar{x}^\lambda)\,v^\lambda\right]\,+\,O(1)
\qqq
where $O(1)$ term is regular when $\lambda\nearrow\lambda_c$.
Assume that $\beta^{\lambda_c}\not=0$ which holds for generic $y$.
Then, when $\lambda\nearrow\lambda_c$, the first term on the right 
hand side diverges like $(\lambda_c-\lambda)^{-1}$ and the second term 
is regular if the purely imaginary eigenvalue $\alpha_{\lambda_c}\not=0$, 
and both terms have the same divergence when $\alpha_{\lambda_c}=0$. 

This completes the simple proof that, in the framework of a dynamical systems 
in the low noise limit undergoing a codimension-one bifurcation, critical 
exponent for two-points correlator is equal to $1$, the mean field 
theory value of the susceptibility exponent $\gamma$.




\nsection{Mean-field systems and the Shinomoto-Kuramoto model}
\label{sec:mean-field}

We consider in this section systems composed of many diffusive particles, 
interacting through a mean-field type two-body potential and driven out of 
equilibrium by external forces. Most of our analysis is valid for any model 
in this class (the two-body interaction could even be non-potential). 
However, in order to go beyond formal results, we discuss in detail the 
example of Shinomoto-Kuramoto system, a simple $1$-d model first introduced
in \cite{shinomoto1986}, which has recently attracted some attention in 
the mathematical literature \cite{luccon2015,giacomin2012,bertini2014,giacomin2015}
as well as in the physical one \cite{ohta2008,pikovsky2011,zaks2003}.

It is known since the $60$s that the evolution of the empirical measure of 
$N$ diffusions with mean-field interaction is described, for $N\to\infty$, by a 
non-linear Fokker-Planck equation known in the mathematics literature as the
McKean-Vlasov equation \cite{mckean1966}, see also 
\cite{sznitman1991,meleard1996} for more modern presentations. We refer 
to this limit as the mean-field behavior of the $N$ particle system.
For the Shinomoto-Kuramoto model, the mean-field behavior exhibits a rich phase
diagram with stationary and periodic phases separated by bifurcation lines,
as first observed in \cite{sakaguchi1988}.

In \cite{dawson1987,dawson1987b}, Dawson and Gartner  
studied the large deviations of the empirical measure for the mean-field
diffusions through a generalization of the Freidlin-Wentzell theory to 
such questions. Extensions to cases where quenched disorder is present, as for 
example in the Kuramoto model \cite{pra1996,hollander2008} or in mean-field 
spin glasses \cite{grunwald1996,arous1995,ben1997}, were considered too. 
Also some large deviation results on a mean-field model for active matter 
were obtained in the physics literature \cite{barre2014}.

In this chapter, we first discuss how the mathematical results 
\cite{dawson1987,dawson1987b} can be obtained formally (i.e., without mathematical 
rigor) by writing an effective evolution for the empirical measure in 
the form of a stochastic PDE, called the Dean equation \cite{dean1996}. 
The noise term in this equation is proportional to $1/\sqrt{N}$ so that it
vanishes as $N\to\infty$. In the latter limit, one recovers the deterministic
McKean-Vlasov equation.

The evolution of the empirical measure is thus formally given by a stochastic 
partial differential equation with weak noise which, at variance with the cases considered in the previous 
sections, is infinite-dimensional. By applying functional integral techniques 
from field theory (the Martin-Siggia-Rose formalism, see \cite{martin1973}), 
we  can write the infinite-dimensional analogue of the Freidlin-Wentzell 
theory. In the physics literature, this extension goes under the name of 
Macroscopic Fluctuation Theory and it has attracted much attention in the 
statistical mechanics community  over last years, see \cite{bertini2014} 
for a review. In the present case, we 
re-obtain formally the rigorous results by Dawson and Gartner 
\cite{dawson1987,dawson1987b}. We also discuss a more general large 
deviation result covering current fluctuations.

We then apply the infinite-dimensional analogue of the 
perturbative scheme discussed in Chapter \ref{sec:Chap2} to obtain explicit 
results about the quasi-potential for the mean-field diffusions. In particular, 
we calculate the quasi-potential perturbatively close to the free particle dynamics 
with no interactions between particles present. This permits also to compute 
perturbatively the rate function for the fluctuations 
of some macroscopic observables. For the Shinomoto-Kuramoto model, explicit
results obtained by implementing a numerical algorithm to compute the quasi-potential 
to the $1^{\rm st}$ order in the coupling are presented, together with an analysis
of the fluctuations of magnetization.

We also discuss how explicit results may be obtained for the Taylor expansion 
of the quasi-potential around a stationary solution of the McKean-Vlasov 
equation. Within this analysis, it is clear that the variance of 
density fluctuations diverges close to bifurcations when external parameters are changed, a result rigorously obtained in \cite{dawson1983,arous1990b}. 
A numerical algorithm to evaluate explicitly the Taylor expansion is also 
discussed but we have not implemented it.

The structure of this part of the article is as follows. 
In Sec.\,\ref{sub:Chap3-mean-field}, we introduce the class of systems 
that will be considered. Sec.\,\ref{sec:Chap3-Empirical-density} 
derives the Dean equation, formally describing the evolution of the empirical 
measure for large but finite $N$. Then, in 
Sect.\,\ref{sec:Chap3-Kuramoto-typical}, the particular case of the 
Shinomoto-Kuramoto model is discussed. We describe the long-time behavior of 
solutions of the associated McKean-Vlasov equation. Despite its simplicity, 
the model, which is a kinetic and non-equilibrium version of a mean-field 
ferromagnet, exhibits a rather complex mean-field behavior
that we describe concentrating on the results that can be 
obtained analytically or semi-analytically. In Sec.\,\ref{sub:Chap3-LD}, 
we discuss how a generalization of the Freidlin-Wentzell theory may 
be formally obtained by applying the Martin-Siggia-Rose formalism  to the 
Dean equation. Finally, in Sec.\,\ref{sub:Chap3-perturbative}, the perturbative 
calculation of the quasi-potential is performed and some explicit 
results for the Shinomoto-Kuramoto model are described.


\subsection{Mean-field diffusions}\label{sub:Chap3-mean-field}

Let us consider a system composed of $N$ particles undergoing an over-damped
diffusion in $\mathbb{R}^d$ or in a torus $\mathbb{T}^d$ and coupled through 
a mean-field 2-body potential $\,V(x)=V(-x)$. The equations of motions 
defining the stochastic evolution are
\begin{equation}\label{eq:Chap3-eq-of-motion}
\dot{x}_n\,=\,  b(  x_n)\,-\,\frac{J}{N}
\sum\limits_{m=1}^N( \nabla V)(  x_n-  x_m)
\,+\,\sqrt{2k_BT}\,\eta_{n}\,,
\end{equation}
where $\eta_{n}(t)$ are independent white noises with zero average and  
covariance $\mathbb{E}\,\eta^i_{n}(t)\,\eta^j_{m}(s)=\delta^{ij}\delta_{n,m}\delta(t-s)$, 
$T>0$ is the temperature, and $k_B$ is the Boltzmann constant.

The quantities which are of central interest for us are the empirical 
density and the empirical current, defined as
\qq
&&\rho_N(t,x)=\frac{1}{N}\sum_{n=1}^N \delta(x-x_n(t))
\label{eq:Chap3-empirical-density}\\
&&j_N(t,x)=\frac{1}{N}\sum_{n=1}^N \delta(x-x_n(t))\circ \dot{x}_n(t)
\label{eq:Chap3-empirical-current}
\qqq
where ''$\,\circ\,$" stands for the product in the Stratonovich convention. It is 
straightforward to show that the following continuity equation holds
\qq\label{eq:Chap3-conservation-law}
\partial_t\rho_N\,+\,\nabla\cdot j_{N}=0
\qqq
(this uses the chain rule which imposes the Stratonovich convention in the definition
of the empirical current). Moreover, substituting the equation of motion
(\ref{eq:Chap3-eq-of-motion}) into the definition of $j_{N}$ and returning
to the Ito convention, we obtain
\qq\label{eq:Chap3-jN}
j_N(t,x)=\rho_N(t,x)\big[b(x)-J\,\nabla (V\ast \rho_N)(t,x) \big] 
-k_BT\nabla \rho_N(t,x)+\frac{\sqrt{_{2k_BT}}}{^N}\sum_{n=1}^N\delta(x-x_n(t))\,\eta_{n}(t)\,,
\qqq
where $(V\ast \rho_N)$ is the convolution between $V$ and $\rho_N(t,\cdot)$ and
the term with $\nabla\rho_N$ was produced by the change of the stochastic convention.
Observe that Eq.\,(\ref{eq:Chap3-conservation-law}) with $j_N$ given by 
(\ref{eq:Chap3-jN}) is not a closed equation for $\rho_N$ because the noise term 
explicitly depends on the particle positions. What we would like to do, instead,
is to write a closed evolution equation for $\rho_N$ into which the particle 
positions enter only through $\rho_N$.


\subsection{Evolution of the empirical density: the Dean equation}
\label{sec:Chap3-Empirical-density}

A closed equation for the evolution of the empirical density was obtained from 
Eqs.\,(\ref{eq:Chap3-conservation-law}) and (\ref{eq:Chap3-jN}) by Dean \cite{dean1996}.
Dean's argument (somewhat brief in the original paper) can be reformulated in the 
following way. The last term on the right hand side of (\ref{eq:Chap3-jN}) may be 
viewed as a white noise in time with values in vector fields on ${\mathbb R}^d$,
\qq\label{white_noise_1}
\frac{\sqrt{_{2k_BT}}}{^N}\sum_{n=1}^N\delta(x-x_n)\,\eta^i_{n}(t)\,,
\qqq
parameterized by the particle positions $(x_n)_{n=1}^N$. White noise (\ref{white_noise_1}) has mean zero
and covariance
\qq\label{1st_covariance}
\frac{2k_BT}{N^2}\,\delta^{ij}\,\delta(t-t')\sum_n\delta(x-x_n)\,\delta(y-x_n)=
\frac{2k_BT}{N}\,\delta^{ij}\,\delta(t-t')\,\delta(x-y)\,\rho(x)\,,
\qqq
where the last expression followed by using the distributional identity
$\,\delta(x-x_n)\,\delta(y-x_n)=\delta(x-y)\,\delta(x-x_n)\,$ and introducing
the particle density $\rho(x)=\frac{1}{N}\sum_n\delta(x-x_n)$. Consider now another 
noise,
\qq\label{white_noise_2}
\sqrt{\frac{_{2k_BT}}{^N}}\, \sqrt{\rho(x)} \,\xi^i(t, x)\,,
\qqq
where $\xi(t,x)$ stands for the vector-valued white noise in space and time satisfying
\qq
\mathbb E \,\,\xi^i(t, x) =0\,,\qquad
\mathbb E\,\,\xi^i(t, x)\,\xi^j(s, y)=\delta^{ij}\,\delta(t-s)\,\delta( x- y)\,.
\qqq    
Random process (\ref{white_noise_2}) may again be viewed as a white noise in time 
with values in vector fields on $\mathbb R^d$, but now parameterized by densities 
$\rho(x)$. It has a zero mean and covariance
\qq
\frac{2k_BT}{N}\,\delta^{ij}\,\delta(t-t')\,\delta(x-y)\,\rho(x)
\qqq 
that coincides with the one of noise (\ref{white_noise_1}) if $\rho$ is related
to particle positions as above. Dean proceeded identifying the two white noises
by writing
\qq
j_N(x,t) =  j_{\rho_N}(t, x)\,+\,\sqrt{\frac{_{2k_BT}}{^N}}\, \sqrt{\rho_N(t, x)} \, 
\xi(t, x)\,,
\qqq
upon which the continuity equation (\ref{eq:Chap3-conservation-law}) became
a closed stochastic PDE in the space of densities for $\rho(t,x)=\rho_N(t,x)$,
\qq\label{eq:Chap3-Dean}
\partial_t\rho(t, x)\,+\,\nabla\cdot j_{\rho}(t, x)\,+
\,\sqrt{\frac{_{2k_BT}}{^N}}\,\nabla\cdot \left( \sqrt{\rho(t, x)} \, 
\xi(t, x)\right)\,=\,0\,,
\qqq
where $ j_\rho$ is the nonlinear functional of $\rho$ given by
\qq
j_\rho(t, x)\,=\,\rho(t, x)\Big( b( x)-J\hspace{-0.07cm}\int
(\nabla V)( x- y)\,\rho(t, y)\,d y\Big)\,
-\,k_BT\,\nabla\rho(t, x)\,.
\label{jrho}
\qqq
We shall call (\ref{eq:Chap3-Dean}) the Dean equation for the empirical density.

Dean's substitution is formal in the infinite-dimensional situation involving the space 
of densities but would be legitimate in a finite-dimensional setup. To explain
what we mean, let us consider the backward Kolmogorov 
equation describing the evolution of averages for functionals of the empirical 
density $\mathcal{S}[\rho_N(t,\cdot)]$. Denoting by $\mathbb{E}$ the average with 
respect to the noises $\eta_n$ and applying the Ito calculus, we infer 
that\footnote{Since the densities are normalized, functional derivatives
$\delta\mathcal{S}/\delta\rho(x)$ are defined only up to a constant, but
such ambiguities drop out in all expressions below where the functional 
derivatives are integrated against functions with vanishing integral.}
\qq\label{eq:Chap-3-dean-derivation1}
\frac{d}{dt}\,\mathbb{E}\,\,\mathcal{S}[\rho_N(t,\cdot)] &=& \int dx \,
\frac{\delta \mathcal{S}}{\delta \rho(x)}[\rho_N(t,\cdot)]\, 
\nabla \cdot\Big(\rho_N(t,x)\big[ - b(x) + J\, \nabla (V\ast \rho_N)(t,x) \big] 
+ k_BT\nabla \rho_N(t,x)\Big) \nonumber\\
&+& \frac{k_BT}{N^2}\sum_{n=1}^N\int dx\,dy\, \frac{\delta^2 \mathcal{S}}{\delta \rho(x)
\delta \rho(y)}[\rho_N(t,\cdot)]\,\nabla_x\cdot\nabla_y
\Big(\delta(x-x_n(t))\,\delta(y-x_n(t))\Big)\nonumber\\
&=& \int dx \,\frac{\delta \mathcal{S}}{\delta \rho(x)}[\rho_N(t,\cdot)]\, 
\nabla \cdot\Big(\rho_N(t,x)\big[ - b(x) + J\, \nabla (V\ast \rho_N)(t,x) \big] 
+ k_BT\nabla \rho_N(t,x)\Big) \nonumber\\
&+&\frac{k_BT}{N}\int dx\int dy\, \frac{\delta^2 \mathcal{S}}{\delta \rho(x)
\delta \rho(y)}[\rho_N(t,\cdot)]\,\nabla_x\cdot\nabla_y\Big(\rho_N(x,t)\,
\delta(x-y)\Big),
\qqq
where the last expression was obtained proceeding as in (\ref{1st_covariance}).
We thus obtain for $\mathbb{E}\left[\mathcal{S}[\rho_N(t,\cdot)]\right]$ the evolution 
equation 
\qq\label{eq:Chap-3-dean-FP}
\frac{d}{dt}\,\mathbb{E}\,\,\mathcal{S}[\rho_N(t,\cdot)] &=& 
\mathbb E\,\,\mathcal{L}\mathcal{S}[\rho_N(t,\cdot)]\,,
\qqq
where $\mathcal{L}$ is the generator given by
\qq\label{Chap3-generator-L}
\mathcal{L}\mathcal{S}[\rho] &=&  \int dx \,\frac{\delta \mathcal{S}[\rho]}
{\delta \rho(x)}\,\nabla \cdot\Big(\rho_N(t,x)\big[ - b(x) + J\,\nabla (V\ast 
\rho_N)(t,x) \big] 
+ k_BT\nabla \rho_N(t,x) \Big)\nonumber\\
&+&\frac{k_BT}{N}\int dx\int dy\, \frac{\delta^2 \mathcal{S}[\rho]}{\delta \rho(x)
\delta\rho(y)} \,(\nabla\cdot\nabla_y)\Big(\rho(x,t)\,\delta(x-y)\Big).
\qqq
Observe now that the above evolution for $\mathbb{E}\left[\mathcal{S}[\rho_N] \right]$ 
is given by the same backward Kolmogorov equation that the one obtained assuming that 
$\rho_N$ solves the Dean stochastic PDE (\ref{eq:Chap3-Dean}) anticipated
at the beginning of the subsection. In finite dimension, it is not enough
to know that $\mathbb E\,f(X(t))=\mathbb E\,Lf(X(t)$ for a generator of a diffusion
process $L$ and each $f$ to deduce that $X(t)$ has the law of the diffusion process 
satisfying the corresponding SDE, but this would follow if one showed that 
$\,f(X(t))-f(X(0))-\int_0^tLf(X(s))\,ds\,$ are martingales \cite{stroock2007}. 
A slight extension of the previous calculation shows that the process
\qq
\mathcal{S}[\rho_N(t,\cdot)]-\mathcal{S}[\rho_N(0,\cdot)]]-
\int\limits_0^t\mathcal{L}\mathcal{S}[\rho_N(s,\cdot)]\,ds
\qqq
possess this property. Nevertheless, since we are in infinite dimension,
the derivation of the Dean equation remains formal.  

To complicate things further, the mathematical status of the Dean equation is unclear: 
it is difficult to give sense to different terms of the equation, including 
the noisy one, in a function space that would contain the (distributional) empirical 
densities of coupled diffusions (\ref{eq:Chap3-eq-of-motion}), not even speaking 
about a theory of solutions of such a stochastic PDE. Although there is a 
considerable mathematical literature about non-linear stochastic PDE's with 
the noise that is delta-correlated in time and space, see 
\cite{faris1982,daprato2014,hairer2013kpz}, it does not cover the case of the Dean 
equation. One may hope, however, that it is possible to give a meaning to this 
equation at least in the large deviation regime for large $N$. Although in what 
follows we completely avoid such rigorous issues and proceed formally, the fact that 
our results agree with those rigorously obtained by Dawson and Gartner 
\cite{dawson1987,dawson1987b} gives support to this assumption.

Before proceeding further, let us observe that the Dean equation can be recast 
in a form similar to that of the finite-dimensional SDE considered in the first part
of the paper by rewriting it as
\qq\label{eq:Chap3-Dean-general}
\partial_t \rho = \mathcal{K}[\rho] + \sqrt{\frac{_2}{^N}}\,\eta[\rho]\,,
\qqq
where 
\qq\label{Dean_drift}
\mathcal{K}[\rho](t,x) = -\nabla\cdot j_{\rho}(t, x)
\qqq
is the drift and
\qq
\eta[\rho](t,x)=-\sqrt{k_BT}\,\,\nabla\cdot\Big(\sqrt{\rho(x)} \,\xi(t, x)\Big)
\qqq
is the white noise in time, parameterized by density $\rho$, with zero mean and 
covariance
\qq
\mathbb E\,\,\eta[\rho](t,x)\,\eta[\rho](s,y) = 
\delta(t-s)\,\mathcal{Q}[\rho](x,y)
\qqq 
for
\qq\label{eq:Chap3-noise-variance}
\mathcal{Q}[\rho](x,y) = k_BT\,\nabla_x\cdot\nabla_y\Big(\rho(x)\delta(x-y)\Big).
\qqq
This form will be useful in the following to formally extend the large deviations 
results that we described in Sec.\,\ref{sec:1} to the present case.

In the Dean equation, the noise term becomes small when $\,N\,$ becomes large.
In particular, in the $\,N\to\infty\,$ limit, one obtains the deterministic non-linear 
Fokker-Planck equation
\qq\label{eq:Chap3-McKean-Vlasov}
\partial_t\rho(t, x)\,+\,\nabla\cdot j_{\rho}(t, x)=0\,
\qqq
known in the mathematical literature as McKean-Vlasov equation \cite{mckean1966}. 
Some authors refer to the above equation as the Vlasov-Fokker-Planck one. In the following,
we call the evolution described by this equation the mean-field dynamics.

The above reasoning showed that the McKean-Vlasov equation should describe the 
evolution of the empirical density in the $N\to\infty$ limit. This is actually 
corroborated by a number of rigorous results, since the original paper of McKean 
\cite{mckean1966}. More precisely, the property of propagation of chaos was proved under 
mild hypothesis on the smoothness of the $b$ and $V$, see \cite{sznitman1991} and 
references therein. Moreover, a bound on a proper distance between the solution to 
the McKean-Vlasov equation and the empirical measure at time $t$, depending on the 
distance at $t=0$ is known. Those results are the analogues in the present context 
of the more famous ones due to Braun-Hepp \cite{braun1977vlasov} and Dobrushin 
\cite{dobrushin1979vlasov} for deterministic particles with mean-field interactions, 
see also \cite{spohn1991}.

The Dean equation suggests that the evolution of the empirical measure for 
finite but large $N$ is described by a weak random perturbation 
(of order $1/\sqrt{N}$) of the McKean-Vlasov equation. We are then in a 
similar context to that of the Freidlin-Wentzell theory discussed in 
Sec.\,\ref{sec:1}, except for the fact that the stochastic dynamical system 
is now an infinite dimensional one. We may, nevertheless, hope to obtain 
large deviation estimates in a similar manner by working at the formal level. 
This will be done in Sec.\,\ref{sub:Chap3-LD}. Before, however, let us further 
investigate the McKean-Vlasov equation by considering a simple model system.


\subsection{Shinomoto-Kuramoto model: the mean-field behavior}
\label{sec:Chap3-Kuramoto-typical}

Models in the class of Eq.\,(\ref{eq:Chap3-eq-of-motion}) can display a very 
rich mean-field dynamics. As a simple example, we consider in this section the 
Shinomoto-Kuramoto model introduced in \cite{shinomoto1986}, describing its 
mean-field behavior by focusing on results that can be obtained analytically 
or semi-analytically.

The Shinomoto-Kuramoto model is a one-dimensional model where $N$ particles 
move on a circle of unit radius and are thus identified by their angular 
coordinate $x_n=\theta_n$ defined modulo $2\pi$, for $n=1,\dots,N$. The model 
is obtained from Eq.\,(\ref{eq:Chap3-eq-of-motion}) by setting 
$b(\theta)=F-h\sin(\theta)$ and 
$V(\theta)=(1-\cos\theta)$, where $F, h, J$ are real constants that we 
shall take non-negative. The equations of motion are
\qq\label{eq:Chap3-eq-of-motion-Kuramoto-Shinomoto}
\dot{\theta}_n\,=\,(F-h\sin\theta_n)\,-\,\frac{J}{N}
\sum\limits_{m=1}^N \sin( \theta_n-  \theta_m)
\,+\,\sqrt{2k_BT}\,\eta_{n}\,,
\qqq
where $\eta_n(t)$ are independent standard scalar white noises.
Observe that negative $F$ is related to positive $F$ and negative $h$
to positive $h$ by the changes of variables
$\theta_n\mapsto-\theta_n$ and $\theta_n\mapsto\theta_n+\pi$,
respectively. One of the  parameters among $F,h,J,T$ is redundant, as it 
can be fixed by rescaling the other three and time. We could, for example, 
fix the coupling strength $J$. We prefer, however, to leave all 
the parameters because we shall be interested in limiting cases where 
one of them vanishes. The Shinomoto-Kuramoto
system is related to the more famous Kuramoto model \cite{kuramoto1975} 
of frequency synchronization phenomena in coupled rotators, from which it is 
obtained by setting all the natural frequencies to the same value $F$ and 
adding white noises acting on each rotator, see \cite{acebron2005} for a review 
on Kuramoto and related models.

For general value of the parameters, the Shinomoto-Kuramoto stochastic dynamics
(\ref{eq:Chap3-eq-of-motion-Kuramoto-Shinomoto}) breaks the detailed 
balance. This is always true except for $F=0$, where the model defines an equilibrium 
dynamics and reduces to a kinetic version of the mean-field ferromagnetic $XY$ model, 
with $\theta_n$ describing the angles of planar spins and $h$ the external magnetic field 
$h$ in the $X$ direction.  
 For $F>0$, the model may be still interpreted within ferromagnetism, except that the planar magnetic field should be taken rotating with angular
velocity $F$ and the spin angles described in the frame rotating with it.
The case $h=0$ is also special since its dynamics only trivially breaks the detailed 
balance that may be restored by returning to the original frame. Thus, the 
Shinomoto-Kuramoto model can be seen as a non-equilibrium version of mean-field 
ferromagnets.

As discussed in Sec.\,\ref{sec:Chap3-Empirical-density}, for $N\to \infty$, 
the evolution of the empirical density is described by the McKean-Vlasov equation 
(\ref{eq:Chap3-McKean-Vlasov}) which, in the present case, reads:
\qq\label{eq:Chap3-McKean-Vlasov-Kuramoto}
\partial_{t}\rho + \partial_{\theta}j_{\rho} &=& 0\,,\\
j_\rho(t, \theta)\,&=&\,\rho(t, \theta)\left( F - h\sin\theta - J\int
\sin( \theta- \vartheta)\,\rho(t, \vartheta)\,d \vartheta\right)\,
-\,k_BT\,\partial_{\theta}\rho(t, \theta)\,.
\label{jrhoSK}
\qqq
The mean-field behavior of the Shinomoto-Kuramoto model was first studied 
in \cite{shinomoto1986}, focusing on long-time behavior. In this work, 
the authors expanded in Fourier modes the stationary McKean-Vlasov equation 
(\ref{eq:Chap3-McKean-Vlasov-Kuramoto})  and numerically integrated the 
resulting coupled ordinary differential equations. 

For $F=0$, the system relaxes to a stationary solution that is unique for 
$h>0$ and for $h=0$ and $k_BT\geq J/2$, with $\rho$ flat in $\theta$ 
(unmagnetized state) in the latter case. For $h=0$ and $k_BT<J/2$ there 
is a one-parameter family of stationary solutions differing by rotation,
with $\rho$ bumped around some value of $\theta$, a well known picture 
for equilibrium ferromagnets of a magnetized state spontaneously 
breaking the planar rotation symmetry. This equilibrium case, also known under the name of Brownian mean field model, is studied in detail in \cite{chavanis1,chavanis2,chavanis3}.

\begin{figure}
\includegraphics[scale=0.8]{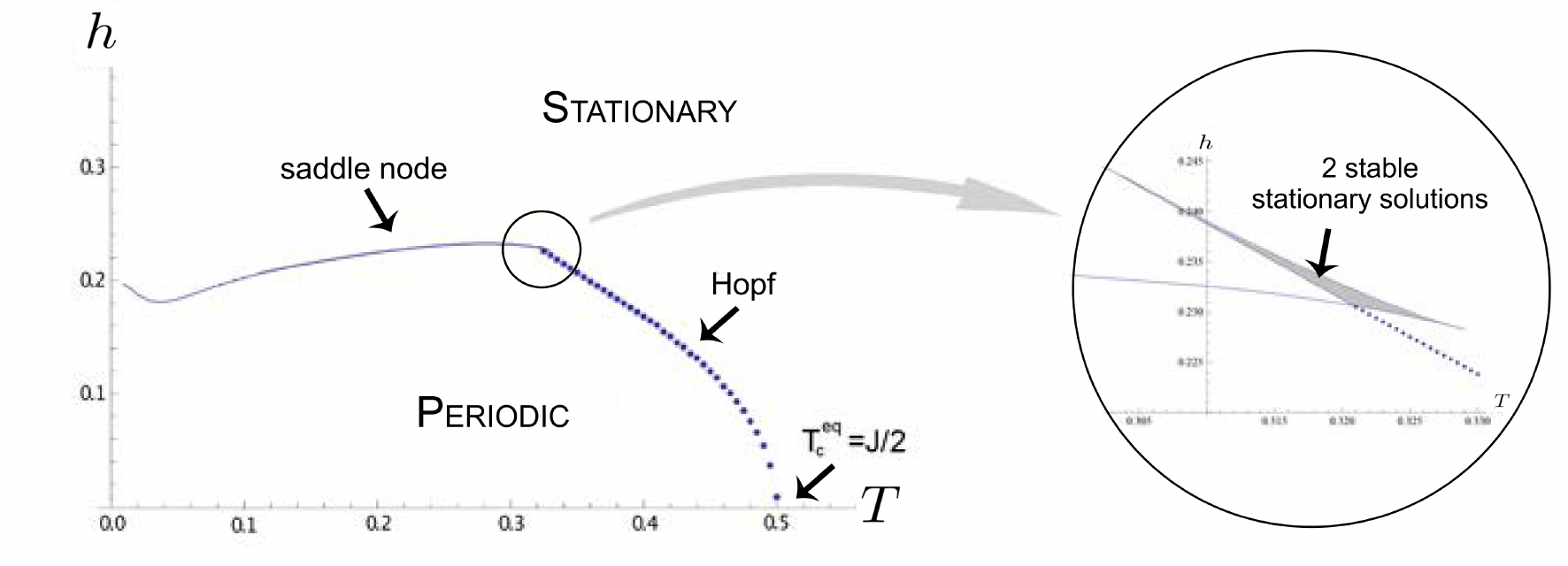}
\caption{{\footnotesize Phase diagram obtained with our semi-analytical results describing 
the long-time behavior of the Shinomoto-Kuramoto model for $J=1$ and $F=0.2$. We have 
checked that the phase diagram for other values of the parameters is similar. Our results 
are in agreement with those first obtained numerically in 
\cite{shinomoto1986,sakaguchi1988}. Qualitatively, the long-time behavior of the 
Shinomoto-Kuramoto model is very different for high $T$ and/or $h$ and for low $T$ and 
$h$. In the first case, the system is stationary at long times and the empirical density 
converges to the unique stationary stable solution of the McKean-Vlasov equation. This 
is true except for a very small region where two stationary stable solutions of the 
McKean-Vlasov equation are present, represented by the shaded region in the inset. 
For low $T$ and $h$, no stationary stable solutions of the McKean-Vlasov equation exists 
and the empirical density converges to a periodic solution. The two regions are separated 
by two lines corresponding to the Hopf (dots) and saddle-node (blue line) bifurcations. 
These two lines merge in a Taken-Bodganov bifurcation. Other stationary but unstable 
states of the McKean-Vlasov equation exist, and will be fully analyzed in the following.}
\label{fig:Kuramoto-phase-diagram}}
\end{figure}

For $F>0$, a more complicated phase diagram for the McKean-Vlasov equation emerges, 
however, see the left part of Figure \ref{fig:Kuramoto-phase-diagram}. For sufficiently 
high $h$ or $T$, the system relaxes to a stationary solution of 
Eq.\,(\ref{eq:Chap3-McKean-Vlasov-Kuramoto}), which for $h=0$ and $k_BT>J/2$ has 
$\rho$ flat in $\theta$, as before. For sufficiently low values of $h$ and $T$, the 
long-time behavior is, instead, periodic. For $h=0$, the periodic phase sets in 
for $k_BT<J/2$, as follows from the relation to the $F=0,\ h=0$ system mentioned before. 
The periodic and stationary regions are separated by bifurcations. The rightmost line 
(blue dots) is a Hopf bifurcation, while the upper line, a saddle-node one. The two 
bifurcations meet forming a Takens-Bogdanov bifurcation. The careful analysis of 
bifurcations occurring in this model has been performed in \cite{sakaguchi1988} and 
is confirmed here. In particular, a tiny region where the system is bistable is found 
around the Takens-Bogdanov bifurcation, see the right part of 
Figure \ref{fig:Kuramoto-phase-diagram}. Here the McKean-Vlasov equation admits two 
stable stationary solutions.

A number of analytical results can be obtained. In particular, we show below that all 
the stationary solutions (stable and unstable) can be studied analytically. Moreover, 
analyzing their stability we can trace the bifurcation curves. On the other hand, 
we were not able to find a closed form for the periodic solutions except for the
trivial cases where $h=0$ or $T=0$.

In Sec.\,\ref{sub:Chap3-typical-F0} below, we consider the case where the 
dynamics respects the detailed balance ($F=0$) or trivially breaks it ($h=0$). Then,
in  \ref{sub:Chap3-typical-generic}, we list all the stationary solutions for generic 
values of the parameters and describe how their stability may be analyzed. Finally, 
we discuss the special case where the particles do not interact ($J=0$), and the 
singular situation of zero temperature ($T=0$) in 
Secs.\,\ref{sub:Chap3-typical-J0} and \ref{sub:Chap3-typical-T0}, respectively.

 
 \subsubsection{Equilibrium dynamics ($F=0$ or $h=0$)}
\label{sub:Chap3-typical-F0}
We analyze in this paragraph the long-time behavior of the Shinomoto-Kuramoto 
model for $F=0$ or $h=0$. We refer the reader to \cite{chavanis2} for a more detailed discussion on the case $F=0$.

Let us start with the case $F=0$, $J\geq0$ and $T>0$. Here, the $N$-body system 
(\ref{eq:Chap3-eq-of-motion-Kuramoto-Shinomoto}) with finite 
$N$ respects the detailed balance with respect to the invariant measure 
given by the Boltzmann-Gibbs distribution
\qq\label{eq:Chap3-Kuramoto-Shinomoto-equilibrium-invariant}
f_N(\theta_1,\dots,\theta_N)=\frac{1}{Z}\exp\bigg[\frac{_1}{^{k_B T}} 
\Big(h\sum_{n=1}^N \cos \theta_n - 
\frac{_J}{^{2N}}\sum_{m,n=1}^N(1-\cos(\theta_n-\theta_m))\Big)\bigg]
\qqq
where $Z$ is the canonical partition function. 
The system is ergodic and the mean of the empirical density converges 
to the one in the Gibbs measure that 
may be easily calculated. Indeed, applying the the Hubbard-Stratonovich 
transformation to 
Eq.\,(\ref{eq:Chap3-Kuramoto-Shinomoto-equilibrium-invariant}),
we get the identity
\qq
f_N(\theta_1,\dots,\theta_N)=\frac{N\ee^{-\frac{NJ}{2k_BT}}}{2\pi J\,k_BT\,Z}
\int\ee^{\frac{1}{k_BT}\Big((m_x+h)\sum\limits_{n=1}^N\cos{\theta_n}+
m_y\sum\limits_{n=1}^N\sin{\theta_n}\,-\,\frac{N}{2J}(m_x^2+m_y^2)
\Big)}\,dm_x\,dm_y\,.\quad
\qqq
The expectation of the empirical density $\rho_N(\theta)$ in the Gibbs
stationary state is equal to the integral of $f_N(\theta,\theta_2,\dots,
\theta_N)$ over $\theta_2,\dots,\theta_N$ that gives
\qq\label{mean_emp_dens}
\mathbb E\,\,\rho_N(x)=
\frac{(2\pi)^{N-2}N\ee^{-\frac{NJ}{2k_BT}}}{J\,k_BT\,Z}\int
\ee^{\frac{1}{k_BT}\Big((m_x+h)\cos{\theta}+m_y\sin{\theta}\,-\,\frac{N}{2J}
(m_x^2+m_y^2)\Big)}\,\Big(I_0\big(\frac{_{m(h)}}
{^{{k_BT}}}\big)\Big)^{\hspace{-0.06cm}N-1}dm_x\,dm_y\,,\quad
\qqq
where $\,m(h)=\sqrt{(m_x+h)^2+m_y^2}\,$ and
\qq
I_0\Big(\sqrt{z_1^2+z_2^2}\Big)\,=\,\frac{_1}{^{2\pi}}\int\limits_0^{2\pi}\ee^{
z_1\cos{\theta}+z_2\sin{\theta}}\,d\theta
\qqq
is the Bessel functions of the first kind. 
From Eq.\,(\ref{mean_emp_dens}), we can 
obtain the equilibrium stationary density in the $N\to\infty$ limit  
by the saddle point calculation: 
\qq
\rho_{inv}^{eq}(\theta)=\lim_{N\to\infty}\,\mathbb E\,\,\rho_N(\theta)
\,=\,\frac{\ee^{\frac{1}{k_BT}\big((m_x+h)\cos{\theta}+m_y\sin{\theta}\big)}}
{2\pi I_0\Big(\frac{m(h)}{{k_BT}}\Big)}\,,
\qqq
where $(m_x,m_y)$ minimizes
\qq\label{function_f}
f(m_x,m_y)=\frac{m_x^2+m_y^2}{2Jk_BT}-\ln I_0\Big(\frac{m(h)}{k_BT}\Big).
\qqq
The corresponding stationarity equations are 
\qq\label{stationarity_eqs}
\frac{m_x}{J}
=\frac{m_x+h}{m(h)}\,\frac{I_1\Big(\frac{m(h)}{k_BT}\Big)}
{I_0\Big(\frac{m(h)}{k_BT}\Big)}\,,\qquad
\frac{m_y}{J}
=\frac{m_y}{m(h)}\,\frac{I_1\Big(\frac{m(h)}{k_BT}\Big)}
{I_0\Big(\frac{m(h)}{k_BT}\Big)}\quad
\qqq
with $I_1=I'_0$. They may be rewritten as the self-consistency equations
\qq
{m_x}=J\int\limits_0^{2\pi}\cos{\theta}\,\rho_{inv}^{eq}(\theta)\,d\theta\,,\qquad
{m_y}=J\int\limits_0^{2\pi}\sin{\theta}\,\rho_{inv}^{eq}(\theta)\,d\theta\,,
\qqq 
so that $(m_x,m_y)$ has the interpretation of the magnetization vector
(in units of $J$). 

For $h>0$, the stationarity equations (\ref{stationarity_eqs}) imply
that $m_y=0$ and $m_x>0$ solves the equation
\qq\label{eq:Chap3-F0-self-consistency}
\frac{m_x}{J}\,=\,
\frac{I_1\big(\frac{m_x+h}{{k_BT}}\big)}
{I_0\big(\frac{m_x+h}{k_BT}\big)}
\qqq
which has a unique solution.
In the limit $h\to 0$, one recovers the $2^{\rm nd}$-order phase transition 
(a pitchfork bifurcation) located at $k_BT_c=J/2$. For higher values of $T$, 
one has $\rho^{eq}_{inv}=\frac{1}{2\pi}$ and the magnetization vanishes 
(i.e. $m_x=0,m_y=0$) while for $T<T_c$ the stationary state is spontaneously 
magnetized ($m_x>0,m_y=0$). Taking $h=0$ directly, any rotation of the 
low temperature solution in $\theta$, with the corresponding rotation 
of the magnetization vector $(m_x,m_y)$, provides a solution of the saddle 
point equations which minimizes (\ref{function_f}), whereas $m_x=0,m_y=0$
corresponding to the flat density gives the maximum of (\ref{function_f})
instead of the minimum. The same limiting densities will be obtained as 
the stationary solutions of the McKean-Vlasov equation for $F=0$, see below.

Let us now consider the second case, where $F,J,T>0$ and $h=0$. 
Strictly speaking, the dynamics does not respect here the detailed balance. 
However, it can be recast as an equilibrium dynamics performing the change 
of variables $\theta_n'=\theta_n+Ft$, which corresponds to sitting on 
the comoving frame with angular velocity $-F$. The empirical density in 
this case can thus be obtained from the equilibrium results. Indicating 
by $\rho^{eq}_N$ the empirical density for $F=0$ and given values 
of $T,J$ and by $\rho^F_N$ the empirical density for the same values 
of  $T$ and $J$ but with $F\neq0$, we have $\rho_N^{F}(t,\theta) 
= \rho^{eq}_N(t,\theta-Ft)$. We conclude that for $T>T_c$, in the limit 
of long times and large $N$, the expectation of $\rho_N^{F}(t,\theta)$ becomes 
stationary and flat, while for $T<T_c$ it becomes periodic. The pitchfork 
bifurcation at equilibrium is thus modified to a Hopf bifurcation.

\subsubsection{Stationary states for generic parameters}
\label{sub:Chap3-typical-generic}

We now consider generic values of the parameter $h,T,J,F$ with $T>0$. 
To find stationary densities $\rho_{inv}$ at $N=\infty$, we consider the 
McKean-Vlasov equation (\ref{eq:Chap3-McKean-Vlasov-Kuramoto}) in the 
stationary form, imposing $\partial_t\rho_{inv}=0$. As the Shinomoto-Kuramoto 
model is one-dimensional, this is equivalent to ask that the mean-field 
current $j_{\rho}$ given by Eq.\,(\ref{jrhoSK}) be a constant,
that we shall denote $c$, for $\rho=\rho_{inv}$. 
We infer that for the stationary densities,
\qq\label{eq:chap3-stat-sol-curr}
j_{\rho_{inv}}(\theta)= \left[F\,-\,(m_x+h)\,\sin{\theta}\,+\,m_y\,\cos{\theta}\right]
\rho_{inv}(\theta)\,-\,k_BT\partial_{\theta}\rho_{inv}(\theta)\,=\,c\,,
\qqq
where $m_x$ and $m_y$ are given by the equations
\qq\label{self_cons}
{m_x}=J\int_0^{2\pi} \cos\theta\,\rho_{inv}(\theta)\,d\theta\,,\qquad 
{m_y}=J\int_0^{2\pi} \sin\theta\,\rho_{inv}(\theta)\,d\theta\,,
\qqq
so that $(m_x,m_y)$ is again the magnetization vector. 
As the current is constant, $j_{\rho_{inv}}(\theta)=j_{\rho_{inv}}(\theta+\theta_0)$ 
for any $\theta_0$ implying that
\qq\label{eq:Chap3-Shinomoto-stationary-eq-0}
&&j_{\rho_{inv}}(\theta)=
\left\{F\,-\,\left[(m_x+h)\,\cos{\theta_0}+m_y\,\sin{\theta_0}\right]
\sin{\theta}\,+\,\left[m_y\,\cos{\theta_0}-(m_x+h)\,\sin{\theta_0}\right]\cos{\theta}
\right\}\rho_{inv}(\theta+\theta_0)\nonumber\\
&&\hspace{1.6cm}-\,k_BT\,\partial_{\theta}\rho_{inv}(\theta+\theta_0)\,=\,c\,.
\qqq
We can now choose $\theta_0$ in such a way that 
\qq\label{eq:Chap3-self-consistency-0}
m_y\cos{\theta_0}-(m_x+h)\sin{\theta_0}\,=\,0\,,\qquad 
(m_x+h)\cos{\theta_0}+m_y\sin{\theta_0}\,=\,y\,\geq\,0\,,
\qqq
so that Eq.\,(\ref{eq:Chap3-Shinomoto-stationary-eq-0}) becomes 
\qq\label{eq:Chap3-Shinomoto-stationary-eq}
j_{\rho_{inv}}(\theta)\,=\,\left(F\,-\,y\,\sin\theta\right)\rho_{inv}(\theta+\theta_0)
\,-\,k_BT\, \partial_{\theta}\rho_{inv}(\theta+\theta_0)\ =\ c\,.
\qqq
For any $y$, the solution to this equation can be written as
\qq\label{eq:Chap-3-rho-inv}
\rho_{inv}(\theta+\theta_0)\,=\,Z^{-1}\,\ee^{\frac{1}{k_BT}\left(F\theta+y
\cos{\theta}\right)}\int\limits_\theta^{\theta+2\pi}
\ee^{-\frac{1}{k_BT}\left(F\vartheta+y
\cos{\vartheta}\right)}\,d\vartheta
\qqq
with $Z$ the normalization factor. Clearly, for Eq.\,(\ref{eq:Chap-3-rho-inv}) 
to be a stationary solution, the self-consistency conditions (\ref{self_cons}) 
have to be satisfied together with 
Eqs.\,(\ref{eq:Chap3-self-consistency-0}).
Such self-consistency condition can be made more explicit by introducing 
\qq\label{eq:Chap3-self-consistency}
f_x(y)\,\equiv\,\int\limits_0^{2\pi}\cos{\theta}\,
\rho_{inv}(\theta+\theta_0)\,d\theta\,=\,
\frac{y}{J}-\frac{h}{J}\,
\cos{\theta_0}\,,\qquad
f_y(y)\,\equiv\,\int\limits_0^{2\pi}\sin{\theta}\,
\rho_{inv}(\theta+\theta_0)\,d\theta\,=
\,\frac{h}{J}\,
\sin{\theta_0}\,.
\qqq
from which we can eliminate $\theta_0$ to obtain an equation for $y$
\qq\label{eq:Chap3-self-consistency-f}
\left(f_x(y)-\frac{y}{J}\right)^2+f_y(y)^2\,=\,\left(\frac{h}{J}\right)^2.
\qqq

\begin{figure}
\includegraphics[scale=0.6]{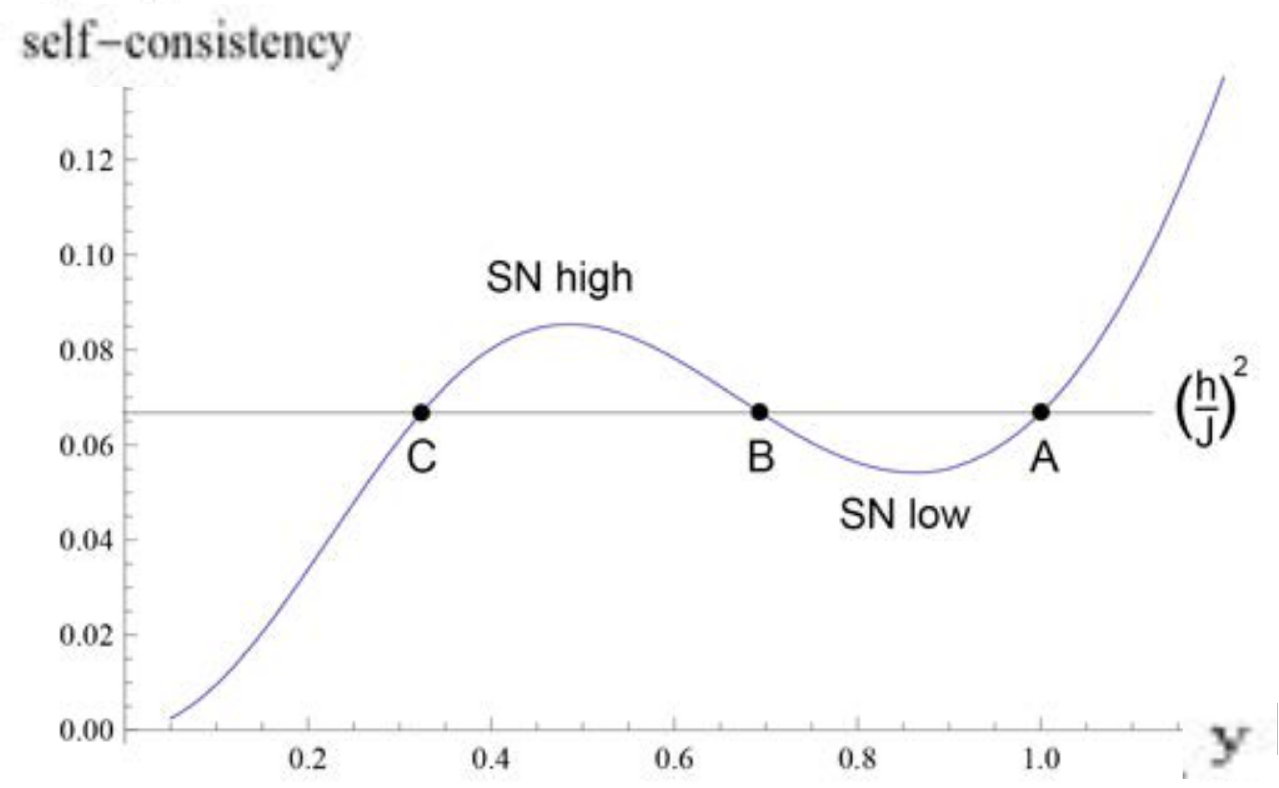}\hspace{1.cm}
\includegraphics[scale=0.6]{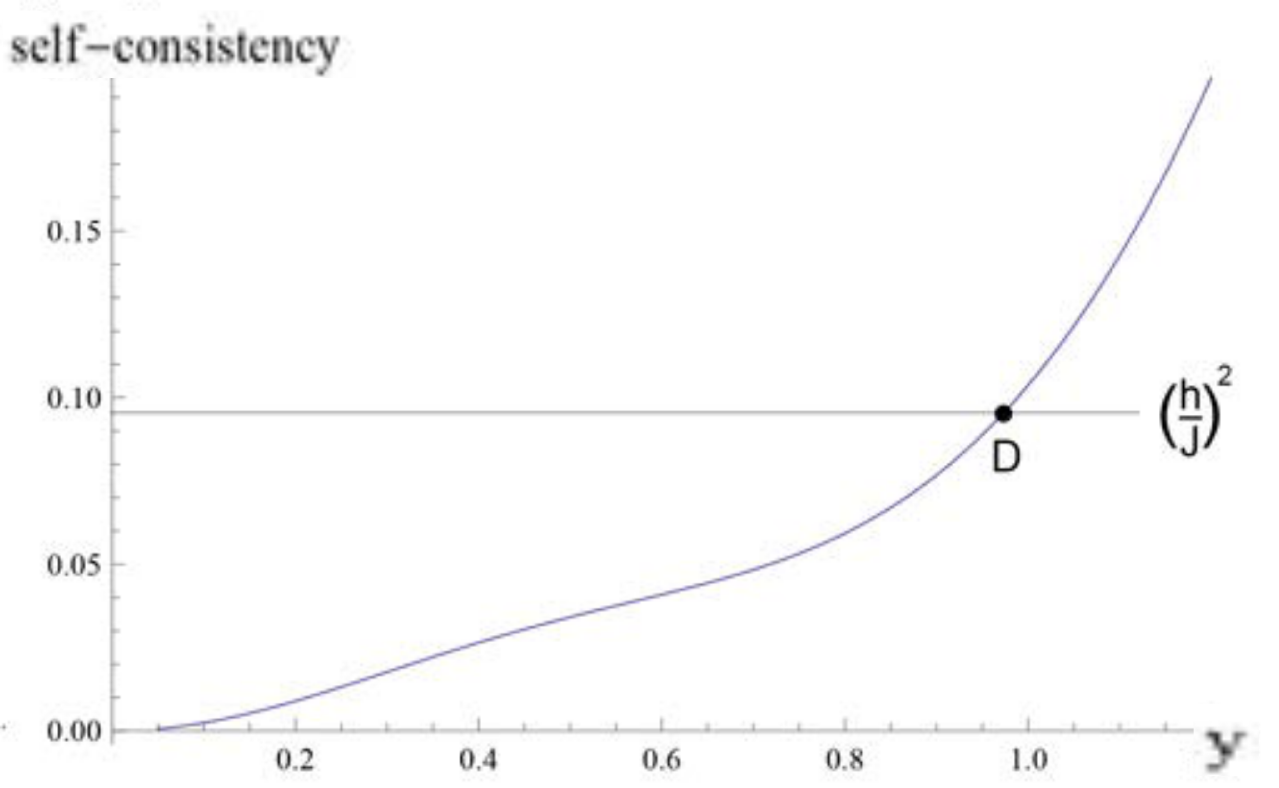}
\caption{{\footnotesize We report here the typical curves that are 
obtained for the left hand side of the self-consistency condition 
(\ref{eq:Chap3-self-consistency-f}) that does not depend on $h$. Two cases 
are observed. If $T>T_{tr}$ (on the right), whatever the value of $h$, only 
one value of $y$ satisfies the self-consistency condition and the 
McKean-Vlasov equation admits only one stationary solution. For $T<T_{tr}$ 
(on the left), depending on the value of $h$, one obtains one or three 
values of $y$ for which the self-consistency is satisfied. Thus, the 
McKean-Vlasov equation admits one or three stationary solutions. Moreover, 
the maximum and the minimum of the curve defines two values of $y$ for 
which a pair of stationary solutions is created or destroyed corresponding
to the values of $h$ where saddle-node bifurcations occur.}
\label{fig:Kuramoto-self-consistency}}
\end{figure}

We have thus showed that, for any $T>0$, stationary solutions of the McKean-Vlasov 
equation associated to the Shinomoto-Kuramoto model 
(\ref{eq:Chap3-McKean-Vlasov-Kuramoto}) are given, up to rotation
by angle $\theta_0$, by Eq.\,(\ref{eq:Chap-3-rho-inv}), where $y$ is chosen 
so that the self-consistency condition 
(\ref{eq:Chap3-self-consistency-f}) is satisfied. At the end, angle $\theta_0$ 
may be found from Eqs.\,(\ref{eq:Chap3-self-consistency}).
The above self-consistency problem can be easily solved numerically. With such 
procedure, we obtained the curves reported in 
Fig.\,\ref{fig:Kuramoto-self-consistency} for the left hand side of 
Eq.\,(\ref{eq:Chap3-self-consistency-f}) as a function
of $y$. Note that such a function does not depend on $h$. Two cases are 
observed. If $T>T_{tr}$, for any value of $h$ only one value of $y$ satisfies 
the self-consistency condition (\ref{eq:Chap3-self-consistency-f}).
In this case, whatever the value of $h$, the McKean-Vlasov equation admits 
only one stationary solution. For $T<T_{tr}$, depending on the value of $h$, 
one obtains one or three values of $y$ for which the self-consistency is 
satisfied. Thus, the McKean-Vlasov equation admits one or three stationary 
solutions. Observe moreover that the maximum and the minimum of the curve 
defines two values of $y$ where a pair of stationary solutions is created 
or destroyed. They thus indicate the values of $h$ where saddle-node 
bifurcations occur. We leave to the reader checking how the self-consistent 
solutions reduce to the ones obtained in the previous
section for the equilibrium case $F=0$. Let us just note that in that case,
$f_x(y)=I_1(\frac{y}{k_BT})/I_0(\frac{y}{k_BT})$ and $f_y(y)=0$.

To finally recover the phase diagram, we must also analyze the stability of 
the stationary solutions just described.
For this, it is enough to look at the linearization of the McKean-Vlasov 
dynamics around a given $\rho_{inv}$,  
\qq\label{eq:chap-3-linearized-FP-may}
\partial_{t}\delta\rho &=& R_{\rho_{inv}}\delta\rho(t,\cdot)\,,\\
 R_{\rho_{inv}}\delta\rho &=& -\,\partial_{\theta}\left[\,\delta\rho(\theta)
\left( F - h\sin\theta - J\int
\sin( \theta- \vartheta)\,\rho_{inv}(\vartheta)\,d \vartheta\right)
\,\right.\\
&&\left.-\,\rho_{inv}(\theta)\,J\int \sin( \theta- \vartheta)\,
\delta\rho(\vartheta)\,d \vartheta
-\,k_BT\,\partial_{\theta}\delta\rho(\theta)\right].\nonumber
\qqq
One should then study the spectra of the linearized Fokker-Planck operator 
$R_{\rho_{inv}}$. This is possible analytically only in very special case when 
$\rho_{inv}$ does not depend on $\theta$, which holds for $h=0$.

\begin{figure}[h]
\includegraphics[scale=0.8]{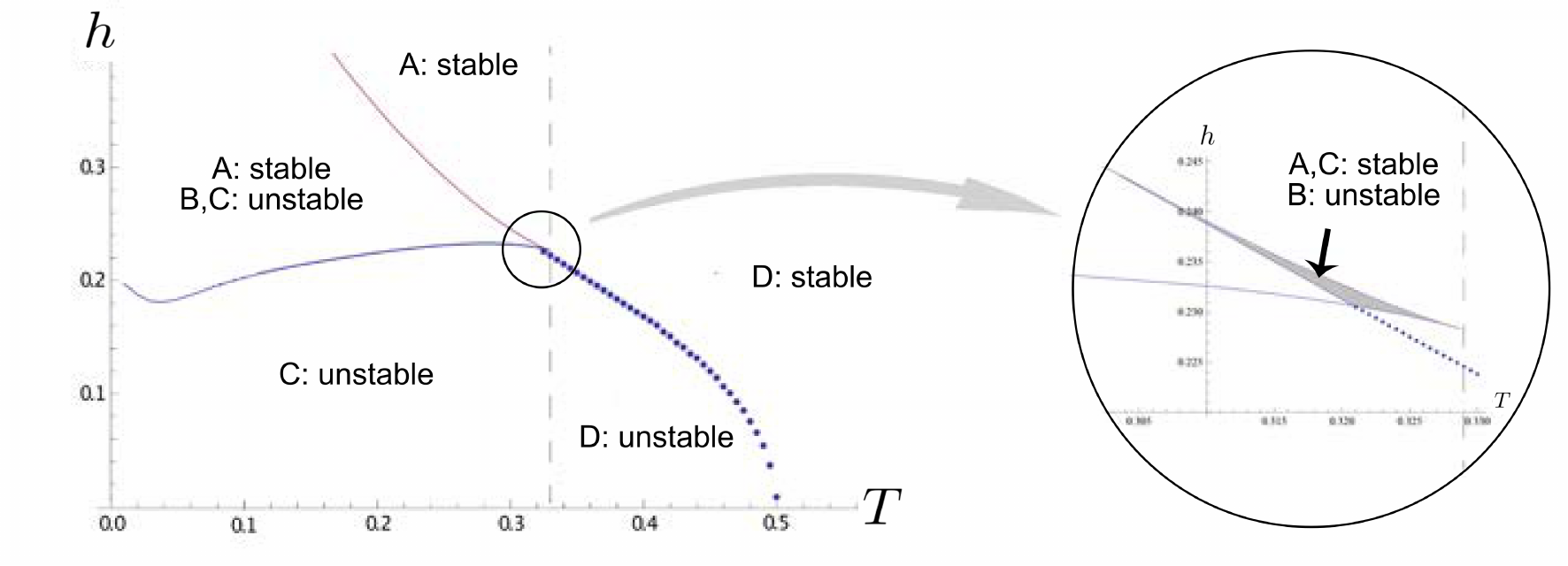}
\caption{\footnotesize Stationary states in the mean-field limit of the 
Shinomoto-Kuramoto model along with their stability. In each region, we report 
which stationary states are present and whether they are stable or unstable, 
where the letters $A,B,C,D$ indicate from which self-consistent solution 
the stationary state is obtained, see Fig.\,\ref{fig:Kuramoto-self-consistency}.
\label{fig:Kuramoto-stationary-states}}
\end{figure}

We checked the stability of the stationary solutions by passing to the Fourier 
transformed picture and truncating the resulting system to modes $k\leq K$, 
in which way, $ R_{\rho_{inv}}$ is reduced to a $K\times K$ matrix. It is then 
a simple numerical task to find the eigenvalues of such a matrix. Only few 
modes ($K\sim 7$) are enough to already obtain very accurate results. The 
stationary states and their stability are summarized in 
Fig.\,\ref{fig:Kuramoto-stationary-states}, from which we 
reconstruct the phase diagram anticipated in 
Fig.\,\ref{fig:Kuramoto-phase-diagram}.

We conclude by observing that it is possible to calculate analytically where 
the shown bifurcation lines end. Indeed, from the equilibrium solution, 
we know that the Hopf bifurcation crosses the $h=0$ axis in $T=J/2$. Moreover, 
it is simple to show that for $T=0$ the bifurcation occurs at $F=h$.

 \subsubsection{Free particles ($J=0$)}\label{sub:Chap3-typical-J0}
 
The stationary solutions for the free particle case ($J=0$) can be obtained 
very easily directly, 
 \qq\label{eq:Chap-3-rho-inv-J0}
\rho_{inv}(\theta)\,=\,Z^{-1}\,\ee^{\frac{1}{k_BT}\left(F\theta+h
\cos{\theta}\right)}\int\limits_\theta^{\theta+2\pi}
\ee^{-\frac{1}{k_BT}\left(F\vartheta+h
\cos{\vartheta}\right)}\,d\vartheta\,,
\qqq
or from Eq.\,(\ref{eq:Chap-3-rho-inv}) since in this case $m_x=m_y=0=\theta_0$
and thus $y=h$.

 
 \subsubsection{Zero temperature}\label{sub:Chap3-typical-T0}

There are two kinds of stationary solutions for $T=0$: delta-function-like 
and smooth solutions. Smooth solutions are physically less relevant because 
they are linearly unstable in the ferromagnetic model ($J>0$) that we 
consider here. We discuss smooth solutions in Appendix \ref{app:T0} and 
consider here only the delta-function-like ones.

The stationary McKean-Vlasov equation in the form
(\ref{eq:Chap3-Shinomoto-stationary-eq}) for $T=0$ is solved by
\qq\label{delta_sol}
\rho_{inv}(\theta+\theta_0)=\delta(\theta-\theta_1)
\qqq
if $\theta_1$ satisfies $F-y\sin\theta_1=0$. 
Clearly, such solutions exist only for $y>F$. The self-consistency 
condition (\ref{eq:Chap3-self-consistency}) reduces in this case to
the relation 
\qq\label{eq:T0-self-consistency}
\Bigg(\pm \sqrt{1-\left(\frac{F}{y}\right)^2} - \frac{y}{J}
\Bigg)^{\hspace{-0.1cm}2}+\left(\frac{F}{y}\right)^{\hspace{-0.07cm}2}
=\left(\frac{h}{J}\right)^{\hspace{-0.07cm}2}
\qqq
that has no solutions for $y$ if $h<F$, one solution if $h=F$ and two 
solutions for $h>F$. They are straightforward to understand.
Indeed, one may show that at such solutions, $F-h\sin(\theta_0+\theta_1)=0$,
so that the delta-function stationary states describe a situation
where all particles sit in the zero of their drift $F-h\sin\theta$,
which clearly provides a solution of the equations of motion 
(\ref{eq:Chap3-eq-of-motion-Kuramoto-Shinomoto}) when $T=0$.
Such solutions exist for all $N$ if and only if $h\geq F$. For large $N$,
their stability depends on the sign of $h\cos(\theta_0+\theta_1)$,
with the positive one corresponding to stable solutions.

\subsection{Large deviations for large but finite $N$}
\label{sub:Chap3-LD}

For large but finite $N$, the dynamics of the empirical density $\rho_N(t,x)$ 
deviates from the mean-field evolution given by the McKean-Vlasov equation 
(\ref{eq:Chap3-McKean-Vlasov}). Formally, $\rho_N$ solves the Dean equation 
(\ref{eq:Chap3-Dean}) which is an infinite-dimensional dynamical system 
perturbed by a weak noise. Below, starting from this equation and employing
a functional integral argument, we derive the results 
obtained rigorously by Dawson and Gartner \cite{dawson1987,dawson1987b} 
that describe the dynamical large deviations of the empirical density 
for mean-field diffusions. We also formulate the functional Hamilton-Jacobi 
equation for the quasi-potential describing the asymptotic form of 
stationary distribution for the empirical densities. Another functional 
integral argument gives the large deviations asymptotics for the
distribution of dynamical fluctuations of the empirical current. 
A generalization of the Freidlin-Wentzell theory to infinite dimensional 
dynamical systems whose deterministic part is given by the Fokker-Planck 
operator with, possibly, non-linear diffusion and drift coefficients 
is known in the physics literature as the Macroscopic Fluctuation 
Theory. It has attracted much attention in last years in non-equilibrium 
statistical mechanics mainly because of its applications to stochastic 
lattice gases, see \cite{bertini2014} for a review.
Our results may be viewed as an application of that theory to diffusions
with mean-field interactions of the general form (\ref{eq:Chap3-eq-of-motion}).
It will be also straightforward to specify the results that follow to 
the case of the Shinomoto-Kuramoto model.

\subsubsection{Dynamical large deviations for the empirical density}

Let us fix two times $t_i<t_f$ and consider the probability that 
the empirical density $\rho_N(t,x)$ is arbitrarily close to a given trajectory
$\hat\rho(t,x)$ for $t_i\leq t\leq t_f$. In fact we shall be interested
only in the large deviations where such probability is described
by the rate function $\mathcal{A}[\rho(\cdot,\cdot)]$ as in
Eq.\,(\ref{eq:Chap1-FW-action}) where trajectories $x(t)$ become
those in the infinite-dimensional space of densities equipped 
with an appropriate norm and the limit $\epsilon\to0$ is 
replaced by $N\to\infty$. Informally,
\begin{equation}
\mathcal{A}[\rho(\cdot,\cdot)]=-\lim_{N\to\infty}\frac{1}{N}\ln 
P[\rho(\cdot,\cdot)]\,,
\end{equation}
where $P[\rho(\cdot,\cdot)]$ is the probability distribution
function(al) describing the distribution of empirical densities in the time
interval $[t_i,t_f]$. This rate function $\mathcal{A}$ is a functional 
of the trajectory $\rho(\cdot,\cdot)$ and will provide a generalization 
of the Freidlin-Wentzell action of Eq.\,(\ref{eq:FW-action})
to the present infinite-dimensional setup. 

We may find the form of $\mathcal{A}[\rho]$ by the Martin-Siggia-Rose
functional integral argument \cite{martin1973}.
A similar derivation to the one presented below may be found in 
\cite{jouvet1979}. Assuming the Dean equation, the distribution 
function $P[\rho]$ is that of its random solutions and may be written 
in the form
\qq
P[\rho]\ =\ 
\mathbb{E}\,\Bigg[\delta\left(\partial_t\rho\,+\,\nabla\cdot j_{\rho}\,
+\,\sqrt{\frac{_{2k_BT}}{^N}}\,\nabla\cdot \left( \sqrt{\rho} \,\, \xi\right)
\right)\,
\det\bigg(\frac{_{\delta\left(\partial_t\rho\,+\,\nabla\cdot j_{\rho}\,
+\,\sqrt{\frac{_{2k_BT}}{^N}}\,\nabla\cdot \left( \sqrt{\rho} \,\, \xi\right)\right)}}
{^{\delta \rho}}\bigg)\Bigg]
\qqq
if we fix the initial value $\rho(t_i,x)\equiv\rho_i(x)$.
The Jacobian factor can be dropped, as it will not contribute to 
the large deviations. Expressing now the delta functional as an oscillatory 
functional integral in the imaginary directions over fields $u(t,x)$ such 
that $\int u(t,x)dx=0$, we obtain
\qq
P[\rho]\,&\asymp&\,\mathbb{E}\,\Bigg[\int  
\exp\bigg[ \frac{_N}{^{2k_B T}}\int_{t_i}^{t_f}\hspace{-0.25cm}dt
\hspace{-0.07cm}\int
\Big(\partial_t\rho\,+\,\nabla\cdot j_{\rho}\,+\,\sqrt{\frac{_{2k_BT}}{^N}}
\,\nabla\cdot \left( \sqrt{\rho} \,\, \xi\right)\Big)\,u\,dx \bigg]
\,\mathcal{D}u\,\Bigg]\\
&=& \int\exp\bigg[ \frac{_N}{^{2k_B T}}\int_{t_i}^{t_f}
\hspace{-0.25cm}dt\hspace{-0.07cm}\int\Big(\left(\partial_t\rho\,
+\,\nabla\cdot j_{\rho}\,\right)\,u\,+\,\frac{_1}{^2}\rho\,(\nabla u)^2\Big)dx 
\bigg]\,\mathcal{D}u\,,
\qqq
where the last equality was obtained by calculating the Gaussian
expectation over the white noise $\xi(t,x)$. The rate function  
$\mathcal{A}[\rho]$ can now be extracted by the saddle point argument:
\qq\label{eq:Chap3-Dawson-Gartner0}
\mathcal{A}[\rho]=-\min_{u}\,\frac{1}{2k_BT}\int_{t_i}^{t_f}\hspace{-0.25cm}dt
\hspace{-0.07cm}\int\Big(\left(\partial_t\rho\,
+\,\nabla\cdot j_{\rho}\,\right)\,u\,+\,\frac{1}{2}\rho\,(\nabla u)^2\Big)dx\,,
\qqq
where the minimum is taken over all functions $u(t,x)$ with  
$t_i\leq t \leq t_f$ and spatial mean zero. The minimum is easy to 
calculate resulting in the formula
\qq\label{eq:Chap3-Dawson-Gartner}
\mathcal{A}[\rho]\,=\,\frac{1}{4k_BT}\int_{t_i}^{t_f}\hspace{-0.25cm}dt
\hspace{-0.1cm}\int
\hspace{-0.15cm}\int\left(\partial_t\rho\,+\,\nabla\cdot j_{\rho}\,\right)(t,x)\,
(-\nabla\cdot \rho(t,\cdot)\nabla)^{-1}(x,y)\,\left(\partial_t\rho\,
+\,\nabla\cdot j_{\rho}\,\right)(t,y)\,dx\,dy\,,
\qqq
where $(-\nabla\cdot \rho(t,\cdot)\nabla)^{-1}(x,y)$ is the kernel of
the inverse of the operator $-\nabla\cdot \rho(t,\cdot)\nabla$ in the action 
on functions of zero mean. Eq.\,(\ref{eq:Chap3-Dawson-Gartner0}) may 
be also rewritten in the form 
\qq\label{eq:Chap3-Dawson-Gartner1}
\mathcal{A}[\rho]\,=\,\min_u\,
\frac{\Big(\int\limits_{t_i}^{t_f}\hspace{-0.05cm}dt\int
\left(\partial_t\rho\,+\,\nabla\cdot j_{\rho}\,\right)\,u\,dx
\Big)^{\hspace{-0.1cm}2}}
{4k_BT\int\limits_{t_i}^{t_f}\hspace{-0.05cm}dt\int\rho(\nabla u)^2\,dx}
\qqq
which was rigorously derived for the large deviations rate function $\mathcal A[\rho]$ 
by Dawson and Gartner in \cite{dawson1987,dawson1987b}. Indeed, 
Eq.\,(\ref{eq:Chap3-Dawson-Gartner1}) 
results from (\ref{eq:Chap3-Dawson-Gartner0}) upon substituting
$u\mapsto\lambda u$ and minimizing over $\lambda\in\mathbb R$. On the other 
hand, expression (\ref{eq:Chap3-Dawson-Gartner}) is completely analogous 
to the Freidlin-Wentzell action (\ref{eq:FW-action}) as it may
be rewritten as
\qq\label{eq:Chap3-Dawson-Gartner-Q}
\mathcal{A}[\rho]\,=\,\frac{1}{4}\int_{t_i}^{t_f}\hspace{-0.25cm}dt
\hspace{-0.1cm}
\int\hspace{-0.15cm}\int\left(\partial_t\rho\,-\,\mathcal{K}[\rho]\right)(t,x)\,
\mathcal{Q}^{-1}[\rho(t,\cdot)](x,y)\,
\left(\partial_t\rho\,-\,\mathcal{K}[\rho]\right)(t,y)\,dx\,dy\,,
\qqq
where $\mathcal{K}[\rho]$ is the drift and $\mathcal{Q}[\rho]$ is the noise 
covariance of the Dean equation that are given by Eqs.\,(\ref{Dean_drift}) and 
(\ref{eq:Chap3-noise-variance}), respectively.

In our arguments, we ignored the fact that the empirical density
$\rho_N(t,x)$ defines a genuine stochastic process only after integrating 
it against a spatial test function. The large deviations for such
stochastic processes will, however, be governed by rate functions
that are contractions of $\mathcal A[\rho]$ obtained by minimizing it 
with imposed integrals against the test functions.

\subsubsection{Quasi-potential and Hamilton-Jacobi equation}

The functional Fokker-Planck equation associated to the Dean equation 
(\ref{eq:Chap3-Dean}) describes the evolution of the probability 
distribution function $P_t[\rho]$ to observe a density profile $\rho(x)$ 
at time $t$ and takes the form
\qq\label{Chap-3-Functional-FP}
\partial_t P = \mathcal{L}^\dagger P\,,
\qqq
where $\mathcal{L}^+$ is the adjoint of the generator 
(\ref{Chap3-generator-L}) calculated with the rule 
$\big(\frac{\delta}{\delta\rho}\big)^\dagger=-\frac{\delta}{\delta\rho}$. 
Analogously to the case of finite-dimensional systems, we define the 
quasi-potential as the functional
\qq
\mathcal{F}[\rho] = -\lim_{N\to\infty}\,\frac{1}{N}\,\ln P_{\infty}[\rho]\,,
\qqq
where $P_{\infty}$ denotes the stationary solution of the above Fokker-Planck 
equation. Proceeding formally, all the structure of Freidlin-Wentzell 
theory generalizes to the present case. We are thus very brief in what 
follows, as the derivation of the statements below is very similar to the 
one for finite dimensional systems, albeit only formal here.

As in the case of finite dimensional systems, we only consider cases where 
the attractor of the deterministic evolution $\partial_t\rho = 
\mathcal{K}[\rho]$ is a fixed point $\rho_{inv}$. The quasi-potential with 
respect to $\rho_{inv}$ can be obtained from the minimization problem 
\qq\label{eq:Chap3-QP-minimisation}
\mathcal{F}[\rho]=\min_{\{ \hat{\rho}(t,x)\,|\, \hat{\rho}(-\infty,x) = 
\hat{\rho}_{inv}(x)\,,\,\hat{\rho}(0,x)=\rho(x) \}}\,
\mathcal{A}[\hat{\rho}]\,.
\qqq
Alternatively, one can try to solve the Hamilton-Jacobi equation 
\qq\label{eq:Chap3-HJ-Q}
\int\frac{\delta \mathcal{F}}{\delta \rho(x)}\,
\mathcal{K}[\rho](x)\,dx \, 
+ \int\hspace{-0.15cm}\int\frac{\delta\mathcal{F}}{\delta \rho(x)} 
\mathcal{Q}[\rho](x,y)\,\frac{\delta\mathcal{F}}{\delta \rho(y)}
\,dx\,dy\,=\,0
\qqq
that can be obtained by inserting Ansatz 
$P_{\infty}[\rho]\sim\ee^{-N\mathcal{F}[\rho]}$ into the functional Fokker-Planck 
equation (\ref{Chap-3-Functional-FP}). Equivalently, we can obtain it from 
the Hamilton-Jacobi equation for finite-dimensional systems 
(\ref{eq:HJ-chap-1}).

Finally, with the same argument as in Sec.\,\ref{subsec:Chap1-HJ}, it 
is simple to show that the minimization in 
Eq.\,(\ref{eq:Chap3-QP-minimisation}) is achieved over the solution of 
the fluctuation or instanton dynamics
\qq
\partial_t \rho = \mathcal{K}_r[\rho]\,,\quad\qquad \mathcal{K}_r[\rho](x)
&=&\mathcal{K}[\rho](x) + 2\,\int\mathcal{Q}[\rho](x,y)\,dy
\frac{\delta \mathcal{F}}{\delta \rho(y)}\cr
&=&\mathcal{K}[\rho](x) + 
2k_BT\,(-\nabla\rho(x)\,\nabla) \frac{\delta \mathcal{F}}{\delta \rho(x)}\,.
\qqq

\subsubsection{Dynamical large deviations for the empirical current} 

Empirical current (\ref{eq:Chap3-empirical-current}) is a more general 
quantity than empirical density (\ref{eq:Chap3-empirical-density}). 
Indeed, $\rho_N$ and $j_N$ satisfy the continuity equation 
(\ref{eq:Chap3-conservation-law}). Given the 
empirical current $j_N$ in a time interval $[t_i,t_f]$ and the initial 
condition for $\rho_N(t_i,x)=\rho_i(x)$, we can 
reconstruct the empirical density in the same time interval as
$\rho_N(x,t)=\rho_i(x) - \int_{t_i}^t \nabla \cdot j_N(x,s)\,ds$. 
The converse is however not true, as two empirical currents that differ 
by a divergent-free quantity give rise to the same empirical density.

Let us consider the probability distribution function $P[j(\cdot,\cdot)]$ 
of empirical current $j_N(t,x)$ in the time interval $t_i\leq t\leq t_f$ 
for fixed initial condition for empirical density. The
corresponding large deviations rate function will be denoted 
by $\mathcal{A}^c$:
\begin{equation}
\mathcal{A}^c[j(\cdot,\cdot)]=-\lim_{N\to\infty}\frac{1}{N}\ln 
\mathbb{P}[j(\cdot,\cdot)]\,.
\end{equation}
It may be found using again the Martin-Siggia-Rose functional integral 
formalism: 
\qq
\mathbb{P}[j] &\asymp&\mathbb{E}\,\bigg[\delta\left(j\,-\,j_{\rho}\,-\,
\sqrt{\frac{_{2k_BT}}{^N}\rho}\,\xi\right)\bigg] \\
&=& \mathbb{E}\,\bigg[\int\exp\left[ -\frac{_N}{^{2k_B T}} 
\int_{t_i}^{t_f}\hspace{-0.25cm}dt\hspace{-0.1cm}\int\left(j\,-\, j_{\rho}\,
-\,\sqrt{\frac{_{2k_BT}}{^N}\rho}\,\xi\right)\cdot A \right]\,\mathcal{D}A\,
\bigg]\,=\\
&=&\int \exp\left[ \frac{_N}{^{2k_B T}}\int_{t_i}^{t_f}\hspace{-0.25cm} dt
\hspace{-0.07cm}\int\left(-(j\,-\, j_{\rho})\cdot A\,+\frac{_1}{^2}\,
\rho\,A^2\right)dx\right]\,\mathcal{D}A\,,
\qqq
where $A$ is vector valued, and evaluating the last functional integral 
with the saddle point method. This gives:
\qq\label{eq:Chap3-fundamental-MFT}
\mathcal{A}^c[j] = \frac{1}{4k_BT}\int_{t_i}^{t_f}\hspace{-0.25cm}dt
\hspace{-0.07cm}\int\frac{(j-j_{\rho})^2}{\rho\ }\,dx\,,
\qqq
where $\rho(t,x)=\rho_i(x)- \int_0^t \nabla \cdot j(x,s)\,ds$.

Eq.\,(\ref{eq:Chap3-fundamental-MFT}) is known in the literature as the 
fundamental formula of Macroscopic Fluctuation Theory. It is often viewed
as giving the rate function for the joint probability of observing 
a trajectory of the empirical density and of the empirical current 
arbitrarily close to $\rho(x,t)$ and to $j(x,t)$, provided that
the latter satisfy the continuity equation, and equal to infinity 
otherwise. This just expresses the fact that density trajectories 
are completely determined once we chose the current trajectory 
and the initial condition for the density. 

Again, we ignored the fact that these are the integrals of the empirical 
currents against test functions, now both over time and space, that make 
sense as random variables. The large deviations for such random variables 
will, however, be governed by the rate functions obtained by contractions 
of $\mathcal A^c[j]$ that minimize it with constraints imposed for the 
space-time integrals of $j$.

Let us conclude by observing that, to the best of our knowledge,  
large deviations for currents in diffusions with mean-field interactions 
have not been discussed in the mathematical literature. We will devote a future 
publication to the investigation of current fluctuations in the 
Shinomoto-Kuramoto model.

\subsection{Perturbative calculation of the quasi-potential}
\label{sub:Chap3-perturbative}

In this final section, we discuss the perturbative calculation of the 
quasi-potential for the diffusions with mean-field interaction. 
In Sec.\,\ref{sec:Chap2}, we developed a perturbative scheme to calculate the 
quasi-potential for finite dimensional systems. The first objective is then 
to translate that perturbative scheme to the infinite-dimensional setting 
described by the Dean equation (\ref{eq:Chap3-Dean-general}), which
we do in Sec.\,\ref{sub:chap4-pert}.

It is possible to obtain explicit results in two cases: close to the 
free-particle dynamics $J=0$, see Sec.\,\ref{sub:chapt4-J0}, and for  
the Taylor expansion of the quasi-potential around stationary solutions 
of the McKean-Vlasov equation, see Sec.\,\ref{sub:chapt4-Taylor}. We shall 
see that in these cases the perturbative expansion of the quasi-potential 
reduces to solving partial differential equations instead of functional 
differential equations.

For the Shinomoto-Kuramoto model, in the case of perturbations close to 
the free-particle dynamics $J=0$, we present explicit results at the $1^{\rm st}$ order 
in $J$, see Sec.\,\ref{sub:chapt4-J0Kur}. Interestingly, the quasi-potential 
that is a local functional of $\rho$ for the unperturbed $J=0$ case, is 
shown to become non-local already at the $1^{\rm st}$ order in $J$. The resulting
analytical expression is evaluated with a simple numerical scheme and we 
also compute the rate function for the fluctuations of single-particle 
observables such as the magnetization. Moreover, we discuss how the numerical 
scheme can be generalized to higher orders stressing that this does not 
increase the computational complexity.

We also discuss the Taylor expansion of the quasi-potential around the 
fixed points of the McKean-Vlasov dynamics. It would be possible to obtain 
explicit results also in this case. We detail how an algorithm can be 
designed for this purpose in Appendix \ref{app:Taylor}. We did not, however,
perform the corresponding calculations, leaving them to future investigations.

Although the perturbative techniques developed in Sec.\,3 in the context 
of finite-dimensional systems are easily transposable to the 
infinite-dimensional setup discussed here, we are unable to say anything
about the nature of the resulting perturbative expansions (are they
convergent? asymptotic?) as, to the best of our knowledges, the techniques used in finite 
dimensions to settle such questions do not extend in a Banach space setup.


\subsubsection{Perturbative schemes in infinite dimensional setting}
\label{sub:chap4-pert}

We generalize here the perturbative schemes for calculating the 
quasi-potential, developed in Sec.\,\ref{sec:Chap2}, to the case of 
infinite dimensional systems at hand. This can be done very easily either 
by starting again from the 
Hamilton-Jacobi equation, now in the infinite dimensional version 
(\ref{eq:Chap3-HJ-Q}), or by formally taking the limit of infinite
dimensions in the expressions of Sec.\,\ref{sec:Chap2}. 
Such a formal limit transforms derivatives into functional 
derivatives and finite-dimensional scalar products into $L^2$ ones.

Both the expansion centered on the attractor $\rho^\lambda_{inv}$ of the 
perturbed system (a stable fixed point of the McKean-Vlasov dynamics) and 
the direct one can be easily obtained. We only consider here the second one, 
as it will be used in the following. We denote
the direct expansion of the quasi-potential as
\qq
\mathcal{F}^{\lambda}[\rho] = \sum_n \lambda^n \mathcal{F}^{(n)}[\rho]\,,
\qqq
dropping the hats of Sec.\,\ref{sec:Chap2-uncentred},
\,and analogously for $\mathcal{K}^{\lambda}[\rho]$ and 
$\mathcal{Q}^{\lambda}[\rho]$.

With the procedures outlined above, the hierarchy 
(\ref{eq:HJ-expanded-F0-naive}) and (\ref{eq:HJ-expanded-naive}) is now 
replaced in the infinite dimensional setting by
\qq\label{eq:Chap3-HJ-expanded-F0-naive}
&&\int\frac{ \delta\mathcal{F}^{(0)}}{\delta\rho(x)} 
\left[ \mathcal{K}^{(0)}[\rho](x) + \int\mathcal{Q}^{(0)}[\rho](x,y) 
\frac{ \delta\mathcal{F}^{(0)}}{\delta\rho(y)}\,dy\right]dx\,=\,0
\quad\,\qquad{\rm for} \qquad n=0\,,\\
\label{eq:Chap3-HJ-expanded-naive}
&&\int\frac{\delta\mathcal{F}^{(n)}}{\delta\rho(x)} \,
\mathcal{K}^{(0)}_r[\rho](x)\,dx\, = \,\mathcal{S}^{(n)}
[\mathcal{F}^{(0)},\dots,\mathcal{F}^{(n-1)}]\hspace{1.3cm}
\quad\qquad{\rm for}\qquad n\neq 0\,,
\qqq
where $\mathcal{K}^{(0)}_r$ is the drift of the fluctuation dynamics for 
the unperturbed problem:
\qq\label{eq:Chap3-zero-order-fluctuation-dynamics}
\mathcal{K}^{(0)}_r[\rho]=\mathcal{K}^{(0)}[\rho] +2 
\mathcal{Q}^{(0)}[\rho]\frac{\delta \mathcal{F}^{(0)}}{\delta \rho}
\qqq
and $\mathcal{S}^{(n)}$ is given by 
\qq\label{eq:Chap3-Sn-1}
\mathcal{S}^{(n)}[\mathcal{F}^{(0)},\dots,\mathcal{F}^{(n-1)}] &=& 
-\,\sum_{k=1}^{n-1} \left[\int\frac{\delta \mathcal{F}^{(n-k)}}
{\delta \rho(x)} \Bigg[ \mathcal{K}^{(k)}[\rho](x) + 
\int\mathcal{Q}^{(0)}[\rho](x,y) \,\frac{\delta \mathcal{F}^{(k)}}
{\delta \rho(y)}\,dy\right]dx\cr
&&+\,\sum_{l=0}^{n-k} \int\hspace{-0.1cm}\int 
\frac{\delta \mathcal{F}^{(n-k-l)}}{\delta\rho(x)}\,
\mathcal{Q}^{(k)}[\rho](x,y)\,\frac{\delta \mathcal{F}^{(l)}}
{\delta\rho(y)}\,dx\,dy\Bigg]\cr
&&-\,\int\frac{\delta \mathcal{F}^{(0)}}{\delta\rho(x)}
\left[ \mathcal{K}^{(n)}[\rho](x) + \int\mathcal{Q}^{(n)}[\rho](x,y) 
\,\frac{\delta \mathcal{F}^{(0)}}{\delta\rho(y)}\,dy\right]dx\,.
\qqq
As it will be clear, there is a full analogy with finite-dimensional systems 
and we refer the reader to Sec.\,\ref{sec:Chap2} for the discussion on 
the perturbative schemes. In particular, we recall that the solution of these 
equations obtained with the methods of characteristics is unique 
and an explicit expression may be written in terms of the solution for 
the $0^{\rm th}$-order fluctuation dynamics. In the actual infinite dimensional 
setting, this solution is given by 
\qq\label{eq:Chap3-solution-HJ-hierarchy}
\mathcal{F}^{(n)}[\rho] = C^{(n)} + 
\int_{-\infty}^0 \mathcal{S}^{(n)}[\mathcal{F}^{(0)},\dots,\mathcal{F}^{(n-1)}] 
[\widetilde{\rho}(t,\cdot)]\,dt\,,
\qqq
where $\widetilde{\rho}(t,x)$ is the solution of the fluctuation dynamics
\qq\label{eq:Chap3-perturbative-fluct-dynamics}
\partial_t\rho = \mathcal{K}^{(0)}_r[\rho]
\qqq
starting at $t=-\infty$ at the stable fixed point $\rho^{0}_{inv}\equiv
\rho_{inv}$ of the unperturbed McKean-Vlasov dynamics and satisfying the 
final condition $\bar{\rho}(0,x)=\rho(x)$, \,compare to 
Eq.\,\eqref{eq:solution-ngr2}.  
It should be clear from the discussion of Sec.\,\ref{sec:Chap2}
that the convergence of the above integral in this infinite-dimensional 
setting is assured once the unperturbed fluctuation dynamics 
(\ref{eq:Chap3-perturbative-fluct-dynamics}) linearized around 
$\rho_{inv}$ escapes fast enough from $\rho_{inv}$. This happens, 
for example, in the typical case in which its generator has a spectral gap.
We finally recall from Sec.\,\ref{sec:Chap2-uncentred} that the constants
$C^{(n)}$ may be iteratively fixed but do not enter the expression for
$\mathcal{S}^{(n)}[\mathcal{F}^{(0)},\dots,\mathcal{F}^{(n-1)}]$ so are
largely irrelevant and may be adjusted conveniently.

\subsubsection{Expansion around the free particles dynamics ($J=0$)}
\label{sub:chapt4-J0}

We now consider the perturbative expansion of the quasi-potential 
$\mathcal{F}$ around the independent particle dynamics, choosing 
the perturbative parameter $\lambda=J$. In this case,
\qq
\mathcal{K}^J[\rho](x) &=& -\nabla\cdot\left[ \rho(x)\Big( b( x)-J\int
(\nabla V)( x- y)\,\rho(y)\,d y\Big)\,
-\,k_BT\,\nabla\rho(x) \right]
\qqq
and 
\qq
\mathcal{Q}^J[\rho](x,y) = k_BT\,\nabla_x\cdot\nabla_y 
\big(\rho(x)\,\delta(x-y)\big)\,.
\qqq
For $J=0$, particles do not interact. Even if for $J=0$ the system breaks 
the detailed balance, the quasi-potential $\mathcal{F}^{(0)}$ is known here
as its explicit form follows from Sanov's theorem:
\qq\label{eq:Chap3-free-unperturbed-QP}
\mathcal{F}^{(0)}[\rho]\,=\,\int\bigg[\rho(x)\,
\ln \frac{\rho(x)}{\rho_{inv}(x)} -\rho(x) + \rho_{inv}(x)\bigg]dx\,,
\qqq
where $\rho_{inv}$ is the stable stationary solution of McKean-Vlasov's 
equation which reduces here to the linear Fokker-Planck equation
for a single-particle diffusion. Hence $\rho_{inv}$ exists and is unique under simple assumptions about the single-particle 
drift $b(x)$. In the case of the Shinomoto-Kuramoto model, we have explicitly 
calculated $\rho_{inv}$ for $J=0$, \,see Eq.\,(\ref{eq:Chap-3-rho-inv-J0}). We observe that, despite the fact that the dynamics breaks in general 
the detailed balance, the quasi-potential for $J=0$ is a local functional 
of $\rho$.

As anticipated, it is convenient to perform the direct perturbative expansion, 
for which we have
\qq
\mathcal{K}^{(0)}[\rho](x) &=& \nabla\cdot\left[\,-\,\rho(x)\,b(x) 
+ k_BT\,\nabla \rho (x)\right],\\
\mathcal{K}^{(1)}[\rho](x) &=& \nabla\cdot\left[ \rho(x)
\int (\nabla V)( x- y)\,\rho(y)\,dy\, \right],\\
\mathcal{K}^{(n)}[\rho](x) &=& 0\ \,\qquad \qquad\quad{\rm for}
\quad\qquad n\geq 2
\qqq
and 
\qq
\mathcal{Q}^{(0)}[\rho](x,y) &=& k_BT\,\nabla_x\cdot\nabla_y 
\big(\rho(x)\,\delta(x-y)\big)\,,\\
\mathcal{Q}^{(n)}[\rho](x,y) &=& 0\qquad \qquad{\rm for}\quad\qquad n\geq 1\,.
\qqq
It should be clear now why, from a practical point of view, it is simpler 
to deal here with the direct expansion than with the expansion centered 
on the attractor of the perturbed dynamics. Indeed, in the former case, almost
all $\mathcal{K}^{(n)}$ and $\mathcal{Q}^{(n)}$ are zero. This would not be 
true in the attractor-centered expansion because the attractor of the 
perturbed dynamics depends on $J$. 

The quasi-potential at the $0^{\rm th}$ order $\mathcal{F}^{(0)}$ is given by 
Eq.\,(\ref{eq:Chap3-free-unperturbed-QP}). We thus have
\qq\label{eq:func_der_0}
\frac{\delta \mathcal{F}^{(0)}}{\delta \rho(x)} = 
\ln \frac{\rho(x)}{\rho_{inv}(x)}\,\,.
\qqq
The $0^{\rm th}$ order fluctuation dynamics takes the form $\partial_t \rho 
= \mathcal{K}_r^{(0)}[\rho]$, where 
\qq\label{eq:K_roperator}
\mathcal{K}_r^{(0)}[\rho]= -\nabla\cdot(\rho\, b) + 2k_BT\, 
\nabla \cdot\left[ \rho \,\nabla \ln \rho_{inv} \right] 
- k_BT\, \nabla^2 \rho\,\equiv K^{(0)}_r\rho\,.
\qqq
Note that $K^{(0)}_r$ is a linear operator.
When detailed balance is respected (i.e. $b=-\nabla U$ for some potential $U$), one can easily check that
\qq\label{fluct_dyn_drift0}
\mathcal{K}_r^{(0)}[\rho] = -\nabla \cdot\left( \rho \nabla U \right) - 
k_BT\,\nabla^2 \rho = -\mathcal{K}^{(0)}[\rho]\,,
\qqq
where we have used that $\rho_{inv}=(1/Z)\exp(-U/k_BT)$ in this case. The above 
relation just reflects the fact that the fluctuation dynamics is the 
time-reversal of the relaxation dynamics for equilibrium problems.

Let us specify the hierarchy (\ref{eq:Chap3-HJ-expanded-naive}) to 
the present situation. The left hand side is clearly unchanged, except that 
$\mathcal{K}^{(0)}_r[\rho]$ takes the particular form
(\ref{fluct_dyn_drift0}),  while the right hand side 
becomes equal to 
\qq
\mathcal{S}^{(1)}[\mathcal{F}^{(0)}]  = - \int 
\frac{\delta \mathcal{F}^{(0)}}{\delta \rho(x)} \, 
\mathcal{K}^{(1)}[\rho](x)\,dx 
\qqq
for $n=1$, and to 
\qq\label{eq:Chap3-Sn-1-J0}
\mathcal{S}^{(n)}[\mathcal{F}^{(0)},\dots,\mathcal{F}^{(n-1)}] =
- \int\frac{\delta \mathcal{F}^{(n-1)}}{\delta \rho(x)} \, 
\mathcal{K}^{(1)}[\rho](x)\,dx - \sum_{k=1}^{n-1} \int\hspace{-0.1cm}\int 
\frac{\delta \mathcal{F}^{(n-k)}}{\delta \rho(x)}\mathcal{Q}^{(0)}[\rho](x,y) 
\,\frac{\delta \mathcal{F}^{(k)}}{\delta \rho(y)}\, dx\,dy\,\,\quad 
\qqq
for $n>1$.
A very important point is that, in this case, the solutions 
$\mathcal{F}^{(n)}[\rho]$ of the hierarchy are for $n\geq 1$, 
homogeneous polynomials of degree $(n+1)$ (up to constants),
\qq\label{eq:Chap3-polyn-J0}
\mathcal{F}^{(n)}[\rho] =C^{(n)}+ \frac{1}{(n+1)!}\int \phi^{(n)}(x_0,\dots,x_n)
\,\rho(x_0)\,\cdots\,\rho(x_n)\,\,dx_0\,\cdots\,dx_n\,,
\qqq
where the kernels $\phi^{(n)}$ are symmetric in $(n+1)$ variables
and the choice of constants $C^{(n)}$ is essentially irrelevant.
Using Eq.\,(\ref{eq:Chap3-polyn-J0}), we infer that the left hand 
side of the hierarchy (\ref{eq:Chap3-HJ-expanded-naive}) can be 
written as\
\qq\label{eq:Chap3-hierarchy-J0-lhs-n1}
&&\hspace{-0.6cm}\int\frac{\delta\mathcal{F}^{(n)}}{\delta\rho(x)} \,
\mathcal{K}^{(0)}_r[\rho](x)\,dx\,=\,
\frac{1}{n!}\,\int\Big(K_{r,x_0}^{(0)\dagger}\,\phi^{(n)}(x_0,\dots,x_n)\Big)
\,\rho(x_0)\,\dots\,\rho(x_n)\,dx_0\,\dots\,dx_n\cr
&&\hspace{1cm}=\,\frac{1}{(n+1)!}\int\big(K_{r,n+1}^{(0)\dagger}
\phi^{(n)}\big)(x_0,\dots,x_n)\Big)\,\rho(x_0)\,\dots\,\rho(x_n)\,dx_0
\,\dots\,dx_n\,.
\qqq
In the above expressions, $K_{r,x_0}^{(0)\dagger}$ denotes, in a slightly 
abusive notation, the adjoint operator to $K_r^{(0)}$,
\qq
K_r^{(0)\dagger} = \big(b 
-2k_BT \left(\nabla \ln \rho_{inv}\right)\big)\cdot\nabla_x - k_BT \nabla^2_x\,,
\qqq
see Eq.\,(\ref{eq:K_roperator}), in the action on a function of variable $x_0$.
Similarly, $K_{r,n+1}^{(0)\dagger}$ denotes the operator
\qq\label{eq:on_tensor_product}
K_{r,n+1}^{(0)\dagger}= 
\sum_{m=0}^n K_{r,x_m}^{(0)\dagger}
\qqq
acting on functions of $n$-tuples $(x_0,\dots,x_n)$.
Note that the differential operator $K_r^{(0)\dagger}(x)$ is 
the generator of a diffusion process with drift 
$-b(x)+2k_BT(\nabla\ln\rho_{inv})$ which is the time-reversal 
of the original single particle diffusion.
This operator has a one-dimensional kernel composed of constants and 
a one-dimensional cokernel composed of functions proportional to $\rho_{inv}$. 
For the same reasons, the operator $K_{r,n+1}^{(0)\dagger}$
has also a one dimensional kernel composed of constants and
a one dimensional cokernel composed of functions proportional to
$\rho_{inv}(x_0)\cdots\rho_{inv}(x_n)$. We shall use those properties
below.  

In order to compute the right hand side of the hierarchy, we shall treat 
separately the cases with $n=1$ and with $n>1$. For $n=1$, we have
\qq\label{eq:Chap3-hierarchy-J0-rhs-n1}
&&\mathcal{S}^{(1)}[\mathcal{F}^{(0)}]\cr
&&=\ -\,\int\left[ (\nabla^2 V)(x-y) 
+ \frac{_1}{^2}(\nabla V)(x-y)\cdot 
\big( \nabla \ln \rho_{inv}(x) - \nabla \ln \rho_{inv}(y) \big) 
\right]\rho(x)\,\rho(y)\,dx\,dy\,,\quad\qquad
\qqq
where we have made the integrand symmetric under the exchange of $x$ and $y$. 
Comparing Eqs.\,(\ref{eq:Chap3-hierarchy-J0-lhs-n1}) and 
(\ref{eq:Chap3-hierarchy-J0-rhs-n1}), we obtain a differential equation 
for $\phi^{(1)}$,
\qq\label{eq:Chap3-perturbative-J0-diff-eq-n1}
\big(K_{r,2}^{(0)\dagger}\phi^{(1)}\big)(x,y) = - 2 
(\nabla^2 V)(x-y) - \, (\nabla V)(x-y)\cdot\big(\nabla \ln \rho_{inv}(x) 
- \nabla \ln \rho_{inv}(y) \big)\,. \quad
\qqq
For $n>1$, the right hand side of the hierarchy can be written in terms of the
kernels $\phi^{(k)}(x_0,\dots,x_k)$ as
\qq\label{eq:Chap3-hierarchy-J0-rhs-ngr1}
&&\mathcal{S}^{(n)}[\mathcal{F}^{(1)},\dots,\mathcal{F}^{(n-1)}]\cr
&&=\ \frac{1}{n!}\int(\nabla_{x_0} V)(x_0-x_n)\cdot (\nabla_{x_0} 
\phi^{(n-1)})(x_0,\dots,x_{n-1})\,\rho(x_0)\,\cdots\,\rho(x_n)\,dx_0\,
\cdots\,dx_n\,\cr
&&-\,\sum_{k=1}^{n-1} \frac{k_BT}{k!(n-k)!}\,\int
(\nabla_{x_k} \phi^{(k)})(x_0,\dots,x_k) \cdot 
(\nabla_{x_k} \phi^{(n-k)})(x_k,\dots,x_n)\,\rho(x_0)\,\cdots\,\rho(x_n)\,
dx_0\,\cdots\,dx_n\,.\qquad\qquad
\qqq
Then, for any $n>1$, we obtain the differential equations 
\qq\label{eq:Chap3-perturbative-J0-diff-eq-ngr1}
&\big(K_{r,n+1}^{(0)\dagger}\phi^{(n)}\big)(x_0,\dots,x_n) = 
(n+1)!\,\,\textrm{Sym}\bigg[ \frac{1}{n!}\,(\nabla_{x_0} V)(x_0-x_n)
\cdot \nabla_{x_0} \phi^{(n-1)}(x_0,\dots,x_{n-1})\cr
&\hspace{2cm}-\,k_BT \sum\limits_{k=1}^{n-1} \frac{1}{k!(n-k)!} (\nabla_{x_k} \phi^{(k)})(x_0,\dots,x_k) 
\cdot (\nabla_{x_k} \phi^{(n-k)})(x_k,\dots,x_n) \bigg],
\qqq
where $\textrm{Sym}\left[\,\cdot\,\right]$ stands for the symmetrization 
of the argument with respect to variables $x_0,\dots,x_n$. 
Eqs.\,(\ref{eq:Chap3-perturbative-J0-diff-eq-n1}) and 
(\ref{eq:Chap3-perturbative-J0-diff-eq-ngr1}) may be solved iteratively
for the kernels $\phi^{(n)}(x_0,\dots,x_n)$ provided that their right
hand sides are orthogonal to the one-dimensional cokernels of operators 
$K_{r,n+1}^{(0)\dagger}$ spanned by the product functions 
$\rho_{inv}(x_0)\cdots\rho_{inv}(x_n)$. These are the solvability conditions 
$\mathcal{S}^{(n)}[\rho_{inv}]=0$ that were discussed in 
Sec.\,\ref{sec:Chap2-uncentred} in the finite-dimensional context,
see Eq.\,(\ref{eq:Chap-2-naive-solvability}). Such conditions are
easy to check in the lower orders but are difficult to prove order by 
order. They are, however, implied by the existence of the perturbative 
expansion centered at the attractors, as discussed in 
Sec.\,\ref{sec:Chap2-uncentred}. Observe finally that $\phi^{(n)}$ 
are defined by the above equations only up to constants that are in 
the kernel of $K_{r,n+1}^{(0)\dagger}$. This ambiguity 
just leads to shifts in constants $C^{(n)}$ appearing in 
Eq.\,(\ref{eq:Chap3-polyn-J0}) which are not relevant
(remember that densities $\rho$ are normalized). 

We have greatly simplified the problem of calculating the quasi-potential 
perturbatively around the free particles dynamics. Instead of solving 
the functional differential equations (\ref{eq:Chap3-HJ-expanded-naive}) 
we have to solve the partial differential equations for the kernels $\phi^{(n)}$ 
defined by Eq.\,(\ref{eq:Chap3-polyn-J0}). 
These differential equations are (\ref{eq:Chap3-perturbative-J0-diff-eq-n1}) 
for the $1^{\rm st}$ order contribution and 
(\ref{eq:Chap3-perturbative-J0-diff-eq-ngr1}) for the higher orders. 
Solving them is a rather simple numerical problem. In the next 
section, we discuss the results obtained in the case of the 
Shinomoto-Kuramoto model.

\subsubsection{Expansion around $J=0\,$: \,results for the Shinomoto-Kuramoto 
model}\label{sub:chapt4-J0Kur}

Let us now specify the discussion to the case of the Shinomoto-Kuramoto model 
introduced in Sec.\,\ref{sec:Chap3-Kuramoto-typical} and describe some
explicit results. In this case, Eq.\,(\ref{eq:Chap3-perturbative-J0-diff-eq-n1})
for the $\phi^{(1)}$ reduces to the relation
\qq\label{eq:Chap3-perturbative-J0-diff-eq-n1-Kuramoto}
\big(K_{r,2}^{(0)\dagger}\phi^{(1)}\big)(\theta,\vartheta) 
= S^{(1)}(\theta,\vartheta),
\qqq
where
\qq
S^{(1)}(\theta,\vartheta) = - 2\cos(\theta-\vartheta) - \, 
\sin(\theta - \vartheta)\cdot\left[ \partial_{\theta} \ln \rho_{inv}(\theta) 
- \partial_{\vartheta} \ln \rho_{inv}(\vartheta) \right] \,,
\qqq
and
\qq
K_r^{(0)\dagger}= \left[F-h\sin\theta -2k_BT 
\left(\partial_{\theta} \ln \rho_{inv}\right) \right] 
\partial_{\theta} - k_BT \partial^2_{\theta}\,.
\qqq
We observe that identity (\ref{eq:Chap3-perturbative-J0-diff-eq-n1-Kuramoto}) 
is a Lyapunov equation and several techniques can be employed to solve it, 
see for example \cite{kitagawa1977,nardini2013,nardini2012kinetic}. The one 
we have chosen is to perform an expansion on the eigenfunctions of 
operators $K^{(0)}_r$ and $K_r^{(0)\dagger}$. 

Let us first set the notations. For  $k=1,\dots$, we denote by $u_k$ 
the eigenfunctions of $K^{(0)}_r$ and by $v_k$ those of 
$K_r^{(0)\dagger}$. As $K^{(0)}_r$ is not self-adjoint 
with respect to the $L^2$ scalar product, these eigenfunctions are not 
connected by complex conjugation but we may assume that the corresponding
eigenvalues are:
\qq\label{eq:Chap4-J0-eigen}
\mathcal{K}^{(0)}_ru_k = \alpha_ku_k
\qquad\qquad \textrm{and}\qquad\qquad  
K^{(0)\dagger}_r\, v_k = \overline{\alpha}_k\,v_k
\qqq
where $ \overline{\alpha}_k$ denotes the complex conjugate of $ \alpha_k$.
$K^{(0)}_r$ is a Fokker-Planck operator. Zero is its simple
eigenvalue and $\rho_{inv}$ is the corresponding eigenfunction. 
We choose it to be equal to $u_1$. Other eigenvalues of $K^{(0)}_r$
have strictly positive real parts. Similarly, we choose $v_1=1$. 
It corresponds to the unique zero mode of $K^{(0)\dagger}_r$. 
We can also assume that the eigenfunctions are mutually 
orthonormal\footnote{This is possible whenever the eigenvalues 
$\alpha_k$ are all distinct. We have checked numerically 
that this is indeed the case.} in the following sense 
\qq\label{eq:Chap4-J0-eigen-orth}
\int\,\overline{u_k(\theta)}\,v_l(\theta)\,d\theta = \delta_{kl}\,.
\qqq
It is then simple to show that 
\qq\label{eq:Chap4-perturbativeJ0-eig-expansion}
\phi^{(1)}(\theta,\vartheta) = \sum_{(k,l)\neq (1,1)} \frac{S^{(1)}_{kl}}{\alpha_k
+\alpha_l}\,u_k(\theta)\,u_l(\vartheta)\,,
\qqq
where 
\qq
S^{(1)}_{kl}= \int \,\overline{v_k(\theta)}\,\overline{v_l(\vartheta)}\,
S^{(1)}(\theta,\vartheta)\,
d\theta\,d\vartheta
\qqq
(note that $\alpha_k+\alpha_l\not=0$ for $(k,l)\not=(1,1))$.
One has $S_{11}=0$ and we have chosen $\phi^{(1)}$ to be orthogonal to the 
kernel of $K_{r,2}^{(0)\dagger}$ composed of
the constant functions. The calculation of constant $C^{(1)}$ of
Eq.\,(\ref{eq:Chap3-polyn-J0}) will be discussed afterwards.

Eq.\,(\ref{eq:Chap4-perturbativeJ0-eig-expansion}) permits to obtain 
numerically the quasi-potential at the $1^{\rm st}$ order in $J$. One 
first looks for an approximation of the eigenfunctions and eigenvalues 
of $K^{(0)}_r$ and of $K_r^{(0)\dagger}$ by expanding 
Eqs.\,(\ref{eq:Chap4-J0-eigen}) in the Fourier modes and truncating the 
hierarchy in order to deal with matrices. The calculation of 
$\phi^{(1)}$ through Eq.\,(\ref{eq:Chap4-perturbativeJ0-eig-expansion}) is 
then easily performed. We checked that very few Fourier modes have to be 
retained ($\gtrsim 20$) to obtain an excellent approximation of 
eigenfunctions and eigenvalues. Even less eigenfunctions ($\gtrsim 5$) 
need to be employed in the calculation in order to obtain quite
accurate results.

Although the scheme proposed here is a bit more involved than other methods, 
such as a direct Fourier expansion of 
Eq.\,(\ref{eq:Chap3-perturbative-J0-diff-eq-n1-Kuramoto}), it is much more 
powerful than the latter. Indeed, in the method proposed here, we only need 
to diagonalize the matrices that approximate operators 
$K^{(0)}_r$ and $K^{(0)\dagger}_r$ that 
act on the space of functions of one periodic variable. Instead, attempting 
to directly solve Eq.\,(\ref{eq:Chap3-perturbative-J0-diff-eq-n1-Kuramoto}) 
by expanding on Fourier modes results in diagonalizing the operator 
$K_{r,2}^{(0)\dagger}$ acting on the space of 
functions of two variables. Even more important is the fact that the method 
proposed here is fully generalizable, without increasing the computational 
complexity to obtain higher perturbative orders (encoded in kernels
$\phi^{(n)}$) for the quasi-potential. In contrast, in the direct Fourier 
expansion, one needs to diagonalize matrices of dimension $(k_{max})^n$ 
where $k_{max}$ is the number of retained Fourier modes, making the problem 
practically intractable already for small $n$.

\begin{figure}
\includegraphics[scale=0.5]{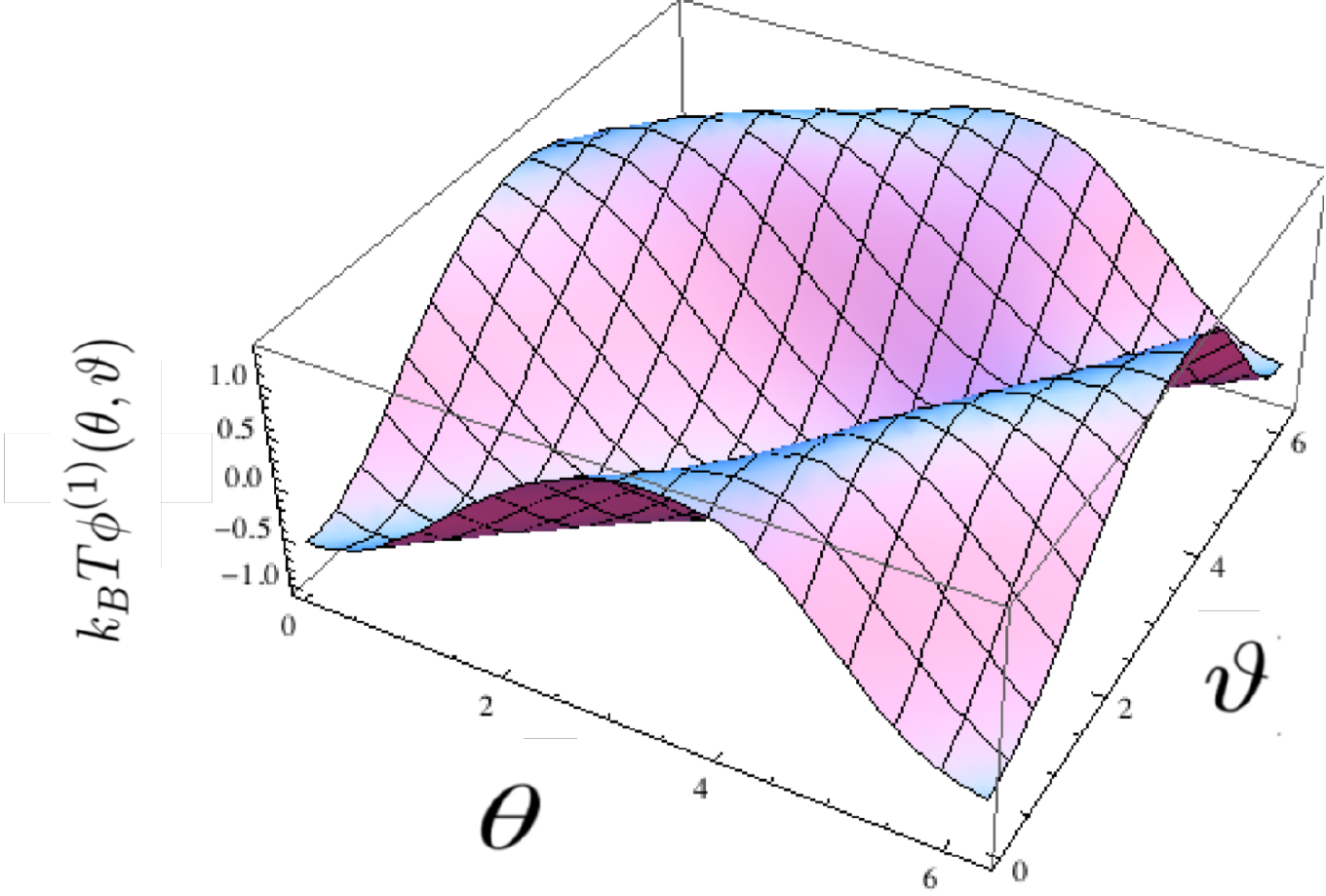}\hspace{2cm}
\includegraphics[scale=0.16]{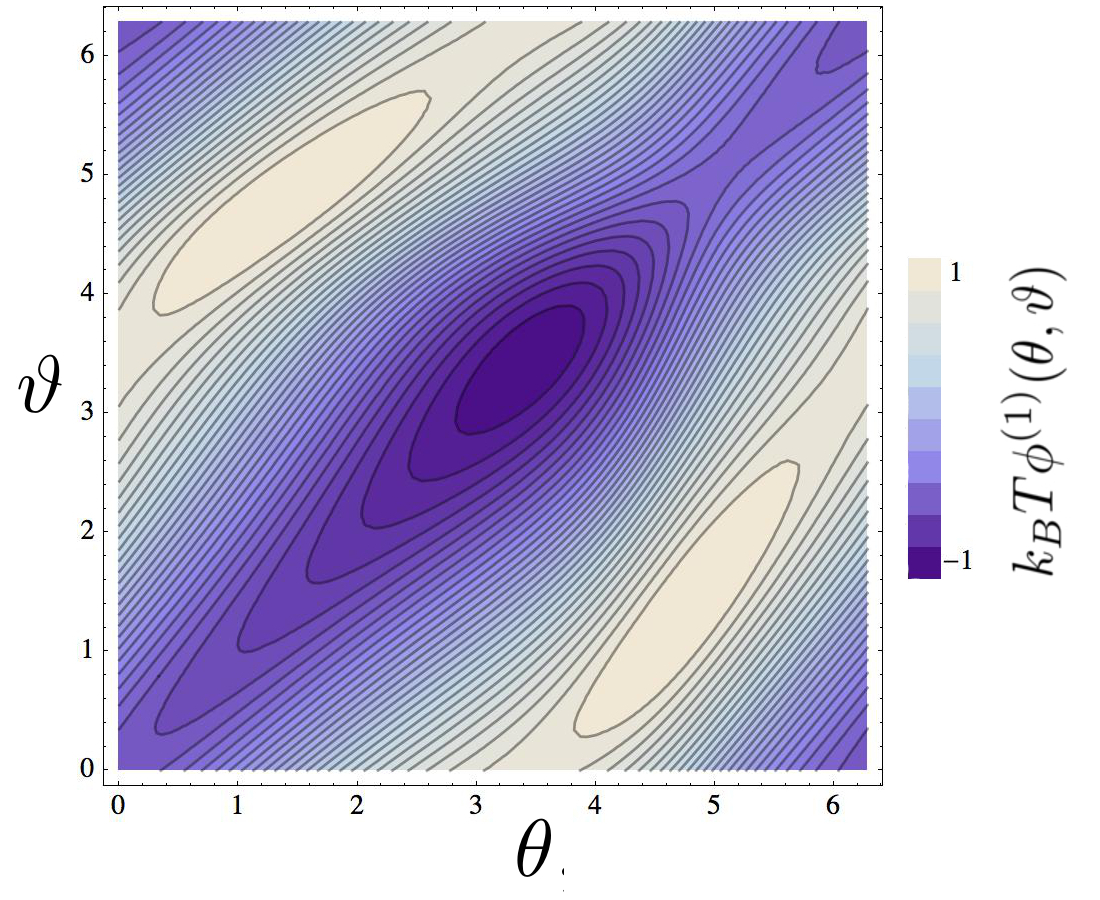}
\caption{On the left: the kernel $k_BT \phi^{(1)}(\theta,\vartheta)$ that 
permits to obtain the first-order correction to the quasi-potential of the 
Shinomoto-Kuramoto model in the perturbative expansion around the free 
particle dynamics with $J=0$. The values of parameters are 
$F=0.2$, $T=0.3$, $h=0.5$. This figure was obtained by numerically solving 
Eq.\,(\ref{eq:Chap3-perturbative-J0-diff-eq-n1-Kuramoto}) as detailed in 
the text. On the right: the same function but in a contour plot.
}\label{fig:Chap3-perturbative-Kuramoto-n1-J0}
\end{figure}

\begin{figure}
\includegraphics[scale=0.2]{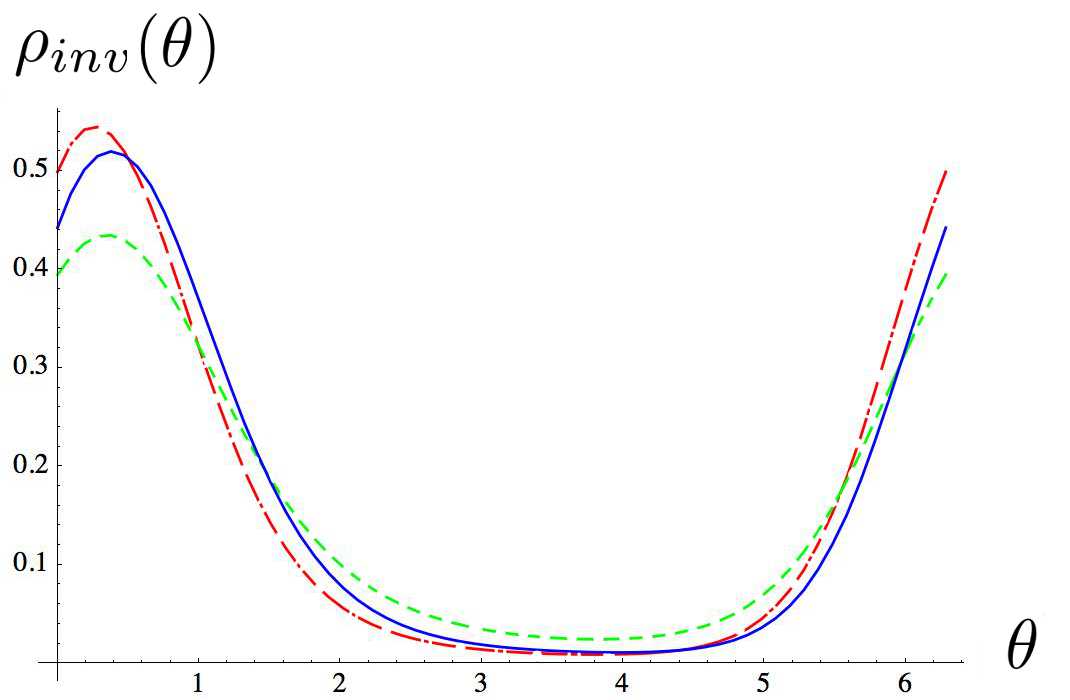}\hspace{0.5cm}
\includegraphics[scale=0.2]{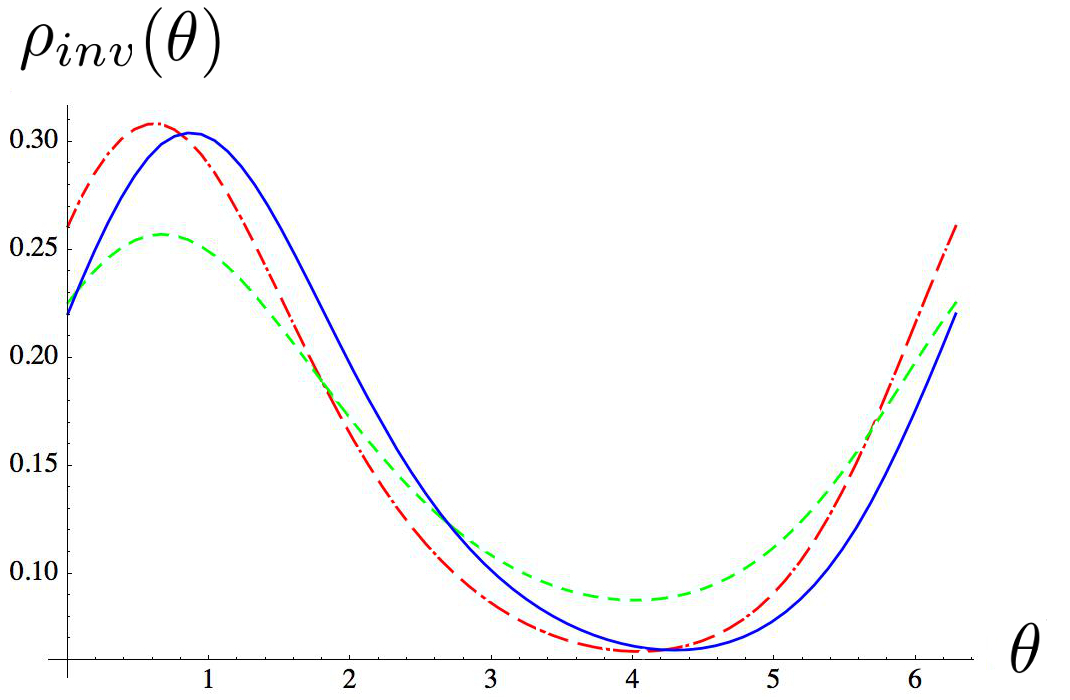}\hspace{0.5cm}
\caption{Stationary state of the Shinomoto-Kuramoto model for $J=0.3$ 
obtained from the perturbative expansion of the quasi-potential (blue continuous curve) 
compared to the exact result (red dashed dotted curve) given by the 
Eq.\,(\ref{eq:Chap-3-rho-inv}) and to the stationary state for the system 
with $J=0$ (green dashed curve). The values of parameters are $F=0.2$, $T=0.3$, 
$h=0.5$ on the left and $F=0.2$, $T=0.2$, $h=0.15$ on the right. 
The plots show that the $1^{\rm st}$ order correction to the stationary state 
strongly improves the prediction.}
\label{fig:Chap3-perturbative-Kuramoto-n1-J0-ss}
\end{figure}

We report the results for the $1^{\rm st}$ order correction $\phi^{(1)}$ to the 
quasi-potential in Fig.\,\ref{fig:Chap3-perturbative-Kuramoto-n1-J0} for
some typical choice of parameters for which the McKean-Vlasov equation 
has a single stationary stable solution (namely $F=0.2$, $T=0.3$, $h=0.5$). 
A rather non-trivial result emerges. As a check of the accuracy of the 
perturbative expansion, we also compared the exact stationary state of 
the system $\rho^J_{inv}$ for $J>0$ to the one, $\rho_{inv,1}$, obtained 
by imposing that the functional derivative of
\qq
\mathcal{F}^{(0)}+J\mathcal{F}^{(1)} + \gamma \left(\int \rho(\theta)\,
d\theta\,-\,1\right)
\qqq
vanishes. This condition implies, with the use of \eqref{eq:func_der_0} and
\eqref{eq:Chap3-polyn-J0}, that
\qq\label{for_rho1}
\rho_{inv,1}(\theta)= \frac{1}{Z}\,
\rho_{inv}(\theta)\exp\left[ -J\int 
\rho_{inv,1}(\vartheta)\,\phi^{(1)}(\theta,\vartheta)\,d\vartheta\right],
\qqq
where $Z=\ee^{\gamma}$ fixes the normalization of $\rho_{inv,1}$.
Recall that the true mean-field stationary density $\rho^J_{inv}$ minimizes
the quasi-potential $\mathcal{F}^J[\rho]$ with the value at the minimum
equal to zero. Eq.\,(\ref{for_rho1}) may be very easily solved numerically 
with an iterative scheme, once $\phi^{(1)}$ is known; we have verified 
that this procedure converges after few iterations. In 
Fig.\,\ref{fig:Chap3-perturbative-Kuramoto-n1-J0-ss}, 
we also compare $\rho_{inv,1}$ with the exact result obtained from 
Eq.\,(\ref{eq:Chap-3-rho-inv}), which shows that already at the $1^{\rm st}$ 
order 
the stationary state for the system with $J>0$ is much better approximated 
by $\rho_{inv,1}$ than by the stationary state at $J=0$. As for the constant
$C^{(1)}$ from Eq.\,(\ref{eq:Chap3-polyn-J0}) with $n=1$, it will
be convenient to fix it so that the quasi-potential calculated to the 
$1^{\rm st}$ order $\mathcal{F}^{(0)}+J\mathcal{F}^{(1)}$ has the minimal value 
equal to zero, as does the complete quasi-potential. This is achieved 
by setting 
\qq
C^{(1)}=-J^{-1}\mathcal{F}^{(0)}[\rho_{inv,1}]-\mathcal{F}^{(1)}[\rho_{inv,1}]\,,
\qqq
where on the right hand side $\mathcal{F}^{(1)}$ is taken homogeneous
in $\rho$ (i.e. with $C^{(1)}$ set to zero).

Once the quasi-potential at order $J$ is known, we can also compute the 
large deviations rate functions for the one-particle observables 
$\frac{1}{N}\sum_{n=1}^N g(\theta_n)$ where $g$ is a given function. 
In the following, we consider the particularly relevant examples of 
the magnetization along $x$ and $y$ axis corresponding, respectively, to 
$g=\cos$ and $g=\sin$. What follows may be straightforwardly generalized 
to any choice of $g$. We denote by $\mathcal{I}_x(\sigma)$ 
(resp. $\mathcal{I}_y(\sigma)$) the rate function for the probability 
of observing a value of magnetization $m_x\sim\sigma$ 
(resp. $m_y\sim \sigma$). Let us consider $\mathcal{I}_x$ (the case of 
$\mathcal{I}_y$ may be analyzed similarly). To the $1^{\rm st}$ order in
$J$, we have
\qq\label{eq:Chapt3-Ix}
\mathcal{I}_x(\sigma)\,
=\,\min_{\rho}\,\left\{\mathcal{F}^{(0)}[\rho]+J\mathcal{F}^{(1)}[\rho]\right\}\,,
\qqq
where the minimum is taken over all positive functions $\rho$ with unit 
integral respecting the constraint $\int \cos(\theta)\,\rho(\theta)\,d\theta
=\sigma$. The profile realizing the minimum will be called $\rho_{opt}$ 
in the following.

\begin{figure}
\includegraphics[scale=0.2]{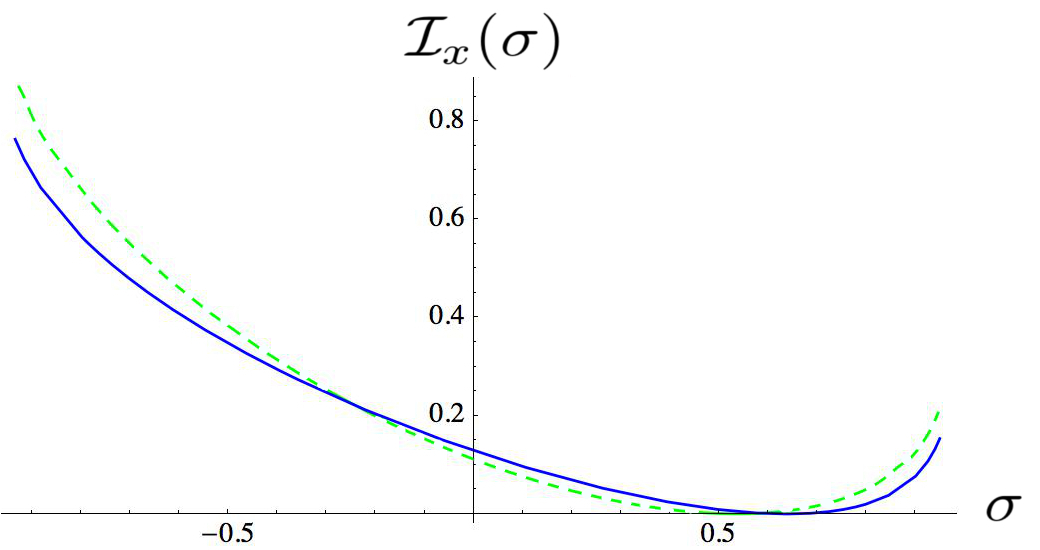}\hspace{1cm}
\includegraphics[scale=0.2]{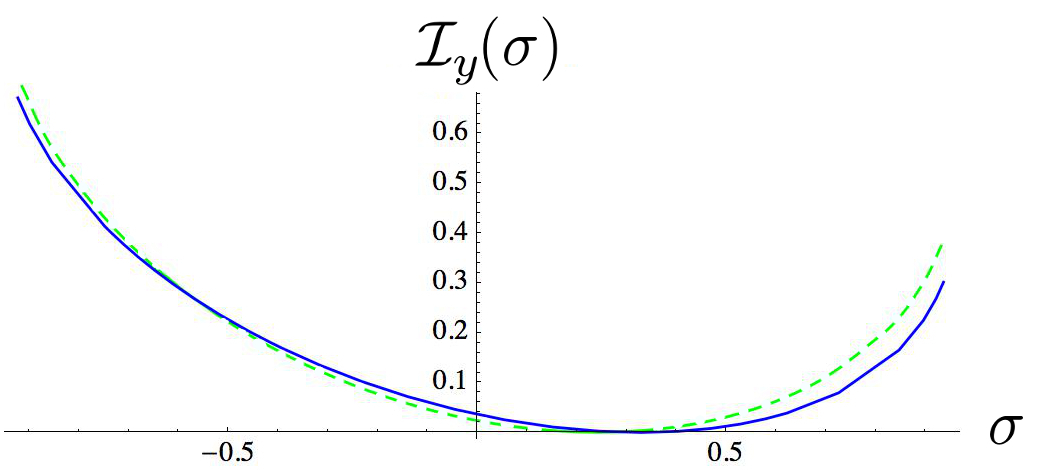}\\

\includegraphics[scale=0.2]{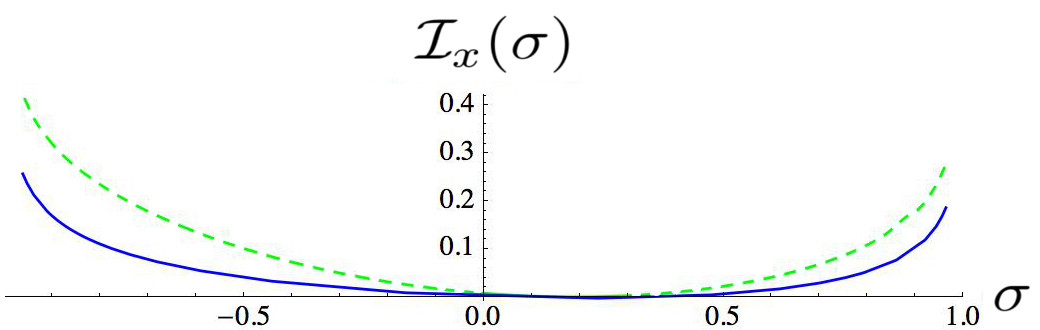}\hspace{1cm}
\includegraphics[scale=0.2]{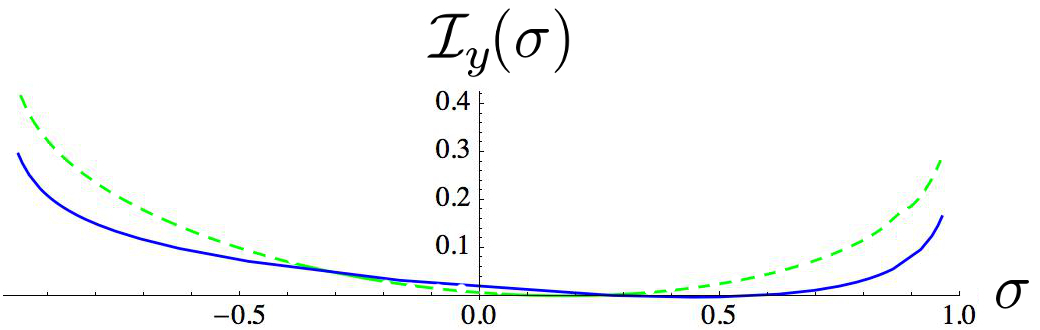}
\caption{Rate functions for the probability of fluctuations of the $x$ 
(on the left) and $y$ (on the right) component of magnetization. 
The results of our perturbative analysis to the $1^{\rm st}$ order in $J$ 
(blue continuous lines) are compared with those of the unperturbed model with $J=0$ 
(green dashed lines). The value of parameters used are the same as in 
Fig.\,\ref{fig:Chap3-perturbative-Kuramoto-n1-J0} in the upper plots 
while they are set to $F=0.2$, $T=0.2$, $h=0.15$ in the lower plots.}
\label{fig:Chap3-perturbative-Kuramoto-m-J0-rate-func}
\end{figure}

The minimization problem in Eq.\,(\ref{eq:Chapt3-Ix}) can be solved 
introducing two Lagrange multipliers $\gamma_1$ and $\gamma_2$ associated, 
respectively, to the total mass and to the magnetization constraints. 
A simple calculation shows that the minimizing profile satisfies the relation
\qq\label{eq:chap3--rate-function-m}
\rho_{opt}(\theta) = \frac{1}{Z}\,\rho_{inv}(\theta)\,
\exp\left[ -\gamma_2\cos\theta - J
\int\phi^{(1)}(\theta,\vartheta)\,\rho_{opt}(\vartheta)\, d\vartheta \right],
\qqq
where $Z$ normalizes $\rho_{opt}$. As for the case of the most probable state, 
the above self-consistent equation can be solved iteratively. We first fix 
$\gamma_2$ and iteratively solve Eq.\,(\ref{eq:chap3--rate-function-m}) 
normalizing the iterated solution to unity at each step. Once the optimal 
profile has been obtained, we then calculate the corresponding value of 
$m_x$. As the initial condition we used the optimal profile for $J=0$. We have 
checked that the iterative scheme converges in few steps.
The results obtained for $\mathcal{I}_x$ and $\mathcal{I}_y$ are plotted in 
Fig.\,\ref{fig:Chap3-perturbative-Kuramoto-m-J0-rate-func} for the same choice 
of parameters as those employed in 
Fig.\,\ref{fig:Chap3-perturbative-Kuramoto-n1-J0-ss}. In the plots, the 
results obtained at the $1^{\rm st}$ order in $J$ are compared with those obtained 
for $J=0$ in the whole range of possible magnetizations.  We observe that 
the $1^{\rm st}$ order corrections are significant. Fox example, the probability 
of fluctuations of $m_x$ smaller than the typical value is diminished 
with respect to the $J=0$ case but the probability of very rare fluctuations 
is instead increased.

We conclude by noting that, as already discussed, it would be simple to 
extend the calculations of the quasi-potential and of rate functions 
of observables to higher orders by computing with the method  
employed in this section the higher kernels $\phi^{(n)}$. 
We do not pursue, however, this direction here, as our scope was rather 
to show that the perturbative approach can lead to explicit results 
than to analyze in details the Shinomoto-Kuramoto model.

\subsubsection{Expansion around the equilibrium dynamics ($F=0$ or $h=0$)}

When the drift $b(x)$ may be written as $b=-\nabla U$ for some $U$, 
the quasi-potential can be given explicitly because detailed balance holds. We note that the Dean equation (\ref{eq:Chap3-Dean-general}) corresponding to this case has also been derived in \cite{chavanis1,chavanis3}, where it is called stochastic Smoluchowski equation.
One has in this case
\qq
\mathcal{F}[\rho]= \frac{1}{{k_BT}}\int\rho(x)\left(U(x)+\frac{J}{2}\int V(x-y)\,
\rho(y)\,dy\right)dx\,+\,\int \rho(x)\ln\rho(x)\,dx\,.
\qqq
We did not develop the perturbative expansion close to the detailed balance 
because obtaining explicit results would be quite complex. Indeed, the 
equations for the quasi-potential at order $n$ are functional differential
equations and not, as in the case of a perturbation close to $J=0$, 
ordinary differential equations. Even though results may be explicitly 
obtained by a discretization, the problem is not computationally 
straightforward.

\subsubsection{Taylor expansion close to a stationary state}
\label{sub:chapt4-Taylor}

Let us discuss now the Taylor expansion of the quasi-potential $\mathcal{F}$ 
around a stable stationary solution $\rho_{inv}$ of the McKean-Vlasov 
equation. As we have seen in Sec.\,\ref{subsec:Chap2-around-attractor} for 
finite dimensional systems, this expansion may be viewed as a particular case 
of the perturbative expansion of the quasi-potential for a parameter-dependent 
system. In this setting, the role of the unperturbed quasi-potential is 
played by the quadratic approximation to $\mathcal{F}$ that, consistently, 
will be denoted in the following by $\mathcal{F}^{(0)}$. Then, once 
$\mathcal{F}^{(0)}$ is known, any higher order correction may, in principle, 
be calculated. After extending here the discussion of 
Sec.\,\ref{subsec:Chap2-around-attractor} to a general diffusion with
mean-field interaction (\ref{eq:Chap3-eq-of-motion}), we discuss in 
Appendix \ref{app:Taylor} how explicit results could be obtained in 
a similar way as for the perturbative expansion around $J=0$.

We would like to calculate the quasi-potential $\mathcal{F}[\tilde\rho]$ for  
$\tilde{\rho}=(\rho_{inv}+\rho)$ as a power series in $\rho$. Note
$\rho$ has then to have the vanishing integral.
In order to use the same notations as in the finite dimensional setting, 
see Sec.\,\ref{subsec:Chap2-around-attractor}, we introduce the 
$\lambda$-dependent system defined by 
\qq
&&\mathcal{K}^{\lambda}[\tilde{\rho}]=\frac{1}{\lambda}\mathcal{K}\left[\rho_{inv}
+\lambda\rho\right]\\
&&\mathcal{Q}^{\lambda}[\tilde{\rho}]=\mathcal{Q}\left[\rho_{inv}
+\lambda\rho\right]\\
&&\mathcal{F}^{\lambda}[\tilde{\rho}]=\frac{1}{\lambda ^2}\mathcal{F}
\left[\rho_{inv}+\lambda\rho\right].
\qqq
For any $n$ we thus have
\qq\label{eq:Chap4-taylor-K}
&&\mathcal{K}^{(0)}[\rho] = -\nabla\cdot\left[ \rho\,b - 
J\rho_{inv} \left(\nabla V\ast \rho\right) - 
J \rho \left(\nabla V \ast \rho_{inv}\right) \right] + k_BT \nabla^2 \rho\,,\\
&&\mathcal{K}^{(1)}[\rho] = J\nabla\cdot\left[ \rho \left(\nabla V\ast \rho\right) 
\right],\\
&&\mathcal{K}^{(n)}[\rho] = 0\,\qquad \qquad\qquad{\rm for}\qquad\quad n\geq 2\,.
\qqq
Clearly,  $\mathcal{K}^{(0)}[\rho]=R_{\rho_{inv}}\rho$, where $R_{\rho_{inv}}$ denotes 
the linearized Fokker-Planck operator. Moreover,
\qq\label{eq:Chap4-taylor-Q}
&&\hspace{-3.9cm}\mathcal{Q}^{(0)}[\rho] = -k_BT\,\nabla\cdot\rho_{inv}\nabla\,,\\
&&\hspace{-3.9cm}\mathcal{Q}^{(1)}[\rho] = -k_BT\,\nabla\cdot\rho\nabla\,,\\
&&\hspace{-3.9cm}\mathcal{Q}^{(n)}[\rho] = 0\qquad\ \qquad\qquad{\rm for} 
\qquad\quad n\geq 2\,.
\qqq
Recall that the computation of the quadratic order of $\mathcal{F}$ 
corresponds, in the above $\lambda$-dependent system, to the calculation 
of $\mathcal{F}^{(0)}$. Moreover, with these notations the hierarchy 
of equations to solve in order to obtain the Taylor expansion  
of the quasi-potential close to $\rho_{inv}$ is given precisely by 
Eqs.\,(\ref{eq:Chap3-HJ-expanded-F0-naive}) and 
(\ref{eq:Chap3-HJ-expanded-naive}) with $\mathcal{K}^{(n)}$ and 
$\mathcal{Q}^{(n)}$ expressed by the above formulae. 

We shall search for the solution of this hierarchy in the 
form of homogeneous polynomials
\qq\label{eq:n+2_exp}
\mathcal{F}^{(n)}[\rho] = \frac{1}{(n+2)!} \, 
\int \varphi^n(x_0,\dots,x_{n+1})\,\rho(x_0)\,\cdots\,\rho(x_{n+1})\, 
dx_0\,\cdots\,dx_{n+1}\,,
\qqq
where $\varphi^n(x_0,\dots,x_{n+1})$ are symmetric kernels on which we shall
impose the relations 
\qq\label{integral_rel}
\int\varphi^n(x_0,\dots,x_{n+1})\,dx_0=0\,.
\qqq

As in the other cases, once the quadratic order $\mathcal{F}^{(0)}$ is known, 
the higher orders can be obtained by using the general solution 
to the hierarchy, see Eq.\,(\ref{eq:Chap3-solution-HJ-hierarchy}). 
We thus concentrate here only on  $\mathcal{F}^{(0)}$. We introduce the 
operator $\Phi$ 
\qq\label{eq:chap3-taylor-Phi-def}
\Phi[\rho](x) = \int \varphi^0(x,y)\,\rho(y)\, dy
\qqq
that will be considered as acting in the space $H_0$ of functions
with vanishing integral and $L^2$ scalar product.  
Note that $\Phi$ is a symmetric operator. From 
Eq.\,(\ref{eq:Chap3-HJ-expanded-F0-naive}), we obtain the operator identity
\qq\label{1st_op_eq}
\Phi R_{\rho_{inv}} + R_{\rho_{inv}}^\dagger \Phi = 2k_BT\,\Phi (\nabla\cdot\rho_{inv}\nabla) 
\Phi\,,
\qqq
where $R_{inv}$ is also viewed as an operator in $H_0$. Assuming that $\Phi$ is 
invertible, as it must be from the fact that $\mathcal{F}^{(0)}$ is positive 
definite, we infer that 
\qq
R_{\rho_{inv}} \Phi^{-1} + \Phi^{-1} R_{\rho_{inv}}^\dagger = 2k_BT(\nabla\cdot\rho_{inv}\nabla)\,,
\qqq
which is the infinite dimensional analogue of Eq.\,(\ref{eq:ABQ}). 
Its solution is given by the relation
\qq\label{eq:Chap3-perturbative-taylor-phi-sol}
\Phi^{-1} = -2k_BT\int_0^{\infty}\ee^{t R_{\rho_{inv}} }
\,(\nabla\cdot\rho_{inv}\nabla)\,\ee^{t R_{\rho_{inv}}^+}\,dt\,,
\qqq
see Eq.\,(\ref{eq:B-1}). Similarly to the finite-dimensional case,
convergence of the integral is assured if $\rho_{inv}$ a non-degenerate
stable stationary solution of the McKean-Vlasov equation, i.e. if
the spectrum of $R_{\rho_{inv}}$ on the space of functions with vanishing integral 
in contained in the complex half-plane with negative real part. Once we know
the invertible operator $\Phi^{-1}$, the function $\varphi^0(x_0,x_1)$ 
is extracted as the kernel of its inverse.

With an argument analogous to the one used in Sec.\,\ref{subsec:Chap2-crot-cal-exponents} 
for the finite dimensional systems, one can show that 
Eq.\,(\ref{eq:Chap3-perturbative-taylor-phi-sol}) implies that the covariance 
of density fluctuations diverges with the mean-field exponent.when approaching 
a codimension-one bifurcations of the McKean-Vlasov dynamics. 
We note that the latter result was obtained rigorously in \cite{arous1990b}.

Let us now consider the $0^{\rm th}$-order fluctuation dynamics. We have
\qq\label{eq:zeroth_order_fl}
\mathcal{K}^{(0)}_r[\rho]=R_{inv}\rho-2k_BT\,\nabla\cdot\rho_{inv}\nabla\Phi\rho
\,=\,-\Phi^{-1}R_{inv}^\dagger\Phi\,\rho\,\equiv\,K_r^{(0)}\rho\,.
\qqq
$K_r^{(0)}$ is again an operator in $H_0$ that is invertible 
and has spectrum in the half-plane with positive imaginary
part (it should not be confused with $K_r^{(0)}$ considered 
in Sec.\,\ref{sub:chapt4-J0}). With Ansatz (\ref{eq:n+2_exp}), the left hand 
side of Eqs.\,(\ref{eq:Chap3-HJ-expanded-naive}) takes now the form
\qq
&&\int\frac{\delta\mathcal{F}^{(n)}}{\delta\rho(x)} \,
\mathcal{K}^{(0)}_r[\rho](x)\,dx\cr
&&\qquad= \,\frac{1}{(n+2)!}\int\big(K_{r,n+2}^{(0)\dagger}\,
\varphi^{n}\big)(x_0,\dots,x_{n+1})\Big)\,\rho(x_0)\,
\cdots\,\rho(x_{n+1})\,dx_0\cdots dx_{n+1}
\qquad
\qqq
in the notation of (\ref{eq:on_tensor_product}). 
As for their right hand side, it follows from Eq.\,(\ref{eq:Chap3-Sn-1}) that
\qq
&&\mathcal{S}^{(n)}[\mathcal{F}^{(0)},\dots,\mathcal{F}^{(n-1)}]\cr
&&\qquad=\frac{1}{(n+2)!}\int
s^{n}[\varphi^0,\dots,\varphi^{n-1}](x_0,\dots,x_{n+1})\,\rho(x_0)\,\cdots\,\rho(x_{n+1})\,
dx_0\cdots dx_{n+1}\,,\qquad\qquad\quad
\qqq
where the symmetric kernels $s^{(n)}[\varphi^0,\dots,\varphi^{n-1}](x_0,\dots,x_{n+1})$ are 
expressed in terms of $\varphi^k$ with $k<n$ and again satisfy the constraint 
(\ref{integral_rel}). Thus in terms of the kernels, 
Eqs.\,(\ref{eq:Chap3-HJ-expanded-naive}) become the identities
\qq\label{eq:for_varphis}
\big(K_{r,n+2}^{(0)\dagger}\,\varphi^{n}\big)(x_0,\dots,x_{n+1})
=s^{n}[\varphi^0,\dots,\varphi^{n-1}](x_0,\dots,x_{n+1})(x_0,\dots,x_{n+1})
\qqq
which may be solved iteratively 
since the operators $\mathcal{K}_{r,n+2}^{(0)\dagger}$ are invertible 
on $H_0^{\otimes(n+2)}$ (note that the constraint (\ref{integral_rel}) 
characterizes the symmetric functions in $H_0^{\otimes(n+2)}$). 
Further details are left to the reader, but in Appendix \ref{app:Taylor}, we 
briefly discuss how $\varphi^0$, as well as higher order kernels, could be 
explicitly calculated numerically, leaving out of this work the explicit 
numerical implementation of the proposed algorithm.


\nsection{Conclusions}\label{sec:conclusions}

The main aim of this paper was to discuss a perturbative approach to 
the calculation of the quasi-potential in parameter dependent stochastic 
dynamical systems. As we have seen, the Taylor expansion of the 
quasi-potential around an attractor of the deterministic dynamics 
was also covered by our theory.
 
The approach developed in the paper may be summarized as it was already
done in the introduction. The perturbative expansion breaks the loop
connecting the quasi-potential and the instanton dynamics, with
one needed to obtain the other. At any perturbative 
order the quasi-potential can be computed just from the knowledge of 
the instanton dynamics for the unperturbed problem and the lower order 
results. Indeed, our approach gives explicit formulae that permit 
to iteratively calculate any order of the power series expansion 
of the quasi-potential in the perturbative parameter. Explicit 
results can be obtained once the instanton dynamics for the unperturbed 
problem may be analytically or numerically solved. 
 
We have first developed the theory for finite dimensional systems, where 
the mathematical details can be easily handled precisely. Our strategy 
was to perturbatively solve the Hamilton-Jacobi equation which, as it is 
well known \cite{freidlin1984}, has the quasi-potential as the unique 
non-trivial solution under the hypothesis that the later is smooth enough. 
It is also known \cite{day1985} that, at least in a neighborhood of the 
attractor, this is indeed the case, while further away singularities may 
occur, see for example \cite{jauslin1987,day1985,maier1996,kamenev2011,graham1985,graham1995}. An investigation on how to deal practically with singularities of the quasi-potential is left for the future. We also note that the perturbative study of transition rates between basin of attraction requires a specific approach, as was done for instance in the reference \cite{smelyanskiy1997}, using Melnikov's method
for perturbations of Hamiltonian systems.

In the second part of the paper, we moved our attention to many body 
systems. In particular, we considered $N$ particles that undergo an
overdamped diffusion, interact with mean field conservative forces, 
and are driven out of equilibrium by an external drift, see 
Eq.\,(\ref{eq:Chap3-eq-of-motion}). In this case it was 
possible to (formally) derive a fluctuating hydrodynamics, with the noise term 
proportional to $1/\sqrt{N}$, that describes in a closed way the evolution 
of the empirical density of particles. This fluctuating 
hydrodynamics is known in the literature under the name of the Dean 
equation \cite{dean1996}. 
In the limit when $N\to \infty$, the empirical density obeys then a 
deterministic PDE, the McKean-Vlasov equation \cite{mckean1966}. This 
provided a formal but quick way to recover the results on propagation of chaos 
that appeared long ago in the mathematical literature.

In the case of a specific mean-field system, the Shinomoto-Kuramoto model
of coupled rotators, we showed that such an approach can be employed to obtain 
explicit results. We first discussed MacKean-Vlasov's dynamics describing
the behavior of the system in the $N\to\infty$ limit, characterizing 
analytically or semi-analytically all the stationary solutions of the system 
and their stability. Already this very simple out-of-equilibrium mean-field 
system displays quite a complex phase diagram with bifurcation lines and 
several attractors (stationary states as well as limit cycles). Albeit 
the Shinomoto-Kuramoto model has been studied in literature by numerical 
techniques, our analytical results characterizing all its stationary 
states are original.

After discussing the $N=\infty$ limit, we concentrated on large deviations 
around it. This was done by studying the Dean equation for $N$ large but 
finite. Applying formal techniques from field theory (the Martin-Siggia-Rose 
representation and the WKB approximation), we showed that fluctuations of the 
empirical density 
are described by a generalization of the Freidlin-Wentzell theory to infinite 
dimensional systems. This permitted to recover formally rigorous results 
that already appeared in mathematical literature \cite{dawson1987,dawson1987b} 
and to make a connection with the Macroscopic Fluctuation Theory 
\cite{bertini2014}.

We subsequently generalized our perturbative approach to the calculation 
of quasi-potential to the case of fluctuating mean-field hydrodynamics. 
For the Shinomoto-Kuramoto model, we explicitly computed the quasi-potential 
close to the uncoupled particle dynamics to the  $1^{\rm st}$ order in the 
mean-field 
coupling strength $J$ and found the corresponding rate function for 
the probability of fluctuations of single-particle observables such as 
the magnetization. Our results seem to provide the first explicit calculation 
of the quasi-potential for diffusions with mean field interaction driven 
out of equilibrium. Higher orders in $J$ could be also computed without 
much difficulty. As our aim here was mainly to develop the general theory and 
to illustrate that it can be employed to obtain explicit results, we have 
not pursed such calculations further but discussed instead how one could 
similarly obtain explicit results for the Taylor expansion of the 
quasi-potential around the stationary states. 

The perturbative approach developed in the present paper seems quite general 
and we expect it to be useful in several other problems. For example, 
it may be employed 
in the future to study other long-range interacting systems which are driven 
out of equilibrium by different mechanisms, such as stochastic forces which 
are spatially correlated. This setting is natural for systems whose 
constituents exhibit long-range interactions (such as plasmas or gravitational 
systems) but are exposed to stochastic external fields (electric for example) 
mainly acting on a given spatial scale. The typical behavior of this class 
of system was analyzed with kinetic theory in 
\cite{nardini2012short,nardini2012kinetic} and numerical simulations 
have shown how the phase transitions are modified when departing from the 
detailed balance \cite{nardini2012kinetic}. The perturbative approach exposed
here may give an analytical insight to such problems. This direction is presently 
pursued by one of the authors.

Similarly, applications of such perturbative methods could be found in the 
study of large deviations for $2d$ turbulence problems, which are relevant 
for climate modelling \cite{nardini2013,bouchet2014stochastic,bouchet2011control,bouchet2014langevin,laurie2015computation}. Even though these are quite 
academic questions \cite{bouchet2014langevin} because realistic models of 
$2d$ turbulence and those arising from climate models are far from the 
perturbative regime, this approach can give valuable physical insight as 
it is difficult to handle the rare events in realistic systems.

Finally, while large deviations for the empirical density of mean-field 
interacting diffusions appeared in literature long ago, mathematical 
literature does not cover fluctuations of the empirical current. The formal 
approach through the Dean equation presented in this paper clearly permits 
to go in this direction. Some results on the current fluctuations that 
we have obtained already, as well as further explicit results on the 
calculation of the quasi-potential for the Shinomoto-Kuramoto model 
and their comparison with results obtained from direct numerical simulations 
are left for a future publication.



\section*{Acknowledgments}
The authors are grateful to J. Barr\'e, P.-H. Chavanis, R. Chetrite 
and H. Touchette for useful discussions and for providing numerous 
references to the literature. The latter were also provided by the 
anonymous referees whom we thank for that contribution and for critical 
comments. C. Nardini acknowledges M. Cates for several discussions in the final stage of this work.
This research, and the position of C. Nardini, were funded 
through the ANR grant ANR STOSYMAP (ANR-2011-BS01-015), and partially 
(C. Nardini) by the EPSRC grant Nr. EP/J007404.  F. Bouchet acknowledges 
funding from the European Research Council under European Union's 
Seventh Framework Programme (FP7/2007-2013 Grant Agreement no. 616811)
 
\appendix
\nsection{Smooth stationary solutions of the Shinomoto-Kuramoto model 
at $T=0$}\label{app:T0}

In Sec.\,\ref{sub:Chap3-typical-T0}, we have discussed simple 
delta-like stationary solutions of the McKean-Vlasov dynamics for 
the Shinomoto-Kuramoto model at $T=0$. They described the particles 
accumulated at the fixed point of single particle drift. These are however 
not the only stationary solutions for vanishing temperature and also 
smooth solutions are present, which, however, are linearly unstable 
in the ferromagnetic model considered in this paper ($J>0$). We briefly discuss 
such smooth stationary solutions at $T=0$ and their stability in this appendix.

For $T=0$, Eq.\,(\ref{eq:chap3-stat-sol-curr}) implies that
\qq\label{eq:smooth_stat_sol_T=0}
\rho_{inv}(\theta+\theta_0)=\frac{\sqrt{1-\left(\frac{y}{F}\right)^2}}
{2\pi\left(1-\frac{y}{F}\sin \theta\right)}
\qqq
provided that $0\leq y<F$ and the self-consistency relation 
(\ref{eq:Chap3-self-consistency-f}) holds taking now the form
\qq
\left(\frac{y}{J}\right)^2+\bigg(\frac{_{1-\sqrt{1-\left(\frac{y}{F}\right)^2}}}
{^{\frac{y}{F}}}\bigg)^{\hspace{-0.05cm}2}=\left(\frac{h}{J}\right)^2.
\qqq
The left hand side of the last equation grows monotonically from $0$ 
at $y=0$ to $(F/J)^2+1$ at $y=F$. Hence there is a unique solution 
for $0\leq y<F$ if and only if $F^2+J^2> h^2$. We infer that there 
is a unique stationary smooth solution at $T=0$ when the latter 
constraint holds and none otherwise.

To examine the stability of such solutions, we examine the eigenfunctions 
$\delta\rho_{\mu}(\theta+\theta_0)$ with vanishing integral of the 
linearized Fokker-Planck operator $R_{\rho_{inv}}$ of 
(\ref{eq:chap-3-linearized-FP-may}) at $T=0$. We may write 
$\delta\rho_{\mu}(\theta+\theta_0)
=\partial_{\theta}\delta f_{\mu}(\theta)$, where $\delta f_{\mu}$ is periodic 
on $[0,2\pi]$. The eigenequation takes the form
\qq
\big(R_{\rho_{inv}}\partial_{\theta}\delta f_{\mu}\big)(\theta)
&=&-\partial_{\theta}\Big((F-y\sin\theta)\,
\partial_{\theta}\delta f_{\mu}(\theta) 
- Y_1\,\rho_{inv}(\theta+\theta_0)\sin\theta + Y_2\,\rho_{inv}(\theta+\theta_0)
\cos\theta\Big)\cr
&=&\mu\,\partial_{\theta}\delta f_{\mu},\,\ 
\qqq
where 
\qq
Y_1=J\int_0^{2\pi}\sin\theta\,\,\delta f_{\mu}(\theta)\,d\theta\,,\qquad  
Y_2=-J\int_0^{2\pi}\cos\theta\,\,\delta f_{\mu}(\theta)\,d\theta\,.
\qqq
We may now define a new operator $S$ such that
\qq
\partial_{\theta}\big(S\,\delta f\big)
=R_{\rho_{inv}}\partial_{\theta}\delta f_{\lambda}\,.
\qqq
Explicitly,
\qq
\big(S\,\delta f\big)(\theta)=-(F-y\sin\theta)\,\partial_{\theta}\delta f(\theta) 
- Y_1\rho_{inv}(\theta+\theta_0)\,\sin\theta + Y_2\rho_{inv}(\theta+\theta_0)
\,\cos\theta\,. 
\qqq
Clearly, the diagonalization of $R_{\rho_{inv}}$ on the subspace of functions 
$\delta g$ with vanishing integral is equivalent to the diagonalization of 
$S$.

A direct calculation shows that the $3$-dimensional subspace $V_3$ spanned 
by functions 
\qq\label{eq:3x3matrix}
f_1(\theta)=\rho_{inv}(\theta+\theta_0)\,\sin\theta\,,\qquad 
f_2(\theta)=\rho_{inv}(\theta+\theta_0)\,\cos\theta\,,\qquad
f_3(\theta)=\rho_{inv}(\theta+\theta_0)
\qqq
is invariant for $S$. In particular, the representation of $S$ on such 
subspace is given by the matrix 
\qq \left( \begin{array}{ccc}
J\,\frac{1-\sqrt{1-\big(\frac{y}{F}\big)^2}}{\big(\frac{y}{F}\big)^2} & F & J\,\frac{1-\sqrt{1-\big(\frac{y}{F}\big)^2}}
{\frac{y}{F}} \\
-F & J\Bigg(1-\frac{1-\sqrt{1-\big(\frac{y}{F}\big)^2}}{\big(\frac{y}{F}\big)^2}\Bigg) & -y \\
0 & -y & 0 \end{array} \right).
\qqq 
A little algebra shows that the above matrix has one zero eigenvalue which 
corresponds to the eigenvector $(-y,0,F)$, i.e. to the constant function
\qq
\delta f(\theta)\,=\,\rho_{inv}(\theta+\theta_0)\,(F-y\sin{\theta})\,=\,
\frac{\sqrt{F^2-y^2}}{2\pi}\,.
\qqq
In this case $\delta g=\partial_\theta \delta f=0$. Consequently, the zero eigenvalue 
does not occur for the operator $R_{\rho_{inv}}$. The other two eigenvalues 
of the matrix (\ref{eq:3x3matrix}) are given by 
\qq
\lambda_\pm\,=\,\frac{J}{2}\,\pm\,\sqrt{\left(\frac{J}{2}\right)^2\,-\,J^2
\frac{_{\sqrt{1-\big(\frac{y}{F}\big)^2}}}
{^{\Big(1+\sqrt{1-\big(\frac{y}{F}\big)^2}\Big)^2}}\,-\,F^2+y^2} \,.
\label{lpm}
\qqq
of which at least one has positive real part if $J>0$.
It follows that for $J>0$, the stationary solution 
(\ref{eq:smooth_stat_sol_T=0}) is linearly unstable.


\nsection{Algorithm for the Taylor expansion of the quasi-potential 
close to a stationary state}\label{app:Taylor}

We discuss briefly in this Appendix how the Taylor expansion of the 
quasi-potential of the Shinomoto-Kuramoto model may be numerically calculated. 
This will be possible, in principle, at all the orders of the Taylor expansion. 

Let us first concentrate on the quadratic term $\mathcal{F}^{(0)}$. 
In order to explicitly calculate kernel $\varphi^0(\theta,\vartheta)$,
we proceed similarly as in Sec.\,\ref{sub:chapt4-J0Kur} and consider
the eigenfunctions of operators $R_{\rho_{inv}}^\dagger$ and $R_{\rho_{inv}}$
acting in space $H_0$: 
\qq\label{eq:R_egenstates}
R_{\rho_{inv}}^\dagger u_{k}=\alpha_k\,u_{k}\qquad 
\textrm{and}\qquad R_{\rho_{inv}} v_{k}=\overline{\alpha}_k\,v_{k}\,.
\qqq
Assuming $\alpha_k$ to be different, we impose the orthogonality relations
(\ref{eq:Chap4-J0-eigen-orth}). 
Eigenfunctions $u_k$ form a basis of $H_0$ but they are not orthogonal.
Similarly for $v_k$. The kernel of the quadratic term of the quasi-potential 
may be represented as
\qq\label{eq:chap3-may-phi2}
\varphi^0(\theta,\vartheta)=\sum_{k,l}\Phi_{kl}\,v_k(\theta)\,
\overline{v_l(\vartheta)}\,
\qqq
where $\Phi_{kl}$ is defined by
\qq
\Phi_{kl}= \int_{0}^{2\pi}\overline{u_k(\theta)}\,d\theta\int_{0}^{2\pi}
\varphi^0(\theta,\vartheta)\,u_l(\vartheta)\,d\vartheta=\langle u_k|\Phi u_l
\rangle\,.
\qqq
The problem is thus reduced to the calculation of $\Phi_{kl}$. Now, 
from Eq.\,(\ref{eq:Chap3-perturbative-taylor-phi-sol}), we can easily 
compute the matrix elements of $\Phi^{-1}$ in the basis $u_k$. 
Indeed, from Eq.\,(\ref{eq:Chap3-perturbative-taylor-phi-sol}),
\qq
(\Phi^{-1})_{kl}\equiv\langle u_k|\Phi^{-1}u_l\rangle
= \frac{2k_BT}{\alpha_k^*+\alpha_l}\int d\theta\,
\overline{(\partial_\theta u_k)(\theta)}\,\rho_{inv}(\theta)\,
(\partial_\theta u_l)(\theta)\,.
\qqq
Note, however, that the matrix $(\Phi^{-1} )_{kl}$ is not the inverse
of $\Phi_{kl}$ because the basis formed by $u_k$ is not orthonormal.
This problem can be handled with simple linear algebra. We introduce 
the matrix of scalar products 
\qq
P_{kl}=\langle u_k|u_l\rangle
\qqq
whose inverse will be denoted by $(P^{-1})_{kl}$. Then, we construct
another matrix 
\qq
(B^{-1})_{kl}\equiv\sum_i(P^{-1})_{ki}\,(\Phi^{-1})_{il}\,
\qqq
with inverse $B_{kl}$. Finally,
\qq
\Phi_{kl}=\sum_i P_{ki}\,B_{il}\,.
\qqq
Once $\Phi_{kl}$ is known, it is straightforward to write the
kernel $\varphi^0$ in the real space using Eq.\,(\ref{eq:for_varphis}).

We conclude by observing that, once $\varphi^0$ is known, one could also 
numerically evaluate the higher order kernels $\varphi^n$ of the Taylor 
expansion  by solving Eq.\,(\ref{eq:for_varphis}) using
the fact that the basis $\Phi v_k$ is composed of eigenstates
of operators $K^{(0)\dagger}_r=-\Phi\,R_{\rho_{inv}}\Phi^{-1}$, see 
Eqs.\,(\ref{eq:zeroth_order_fl}) and (\ref{eq:R_egenstates}).



\end{document}